\documentclass[usenatbib,useAMS]{mn2e}

\usepackage{natbib}
\usepackage{epsfig}

\usepackage{amsmath}
\usepackage{amssymb}

\usepackage{subfigure}

\newcommand{\msun}{{\rm M_{\odot}}}

\newcommand{\fpt}{\text{fp}}
\newcommand{\co}{0}

\newcommand{\OmegaF}{\Omega}

\newcommand{\RedeclareMathOperator}[2]{\renewcommand{#1}{}\let#1\relax\DeclareMathOperator{#1}{#2}}

\newcommand{\const}{\text{const}}

\newcommand{\myvec}{\boldsymbol}

\newcommand{\htau}{{{\hat\tau}}}
\newcommand{\hn}{{{\hat n}}}
\newcommand{\hphi}{{{\hat\varphi}}}

\newcommand{\p}{\partial}

\newcommand{\rhs}{r.h.s.\hbox{}}

\newcommand{\simplefrac}[2]{#1/#2}

\newcommand{\abs}[1]{\ensuremath{\lvert#1\rvert}}

\newcommand{\Abs}[1]{\ensuremath{\left\lvert#1\right\rvert}}

\newcommand{\ie}{i.e.\hbox{}}
\newcommand{\eg}{e.g.\hbox{}}

\newcommand\aj{\rmfamily{AJ}}%
\newcommand\apj{\rmfamily{ApJ}}%
\newcommand\apjl{\rmfamily{ApJ}}%
\newcommand\apss{\rmfamily{Ap\&SS}}%
\newcommand\aap{\rmfamily{A\&A}}%
\newcommand\mnras{\rmfamily{MNRAS}}%
\newcommand\prd{\rmfamily{Phys.~Rev.~D}}%
\newcommand\nat{\rmfamily{Nature}}%
\newcommand\physrep{\rmfamily{Phys.~Rep.}}%
\title[Simulations of Ultrarelativistic Magnetodynamic GRB Jets]
      {Simulations of Ultrarelativistic Magnetodynamic Jets from Gamma-ray Burst Engines}

\author[A. Tchekhovskoy, J.~C. McKinney, \& R. Narayan]
   {Alexander Tchekhovskoy,$^1$\thanks{\hbox{E-mail: atchekho@cfa.harvard.edu~(AT);} \hbox{jmckinne@stanford.edu~(JCM);~~~~~~} \hbox{rnarayan@cfa.harvard.edu~(RN)}}
    Jonathan C. McKinney,$^2$\footnotemark[1]
    Ramesh Narayan$^3$\footnotemark[1]\\
  $^1$Harvard-Smithsonian Center for Astrophysics, 60 Garden Street, MS 10,
    Cambridge, MA 02138, USA \\
  $^2$Kavli Institute for Particle Astrophysics and Cosmology, Stanford University, P.O. Box 20450, MS 29,
Stanford, CA 94309; Chandra Fellow \\
  $^3$Institute for  Theory and Computation, Harvard-Smithsonian Center for Astrophysics,
  60 Garden Street, MS 51, Cambridge, MA 02138, USA
  }

\date{Accepted . Received ; in original form }

\pagerange{\pageref{firstpage}--\pageref{lastpage}} \pubyear{2008}

\begin{document}
\label{firstpage}

\maketitle

\begin{abstract}

Long-duration gamma-ray bursts (GRBs) require an engine capable of driving a jet of
plasma to ultrarelativistic bulk Lorentz factors of up to several
hundred and into narrow opening angles of a few degrees.
We use
global axisymmetric stationary solutions of magnetically-dominated
(force-free) ultrarelativistic jets to test whether the popular magnetic-driving paradigm can
generate the required Lorentz factors and opening angles.
Our global solutions are obtained via time-dependent
relativistic ideal magnetodynamical numerical simulations which follow the
jet from the central engine to beyond six orders of magnitude in
radius. Our model is primarily motivated by the collapsar model, in
which a jet is produced by a spinning black hole or neutron star and
then propagates through a massive stellar envelope.

We find that the size of the presupernova
progenitor star and the radial profile of pressure inside the star
determine the terminal Lorentz factor and opening
angle of the jet.
At the radius where the jet breaks out of the star, our well-motivated fiducial model
generates a Lorentz factor $\gamma\sim 400$ and a half-opening angle
$\theta_j\sim 2^\circ$, consistent with observations of many
long-duration GRBs. Other models with slightly different parameters
give $\gamma$ in the range $100$ to $5000$ and $\theta_j$ from
$0.1^\circ$ to $10^\circ$, thus reproducing the range of properties
inferred for GRB jets.
A potentially observable feature of some of our
solutions is that the maximum Poynting flux in the jet is found at
$\theta \sim \theta_j$ with the jet power concentrated in a hollow
cone, while the maximum in the Lorentz factor occurs at an angle
$\theta$ substantially smaller than $\theta_j$ also in a hollow cone.
We derive approximate
analytical formulae for the radial and angular distribution of $\gamma$
and the radial dependence of $\theta_j$.  These formulae reproduce the simulation
results and allow us to predict the outcome of models beyond
those simulated.  We also briefly discuss applications to active
galactic nuclei, X-ray binaries, and short-duration GRBs.
\end{abstract}


\section{Introduction}\label{sec_intro}

Models of long-duration gamma-ray bursts (GRBs) require the ejected plasma to move
at ultrarelativistic speeds in order to avoid the compactness problem
\citep{pir05}.  The Lorentz factor required can be as high as
$\Gamma\sim 400$ \citep{lithwick_lower_limits_2001}, which
necessitates a relativistic engine capable of launching plasma with an
enormous amount of energy per particle.  Achromatic `jet breaks' in
the GRB afterglow imply a finite geometric opening angle
$\theta_j\sim$ a few degrees for a typical long-duration GRB
\citep{frail01,pir05,zeh_long_grb_angles_2006}.  Combined with the observed fluence and the
known distance to the source, this gives a typical event energy of
$\sim 10^{51}$ ergs, comparable to the kinetic energy released
in a supernova explosion.

An ideal engine for producing ultrarelativistic jets with small
opening angles, low baryon contamination, and high total energies is a
rapidly rotating black hole threaded by a magnetic field and accreting
at a hyper-Eddington rate \citep{nar92,le93,mr97}.  In such a model,
the black hole launches an electromagnetically pure jet via the
Blandford-Znajek effect \citep{bz77}.  More recently, millisecond
magnetars have been seriously considered as another possible source of
magnetically-dominated outflows \citep{usov_magnetar_grbs_1992,lyutikov_em_model_grbs_2006,um07,buc07}.  The standard
alternative to this magnetic-driving paradigm is neutrino annihilation
\citep{woosley_gamma_ray_bursts_1993,mac99}, but this mechanism
probably does not produce sufficient luminosity to explain most GRBs \citep{pwf99,dm02}.

Rapidly rotating black holes or millisecond magnetars are thought to
be the products of core-collapse
\citep{woosley_gamma_ray_bursts_1993,pac98} or binary collisions of
compact objects \citep{nar92,narpirankumar01}.  For failed supernovae, the
black hole or magnetar is surrounded by an accretion disc whose corona
and wind affect the jet structure through force-balance between the
jet and the surrounding gas.  In any core-collapse event the jet
must penetrate the stellar envelope which can significantly modify the
structure of the jet~\citep{woosley_gamma_ray_bursts_1993,
mac99,alo00,narpirankumar01,zhang_relativistic_jets_2003,
aloy_relativistic_outflows_2007}.  Indeed, as we demonstrate in this paper, it is
likely the case that the properties of the stellar envelope
\emph{determine} the Lorentz factor and opening angle of the jet.

We seek to understand how magnetized rotating compact objects can
launch jets that become sufficiently ultrarelativistic and narrow in
opening angle to produce long-duration GRBs.  To achieve this goal we use the
relativistic ideal magnetohydrodynamical (MHD) approximation, which is
a valid approximation for much of the GRB jet (e.g., see
\citealt{mckinney2004}).  The primary difficulty has been in obtaining
a self-consistent global model of the jet that
connects the compact object at the center to large distances where the
observed radiation is produced.  In the context of the collapsar
model, this means we need a model that goes all the way from the black hole or neutron
star at the center to beyond the outer radius of the Wolf-Rayet progenitor
star.

In the past the MHD approximation has been used in numerous analytical
efforts aimed at understanding the physics behind acceleration and
collimation of relativistic jets.  The MHD equations for stationary
force balance are highly non-linear, and so analytical studies have
been mostly confined to special cases for which the equations can be
simplified, \eg\ for particular field
geometries~\citep{bz77,bes98,bes06}, for asymptotic
solutions~\citep{appl_camenzind_asymptotic_1993,begelman_asymptotic_1994,lovelace_poynting_jets_2003,fo04}, or for
self-similar solutions that allow
variable separation~\citep{cl94,con95,vla03a,vla03b,nar07}.
Semi-analytical methods using finite element, iterative relaxation, or shooting techniques
have also been used to find jet solutions, such as for spinning
neutron stars~\citep{camenzind_finite_element_1987,lovelace_jet_pulsar_2006} and black holes~\citep{fendt97}.
Such analytical studies are useful since sometimes one finds families of solutions that
provide significant insight into the general properties of
solutions~\citep[\eg,][]{nar07}.

Time-dependent simulations complement analytical studies by allowing
one to investigate a few models with much less restrictive
assumptions.  In particular, much recent insight into the
accretion-jet phenomenon has been achieved within the framework of
general relativistic magnetohydrodynamics (GRMHD) via time-dependent
numerical simulations~\citep{dev03,mck04,mck05, kom05,
aloy_relativistic_outflows_2007}. Indeed, numerical simulations of accretion
have successfully reproduced collimated relativistic outflows
with Lorentz factors reaching 10~\citep{mck06jf}.  Within the
collapsar model, GRMHD simulations show that magnetized jets can be
produced during core-collapse~\citep{mizuno04,liu_shapiro_collapse_2007,
barkov_stellar_2007,stephens_shapiro_collapse_2008}.
However, no MHD simulation of core-collapse has yet demonstrated the
production of an ultrarelativistic jet.  Computationally, such simulations are prohibitively expensive due to
the need to resolve vast spatial and temporal scales while at the
same time modeling all the physics of the black hole, the accretion
disc, the disc wind, and the stellar envelope.  A more practical approach,
one that we take in this paper,
is to replace the real problem with a simplified and
idealized model and to explore this model over the large spatial and
temporal scales of interest for long-duration GRBs.
Such an approach will hopefully
demonstrate how ultrarelativistic jets can be produced and will help
us assess the applicability of the mechanism to the collapsar model.

In the present work we obtain global solutions of ultrarelativistic
magnetically-dominated jets via time-dependent numerical MHD
simulations in flat space-time (no gravity).  We focus on the relativistic \emph{magnetodynamical}, or \emph{force-free},
regime~\citep{gol69,okamoto_magnetic_braking4_1974,
blandford_accretion_disk_electrodynamics_1976,lov76,bz77,macdonald_thorne_bh_forcefree_1982,
fendt_collimation1_1995,kom01,kom02,mck06ffcode}, which corresponds to
a magnetically-dominated plasma in which particle rest-mass and
temperature are unimportant and are ignored.  This is a reasonable
model for highly magnetized flows~\citep{bz77, mck06jf}.
The model parameters we consider are motivated by presupernova stellar
models~\citep{mac99,alo00,zhang_relativistic_jets_2003,
  heger_presupernova_2005, zhang_fallback_2007} and GRMHD simulations
of turbulent accretion discs~\citep{mck04,mck07a,mck07b}.  We compare
our numerical solutions against self-similar solutions derived
by~\citet{nar07} and obtain simple physically-motivated formulae for
the variation of the Lorentz factor, collimation angle, and Poynting flux
along the
axis of the jet and across the face of the jet. Based upon our
simulations and analytical scalings, we suggest that the terminal
Lorentz factor of GRB jets is determined by the size and radial
pressure profile of the progenitor star rather than the initial
magnetization, for a large range of initial magnetizations.

In~\S\ref{sec_motivation} we discuss the problem setup and give a
brief overview of our numerical method.
In~\S\ref{sec_results} we present the numerical results and interpret
them in terms of analytical scalings.  In~\S\ref{sec_comparison_to_other_work}
we make a comparison to other models.  In~\S\ref{sec_astrophysical_applications} we
discuss astrophysical applications of our models, and
in~\S\ref{sec_conclusions} we give a brief conclusion.  Readers
who are not interested in the details may wish to look at
Figures~\ref{fig_collapsarcartoon}~--~\ref{fig_nu075m025a15field}
and to read~\S\ref{sec_astrophysical_applications}. In
Appendix~\ref{sec_approximate_analytic_jet_solution} we introduce an
approximate model of force-free jets and present a comprehensive
discussion of the analytical properties of these jets. In Appendix~\ref{sec_fluid_speed}
we discuss the kinematics of any (dynamically unimportant) plasma that
may be carried along with a force-free jet.

\section{Motivation, Problem Setup And Numerical Method}
\label{sec_motivation}

\begin{figure}
\begin{center}
\epsfig{figure=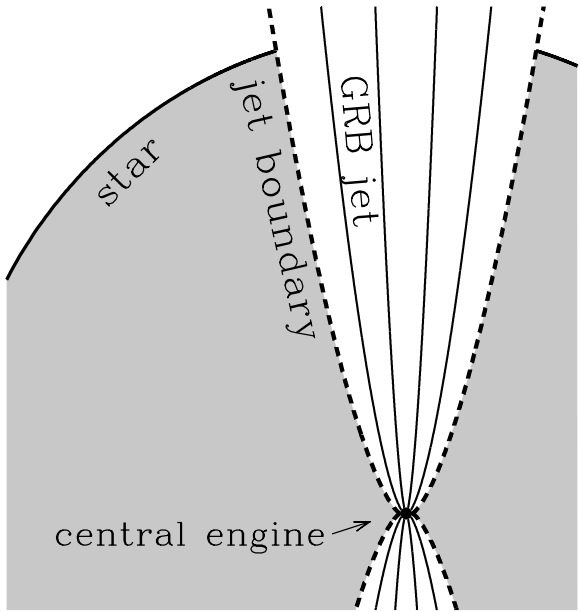,width=0.9 \columnwidth}
\end{center}
\caption{Cartoon of the large-scale structure of a GRB source (not
to scale).  The
major elements are a central engine which launches a polar
magnetically-dominated ultrarelativistic jet, and a gaseous stellar
envelope (gray shading) which confines the jet.  The central engine may be an
accreting rapidly rotating black hole or a millisecond magnetar.  For a failed supernova,
there could also be a disc wind which may additionally confine the jet.  }
\label{fig_collapsarcartoon}
\end{figure}

As depicted in Figure~\ref{fig_collapsarcartoon}, a crucial aspect of
the collapsar GRB model is that the central engine must produce a
jet powerful enough to penetrate the stellar envelope.  The
interaction between the stellar envelope and the jet is found to be quite
complex in time-dependent hydrodynamical numerical simulations of jets
injected at an inlet within the presupernova core
\citep{alo00,zhang_relativistic_jets_2003,mor06,wang_relativistic_2007}.  These
simulations show that the jet collimates and accelerates as it pushes
its way through the confining stellar envelope, thus suggesting that the envelope
plays a crucial role in determining the opening angle and Lorentz
factor of the flow that emerges from the star.  If the collapsar
system forms an accreting black hole, then the ultrarelativistic jet
may be accompanied by a moderately relativistic disc wind that may
provide additional collimation for the jet \citep{mck05,mck06jf}.
We note that the larger the radius of the progenitor star and/or the denser
the stellar envelope, the more energy is required for the jet to
have to penetrate the stellar envelope and reach
the surface of the star.
\citet{burrows_snjets_2007} find that a relativistic jet
in the collapsar scenario may be preceded by a non-relativistic
precursor jet that might clear the way for the second,
relativistic jet.
In the magnetar
scenario, the stellar envelope is the primary
collimating agent \citep{um07}.  Eventually, one
would like to study individually the collapsar and
other models of GRBs.  However, at the basic level, all models are fundamentally
similar, since they involve a central magnetized rotating compact
object that generates a jet confined by some medium (\eg, Fig.~\ref{fig_collapsarcartoon}).

Figure~\ref{fig_bhcartoon} shows our idealized approach to this problem.
We reduce the various scenarios to a rigidly rotating star of unit radius
surrounded by a razor-thin differentially rotating disc.  Magnetic
field lines thread both the star and the disc. We identify the field
lines emerging from the star as the `jet' and the lines from the disc
as the `wind.'  Both components are treated within the magnetodynamical, or force-free,
approximation.  That is, they are taken to be perfectly conducting,
and we assume that the plasma inertia and thermal pressure can be neglected.
In terms of the standard magnetization parameter
$\sigma$~\citep{mic69,goldreich_julian_stellar_winds_1970}, we assume $\sigma \rightarrow \infty$.
In this idealized model, the force-free disc wind plays the
role of the stellar envelope (plus any gaseous disc wind) that
collimates the jet in a real GRB (Fig.~\ref{fig_collapsarcartoon}).

In the context of the collapsar picture,
the `wind' region of our idealized model can be considered as a freely moving pressure boundary for the jet.
The jet boundary in our simulations is able to self-adjust in response to pressure changes within the jet, and thus
the boundary is able to act like the stellar envelope in a real collapsar.
Notice that replacing the stellar envelope with our idealized magnetized `wind' is
a good approximation because in ideal MHD the wind region could be cut
out (along the field line that separates the jet and wind) and
replaced with an isotropic thermal gas pressure.  The problem would be
mathematically identical if the material in the wind region were slowly moving,
as is true for the stellar envelope.  Even when the material in the wind region is
rapidly rotating, we show later that the pressure in the wind region changes very little,
so the approximation still remains valid. Since the only importance of the wind in
our model is to provide pressure support for the jet, we adjust our
disc wind to match the expected properties of the confining medium in
a collapsar.

We work with spherical coordinates $(r, \theta, \varphi)$, but we also
frequently use cylindrical coordinates $R=r\sin\theta$,
$z=r\cos\theta$. We work in the units $c = r_0 = 1$, where $c$ is the
speed of light and $r_0$ is the radius of the compact
object. Therefore, the
surface of the compact object is always located at $r = 1$, and
the unit of time is~$r_0/c$.

\subsection{Jet Confinement}
\label{sec_jetconfinement}

\begin{figure}
\begin{center}
\epsfig{figure=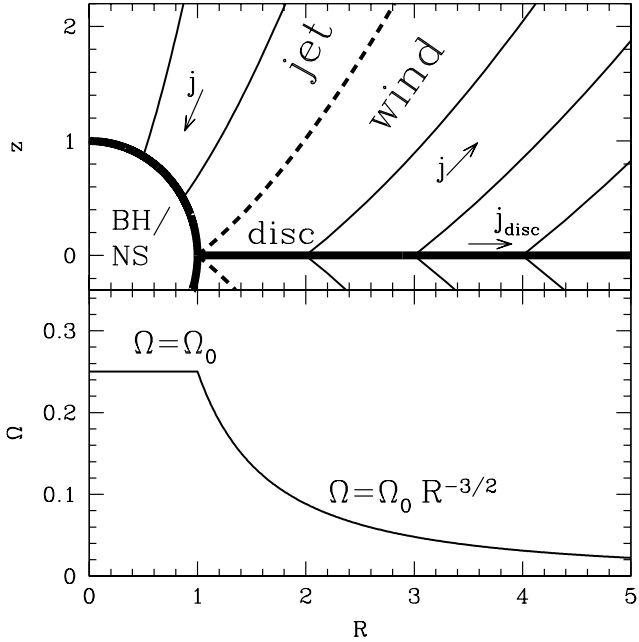,width=\columnwidth}
\end{center}
\caption{Idealized model studied in this paper.  The thick solid lines
in the upper panel show an azimuthal cut through a compact star
surrounded by a razor-thin disc.  The star and the disc are threaded
by magnetic field lines, which are shown as thin solid lines.  The
magnetized plasma above the star and the disc is assumed to be
perfectly conducting and to have an ultra-high magnetization
parameter.  Arrows show the direction of the poloidal electric
current.  The thick dashed line indicates the field line that
separates the jet from the disc wind.  The disc wind provides pressure
support for the jet and plays the role of the gaseous stellar envelope
in Fig.~\ref{fig_collapsarcartoon}.  The degree of pressure support is
adjusted by varying the magnetic field strength profile in the disc.
The lower panel shows the angular frequency of rotation of field lines
as a function of the cylindrical radius of their foot-points.}
\label{fig_bhcartoon}
\end{figure}

Most of the jet power output from a BH accretion system moves along
field lines that originate from the compact object~\citep{mck04,
  mck06jf}, i.e., along the zone that we identify as the `jet' in our
model~(Fig.~\ref{fig_bhcartoon}). A crucial factor which determines
the degree of acceleration and collimation of the jet is the total
pressure support provided by the surrounding medium through gas pressure,
magnetic pressure, ram pressure, or other forces.
We parameterize the initial total confining pressure (at time $t=0$,
which is the starting time of the simulation) as a power-law:
\begin{equation}
p_\text{amb} \propto r^{-\alpha}.
\label{eq_jetpscaling}
\end{equation}

Near the compact object the accretion disc itself can provide the jet
with support. For example, \citet{mck07a, mck07b} found via GRMHD
simulations of magnetized accreting tori that the wind from their
torus has an ambient pressure support that approximately follows a
simple power-law with $\alpha\sim2.5$ for two decades in radius.  At a
larger distance from the compact object the disc wind is expected to
become less effective, and the ambient pressure in the case of a GRB
is presumably due to thermal and ram pressure of the stellar
envelope. According to a simple free-fall model of a collapsing
star~\citep{bethe_supernova_mechanisms_1990}, for which density and
velocity scale as $\rho\propto r^{-3/2}$ and $v\propto r^{-1/2}$, the
ram pressure varies with radius as $\sim r^{-2.5}$, identical
to the GRMHD disc wind result. Moreover, hydrodynamic simulations of
GRB jets show that the internal thermal pressure also has the same
scaling, $\alpha\sim 2.5$ (see, e.g., Model JA-JC in
\citealt{zhang_relativistic_jets_2003}).

In our model, the vertical component of the magnetic field at the surface of the disc
is taken to vary as a power-law with radius,
\begin{equation}
B_z(R) \propto R^{\nu-2}, \quad \nu = \const.
\label{eq_phidiscscaling}
\end{equation}
This is our boundary condition on the field at $z=0$,
$R\ge1$.  If $\nu = 1$, the wind has a paraboloidal shape and the
magnetic pressure has a power-law scaling $r^{-2}$, whereas if
$\nu=0$, the wind corresponds to a split monopole with pressure varying as
$r^{-4}$~\citep{bz77,mck07a,mck07b,nar07}.  For a general value of
$\nu$, the magnetic pressure in the wind is very close to a power-law,
$r^{-\alpha}$,
with
\begin{equation}
\alpha = 2(2-\nu).
\label{eq_pmagjet}
\end{equation}
Since we wish to have $\alpha=2.5$, therefore for our fiducial model
we choose
\begin{equation}
\nu = 2 - \alpha/2 = 0.75.
\end{equation}
We have considered many other values of $\nu$, but focus on two other cases: $\nu=0.6,\;1$.

\subsection{Model of the Central Compact Object}

We treat the central compact object as a conductor with a uniform
radial field on its surface, i.e., as a split monopole.  The compact object
and the field lines rotate at a fixed angular frequency $\Omega_0$,
and it is this magnetized rotation that launches and powers the
jet. We neglect all gravitational effects.
As shown in~\citet{mck07a, mck07b}, this is a
good approximation since (relativistic) gravitational effects do not
qualitatively change the field geometry or solution of the
magnetically-dominated jet even close to the BH. Also, jet
acceleration is known to occur mostly at large distances from the
compact object for electromagnetically-driven
jets~\citep[\eg,][]{bes06}.

For a spinning BH the angular frequency of field lines in the
magnetosphere is determined by general relativistic frame dragging in
the vicinity of the hole.  This causes the field lines threading
the BH to rotate with a frequency approximately equal to half the
rotation frequency of the BH horizon,
\begin{equation}
\OmegaF(a) \approx 0.5 \, \Omega_\text{H}(a)
           = \frac{a c}{4 r_\text{H}} = \frac{a}{4},
\label{eq_omegastar}
\end{equation}
where the dimensionless Kerr parameter $a$ describes the BH spin and
can take values between $-1$ and~$1$.  In our chosen
units the radius of the BH horizon, $r_\text{H} = (1+\sqrt{1-a^2})GM/c^2$,
is unity (see~\S\ref{sec_motivation}).
Equation~\eqref{eq_omegastar} is for a monopole field threading the
horizon~\citep{mck07a,mck07b}. For field geometries other than a
monopole, the field rotation frequency does not remain exactly
constant on the BH horizon~\citep{bz77, mck07b}. For example, for a
parabolical field geometry, $\OmegaF$ near the poles is smaller than in
the monopole case by a factor of two. We do not consider this effect
in the current study.  According to \citet{mck07b} it should not
change our results significantly.

Various studies of BH accretion systems
suggest that rapidly spinning BHs ($a\gtrsim0.9$) are
quite common~\citep{gammie_bh_spin_evolution_2004,shafee_bh_spin_2005,
mcclintock_grs1915_2006}.
Therefore, for our fiducial model, we consider a maximally spinning BH with Kerr
parameter $a = 1$, so we choose
\begin{equation}
\Omega_\co = \OmegaF(a=1) = 0.25.  \label{eq_maxspinfreq}
\end{equation}
This is the maximum frequency that field lines threading a BH can have
in a stationary solution.

Even though we primarily focus on the case of a maximally spinning BH,
we also apply our model to
magnetars. A magnetar with a characteristic spin period
of $1$ ms and a radius of $10$~km has spin frequency
$\Omega_\text{NS} \approx 0.21$ in the chosen units (unit of length
$r_0 = 10^6$~cm and unit of time $r_0/c \approx 3.3\times10^{-5}$~s),
and so is comparable to a maximally rotating black hole. Thus,
$\Omega_\co=0.25$ is a reasonable approximation for either rapidly
rotating black holes or millisecond magnetars.

\subsection{Astrophysical Problem Setup: Models A, B, and C}
\label{sec_fiducial_model_setup}

Since we study axisymmetric magnetic field configurations, it is
convenient to separate poloidal and toroidal field components,
\begin{equation}
\myvec{B} = \myvec{B}_p + {B}_\varphi \hat\varphi.
\label{eq_Btot}
\end{equation}
It is further convenient to introduce a magnetic field stream function
$P$ to describe the axisymmetric poloidal field
$\myvec{B}_p$~\citep{okamoto1978,tpm86,bes97,nar07},
\begin{equation}
\myvec{B}_p = \frac{1}{r^2\sin\theta} \frac{\p P}{\p \theta} \,\hat r
              - \frac{1}{r\sin\theta} \frac{\p P}{\p r} \, \hat\theta.
\label{eq_deffluxfunction}
\end{equation}
This representation automatically guarantees $\myvec\nabla \cdot \myvec B = 0$.  Here
$\hat r$, $\hat \theta$, and $\hat\varphi$ are unit vectors in our
spherical coordinate system.  The stream function gives the magnetic
flux $\Phi$ enclosed by a toroidal loop passing through a point
$(r,\theta)$~\citep{nar07},
\begin{equation}
\Phi(r,\theta)=2\pi P(r,\theta).
\end{equation}

We perform the simulations over the region $(1,r_\text{max}) \times
(0, \pi/2)$.  We initialize the simulation with a purely poloidal initial
magnetic field,
\begin{equation}
P = r^\nu (1-\cos\theta).
\label{eq_aphi0text}
\end{equation}
This initial field corresponds to a split monopole field configuration at the
compact object (constant $\abs{B_r}$) and has a power-law profile for the vertical
component of the field on the disc.  For our fiducial model A, we take
$\nu = 0.75$, as explained in \S\ref{sec_jetconfinement}. This magnetic
field configuration has a confining pressure varying as $r^{-2.5}$ and
is approximately
an equilibrium nonrotating jet solution as we show in
Appendix~\ref{appendix_fluxfunction}.

We consider both the surface of the compact object, $r=1$, and the
surface of the disc, $\theta = \pi/2$ ($z = 0$), to be ideal
conductors.  The number of quantities we fix at these boundaries is
consistent with the counting argument of~\citet{bog97}. A paraphrasing
of this argument is that the number of quantities that one should
relax at the boundary of a perfect conductor equals the number of
waves entering the boundary. In our case of a sub-Alfv\'enic flow
there are two waves entering the boundary: an incoming Alfv\'en wave
and an incoming fast wave.  Thus, we leave the two components of the
magnetic field parallel to the conductor unconstrained, and we only
set the normal component of the field.

\begin{figure*}
\begin{center}
\subfigure{
\epsfig{figure=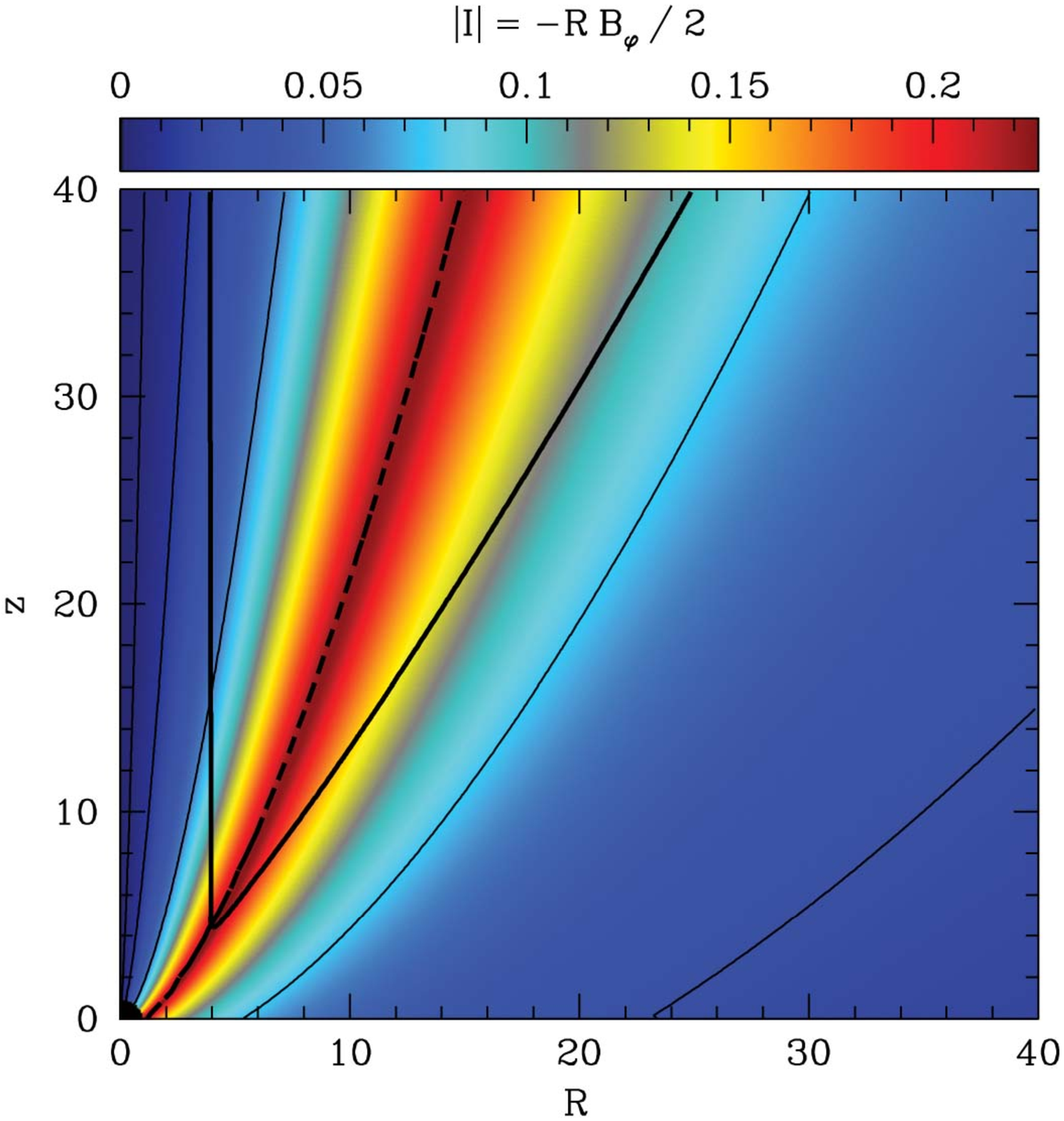,height=0.9 \columnwidth}
}
\subfigure{
\epsfig{figure=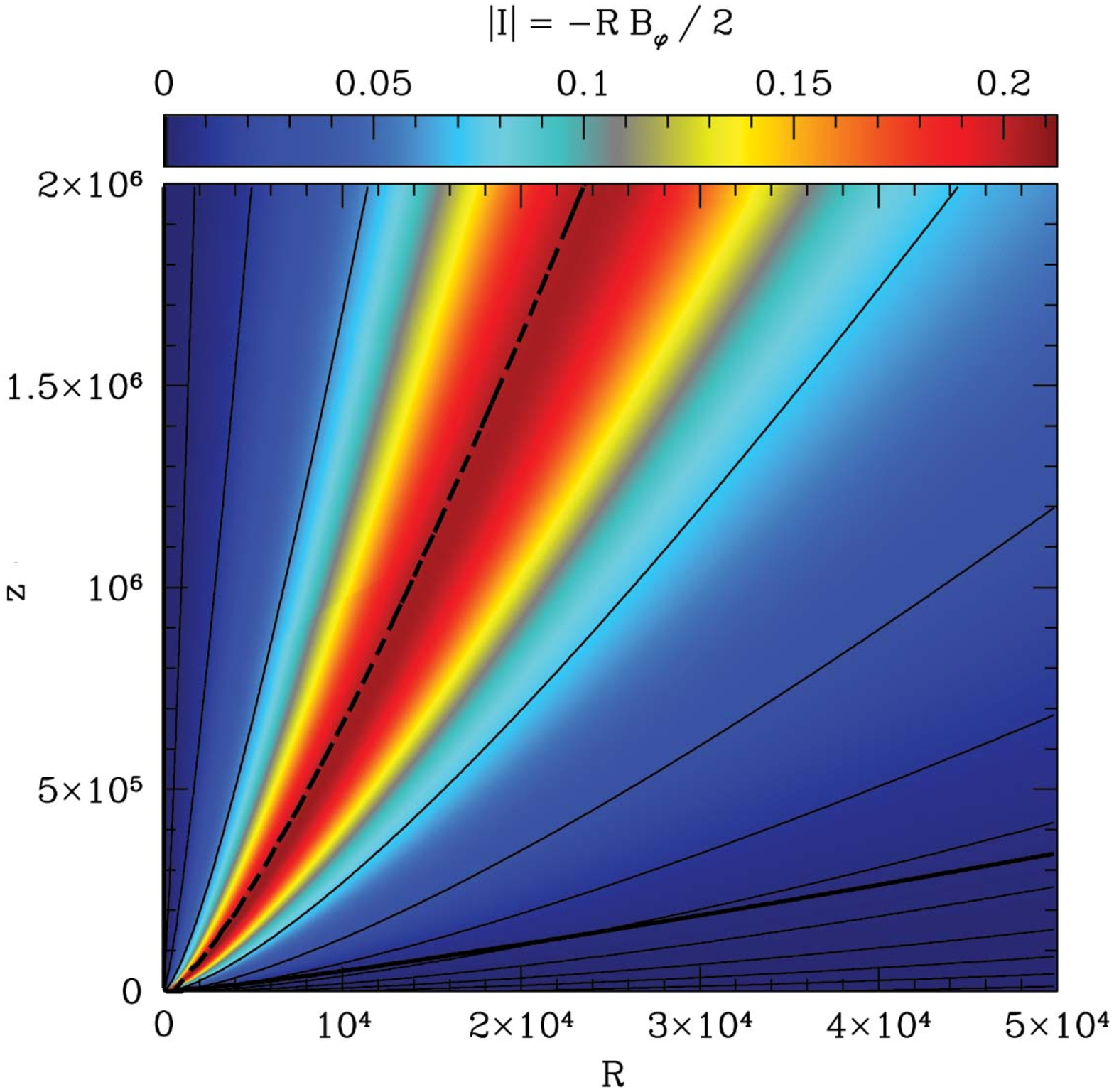,height=0.9 \columnwidth}
}\\
\subfigure{
\epsfig{figure=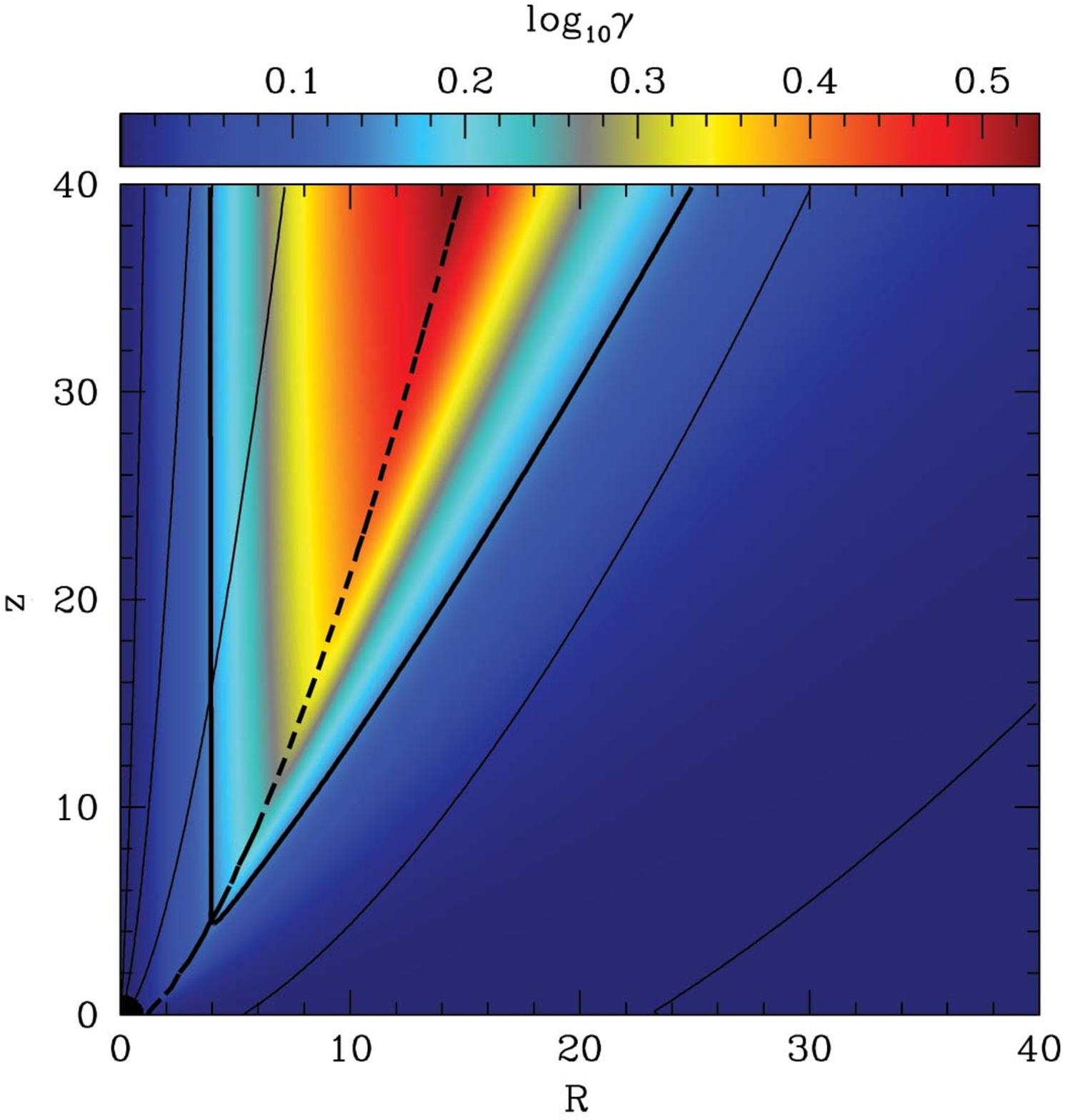,height=0.9 \columnwidth}
}
\subfigure{
\epsfig{figure=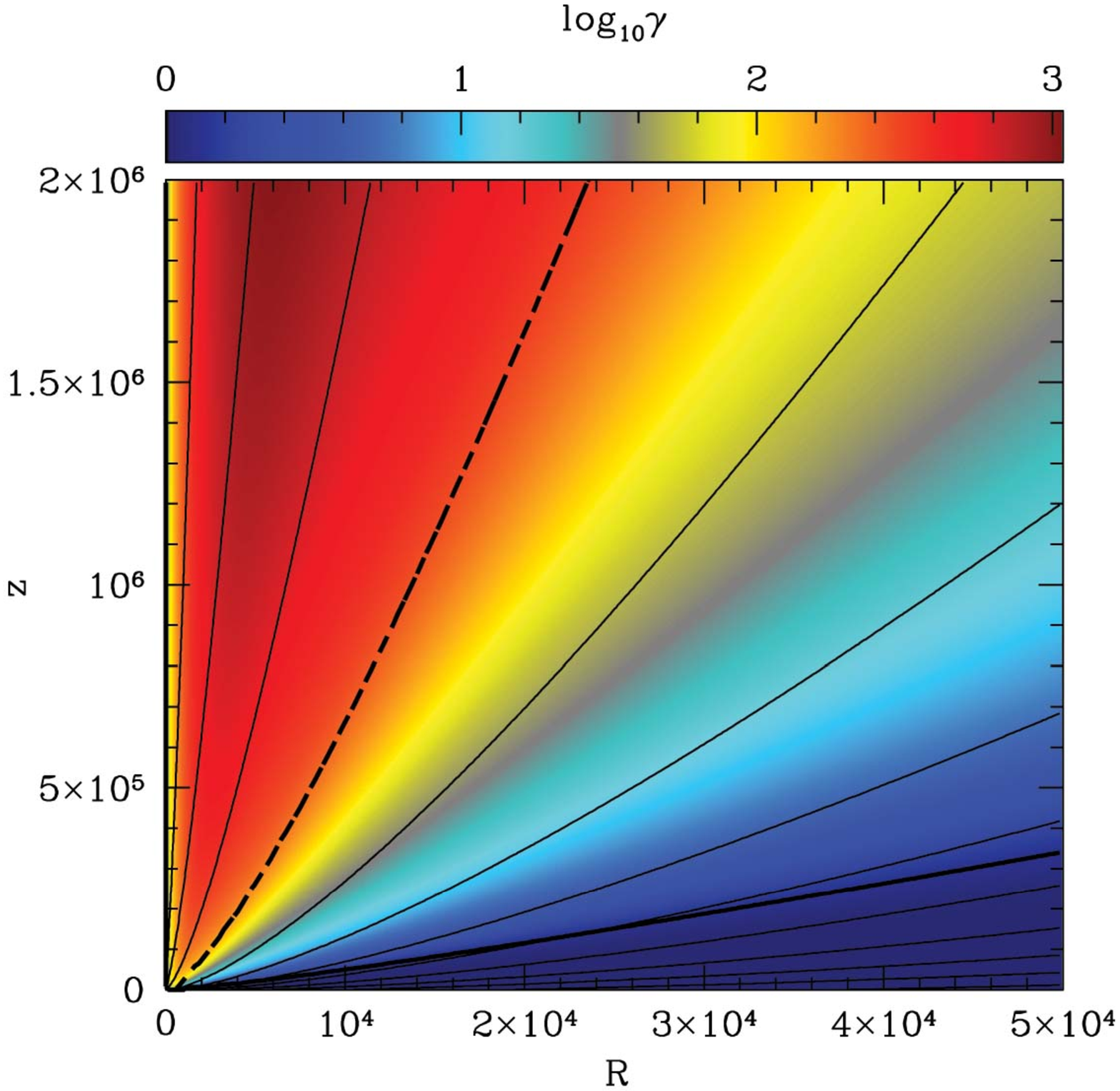,height=0.9 \columnwidth}
}
\end{center}
\caption{Poloidal magnetic field lines, shown as solid lines, overlaid
on the colour-coded absolute magnitude of the enclosed poloidal current
$\abs{I} = \abs{R B_\varphi/2}$ (upper panels), and the colour-coded
Lorentz factor $\gamma$ (lower panels).  The results are for the
fiducial model~A, which corresponds to $\nu = 0.75$, $\Omega_\co =
0.25$, $\beta = 1.5$.  Left (right) panels show the near (far) region
of the jet solution.  The thick dashed lines indicate the position of
the field line that separates the jet from the disc wind
(see Fig.~\ref{fig_bhcartoon}), and the
thick solid lines show the position of the Alfv\'en surface, $\Omega R=1$. Red
(blue) colour corresponds to high (low) values of the plotted
quantities. Note that, for any distance from the compact object,
the maximum $\abs{I}$ is nearly coincident with the jet-wind
boundary.  However, the maximum Lorentz factor is found
inside the jet, closer to the axis (see Fig.~\ref{fig_faceon_jetview}).}
\label{fig_nu075m025a15field}
\end{figure*}

We set the values of two magnetic field drift velocity components
(perpendicular to the magnetic field) through the stationarity
condition~\citep{nar07}. For the compact object we choose a constant angular velocity
rotation profile $\Omega_\co=0.25$ (eq.~\ref{eq_maxspinfreq}),
and for the disc we choose a Keplerian-like rotation profile,
\begin{equation}
\Omega^\text{disc}(R, z=0) = \Omega_\co R^{-\beta}, \quad\beta=3/2.
\label{eq_omegadisc}
\end{equation}
Where the compact object meets the disc, the magnitude of the disc
angular frequency per unit Keplerian rotation frequency is
$\Omega/\Omega_K\approx 1/2$ as consistent for millisecond magnetars of $10$~km size
and mass $1.4\msun$ and consistent with GRMHD simulations for near an $a=1$
black hole~\citep[see, \eg,][]{mck07a,mck07b}.
Therefore, near the compact object our model is more accurate
than a precisely Keplerian model while keeping $\Omega$ everywhere continuous.
We use an antisymmetric boundary condition at the polar axis $\theta =
0$ and an outflow boundary condition at the outer boundary $r =
r_\text{max} = 10^8$.  Since our time-dependent solutions never reach this artificial outer
boundary, our results do not depend upon the details of the boundary condition there.

We discuss results obtained with this fiducial model~A in
\S\ref{sec_fiducialmodel}.  We also discuss two other models
that differ from model~A only by the value of $\nu$:
model~B, which has $\nu=1$ (parabolic field lines), and model~C, which
has $\nu=0.6$.  We have run a number of other models that we mention
briefly in \S\ref{sec_othermodels}.

\subsection{Numerical Method}
\label{sec_nummethod}

We use a Godunov-type scheme to numerically solve the time-dependent
force-free equations of motion~\citep{mck06ffcode}.  Our code has
been successfully used to model BH and neutron star
magnetospheres~\citep{mck06ffcode,mck06pulff,mck07a,mck07b,nar07}.

To ensure accuracy and to properly resolve the jet, we use a numerical
grid that approximately follows the magnetic field lines in the jet
solution~\citep{nar07}. We are thus able to simulate the jet out to
large distances without making significant errors in the solution.
For the three models, A, B, C, discussed in this paper we used a
resolution of $2048$x$256$.  Since our grid
follows the poloidal field lines, the above resolution corresponds to an
effective resolution of about $2048$x$100,000$ in spherical polar coordinates.
A comparison with lower-resolution runs shows that these models are
well converged. In particular, over $6$ orders of magnitude in
distance from the compact object, the shape of poloidal field lines is
accurate to within~$\delta\theta/\theta < 10$\% (see \S\ref{sec_results}).

In order to speed up the computations, we use a time-stepping technique
such that only the non-stationary region is evolved.  This is achieved
by defining the active section, where the evolution is performed, to be the exterior to a
sphere of radius $r_\text{stat} = \xi_\text{stat} c t$, where $t$ is
the time of the simulation, $c$ is the speed of light, and
$\xi_\text{stat} = 0.01$. We set the electromotive forces at all
boundaries of the active section to zero. If the initial condition is a
force-free solution, then this procedure is mathematically justified even within the fast critical surface since the
time-dependent solution only contains outgoing waves, so the solution
rapidly settles to its final state behind the wave.  In all cases our initial condition
is an exact solution or is close enough to the exact solution to avoid significant ingoing waves.  We also
experimentally verified that by not evolving the solution interior to
the active section, we make an error of less than one part in ten
thousand. The use of grid sectioning speeds up the simulations by a
factor of up to a thousand since it allows us to use a larger time step.
\citet{kom07} used a similar approach.

\section{Simulation Results}
\label{sec_results}
We first present simulation results for our fiducial model~A and
analyze them morphologically.  Then, we develop an analytical model of
the jet structure and use it to gain insight into jet acceleration and
collimation. Finally, we consider the other two jet models, B and C,
and discuss the variation of jet properties with $\theta$.

\subsection{Fiducial model A}
\label{sec_fiducialmodel}

\begin{figure}
\begin{center}
\epsfig{figure=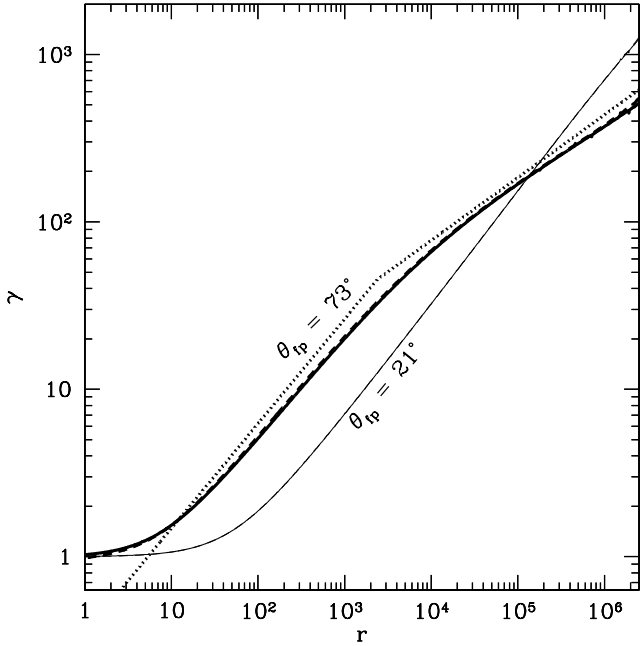,width=\columnwidth}
\end{center}
\caption{ Radial dependence of the Lorentz factor $\gamma$ in the
fiducial model A for two field lines.  One field line starts from the
compact object at an angle $\theta_\text{fp}\approx73^\circ$
(indicated with thick lines), and the other starts at
$\theta_\text{fp}\approx21^\circ$ (thin lines). Solid lines show the
numerical solution, dashed lines show the analytical
approximation~\eqref{eq_asymalltext} with $C = \sqrt{3}$ (the solid
and dashed lines are virtually indistinguishable for $\theta_\fpt = 21^\circ$),
and dotted lines
show the individual scalings given
in~\eqref{eq_asym1thetascalingtextalongfieldline}
and~\eqref{eq_asym2thetascalingtextalongfieldline}.  Note that the
field line with $\theta_\fpt=73^\circ$ accelerates quickly as it moves
away from the compact object but it then switches to a slower second
regime of acceleration.  In contrast, the field line with
$\theta_\fpt=21^\circ$ begins accelerating only after it has moved a
considerable distance from the compact object.  However, it then
maintains a rapid rate of acceleration without switching to the second
acceleration regime. When the jet reaches the outer edge of the
simulation at $r\sim2\times10^6$, this field line has a very large
bulk Lorentz factor $\gamma>1000$, whereas the field line with
$\theta_\fpt=73^\circ$ has a smaller $\gamma\sim500$.  Thus, the jet
develops a fast core surrounded by a slower sheath.
}
\label{fig_nu0.75m0.25gamma}
\end{figure}

\begin{figure}
\begin{center}
\epsfig{figure=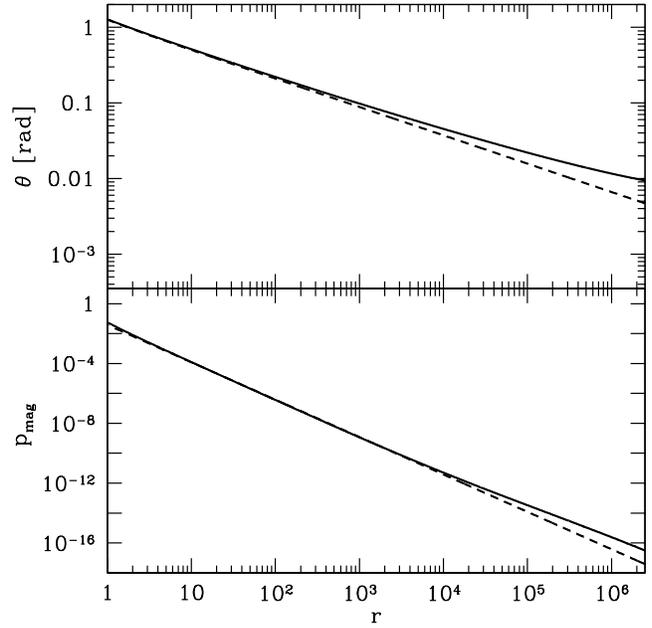,width=\columnwidth}
\end{center}
\caption{ Field line shape (upper panel) and comoving magnetic
pressure (lower panel) in the fiducial model~A for a field line that
starts from the compact object at $\theta_\text{fp} \approx 73^\circ$.
Solid lines show the results from the numerical simulation, and dashed
lines show the analytic scalings for the non-rotating solution.  Field lines with foot-points
in the disc (wind) show better agreement with the analytic scalings. For example, the
magnetic pressure along a field line with
$r_\text{fp} \gtrsim 500$ (not shown) is indistinguishable from
the dashed line in the lower panel.}
\label{fig_nu0.75m0.25all}
\end{figure}

Model A consists of a compact object of unit radius rotating with an
angular frequency $\Omega_0=0.25$, surrounded by a Keplerian-like disc
($\beta=3/2$).  On the surface of the compact object, the
radial component of the magnetic field is uniform,
{}$B_r=1$, while at the disc,
the vertical component of the field varies with radius in a
self-similar fashion with index $\nu=0.75$, i.e., $B_z (R,z=0) \propto
R^{\nu-2} =R^{-1.25}$  (eq.~\ref{eq_aphi0text}).  Starting with this purely poloidal initial
field configuration, we have run the force-free simulation for a time
equal to $10^7 r_0/c$.
At the end of the calculation we obtained a time-steady
solution out to a distance of $2\times10^6 r_0$.  We describe below the
properties of this steady solution.

The panels in Fig.~\ref{fig_nu075m025a15field} show the poloidal field
structure of model~A in steady state.  The poloidal field in the final rotating
state is nearly the same as in the initial non-rotating state, just as
was seen for the self-similar solutions discussed in \citet{nar07}.
This is despite the fact the final steady solution has a strong
axisymmetric toroidal field $B_\varphi(r,\theta)$, which is generated
by the rotating boundary conditions at the star and the disc.

The toroidal field at any point is related to the total enclosed
poloidal current $I(r,\theta)$ at that point by Ampere's Law,
\begin{equation}
I(r,\theta) = RB_\varphi(r,\theta)/2 \le 0, \quad R=r\sin\theta.
\end{equation}
The enclosed current is negative because we have a
positive~$B_z$ and positive~$\Omega$, so that $B_\varphi$ is negative.
The colour-coding in the upper panels of
Fig.~\ref{fig_nu075m025a15field} indicates the absolute magnitude of
the enclosed current as a function of position.  As expected, we see
that $I$ is constant along field lines, which corresponds to
$RB_\varphi$ being constant.  More interestingly, we see that at any
$r$, the absolute value of the enclosed current starts at zero,
increases as we move away from the axis, reaches a maximum value, and
then decreases back to zero.  The maximum in the absolute enclosed
current corresponds to a transition from a negative current density
(inward current) to a positive current density (outward current).
This transition is coincident with the field line that originates at
$r_{\rm fp}=1$, $\theta_{\rm fp} =\pi/2$ and that defines the boundary
between the `jet' and the `wind' in our model (see
Fig.~\ref{fig_bhcartoon}).

As a result of rotation, the solution develops a poloidal electric
field in the lab-frame (or inertial frame).  The electric field
strength at each point is equal to $\Omega R B_p$, where
$\Omega$ is equal to the angular frequency at the foot-point of the
local field line.  The electric field gives an outward Poynting flux
$\myvec{S}= \myvec{E}\times\myvec{B}/4\pi$ which we discuss later.  It
also gives a drift speed $v=E/B$, and a corresponding Lorentz factor
$\gamma$.  The colour-coding in the lower panels of
Fig.~\ref{fig_nu075m025a15field} indicates the variation of $\gamma$
with position in the steady solution.  The Lorentz factor reaches up
to a maximum $\sim1000$ in this particular model.  As
Fig.~\ref{fig_nu075m025a15field} shows, the acceleration proceeds
gradually and occurs over many decades in distance from the compact
object.

Note that, at a given distance from the compact object, the maximum Lorentz
factor is not achieved at either the jet-wind boundary or on the axis,
but occurs at an intermediate radius inside the jet.  For instance, at
$z=5\times10^5$, $\gamma$ is maximum at $R\sim3\times10^3$, whereas
the jet-wind boundary is located at $R\sim8\times10^3$.  Thus, the
jet consists of a slow inner spine, fast edge, and a slow outer sheath which actually contains most
of the power density.
\citet{kom07} apparently observed this `anomalous' effect in one of their solutions.
In the next subsection we explain the origin of the effect and
quantify it.

Figure~\ref{fig_nu0.75m0.25gamma} shows the variation of the Lorentz
factor with distance along two field lines emerging from the compact object.
The field line that starts closer to the equator, with $\theta_{\rm
  fp}=73^\circ$, undergoes rapid acceleration once it is beyond a
distance $\sim10$.  However, at a distance $\sim10^3$ it switches to a
different and slower mode of acceleration, reaching a final
$\gamma\sim500$ at $r=2\times10^6$.  In contrast, the field line that
starts closer to the axis at $\theta_{\rm fp}=21^\circ$ does not
begin accelerating until a distance $\sim100$.  It then accelerates
rapidly almost until it reaches the outer radius (there is a hint of a
transition to the slower acceleration mode near the end), by which
point it has a larger Lorentz factor $\sim1000$ than the other field
line.  This inverted behaviour is what causes the natural development
of a fast structured spine and slow sheath that
contains most of the power density.

The upper panel in Figure~\ref{fig_nu0.75m0.25all} shows, for the
steady state solution, the variation of polar angle $\theta$ as a
function of distance along the field line that starts at $\theta_{\rm
  fp}=73^\circ$.  The dashed line shows the corresponding quantity for
the initial purely poloidal field with which the simulation was
started.  We see that the final field shape is mildly perturbed by
rotation.  However, even at a distance of $2\times10^6$ the change in
$\theta$ is no more than a factor of 2.

The lower panel in Figure~\ref{fig_nu0.75m0.25all} shows the variation
of the comoving pressure with distance along the same field line.  The
pressure varies as $r^{-2.5}$ (the dashed line), the desired
dependence, for distances up to $r\sim10^3$ or so.  Beyond
this distance, the pressure variation in the jet becomes a little shallower.  The change
occurs in the region where the slower mode of acceleration operates
(see Fig.~\ref{fig_nu0.75m0.25all}).  We explain this behaviour in the
next subsection.  We note that this change in pressure
inside the jet does not affect the confining pressure
profile of the external medium/wind which stays the same as in the initial configuration,
and varies as $r^{-2.5}$.

\begin{figure}
\begin{center}
\epsfig{figure=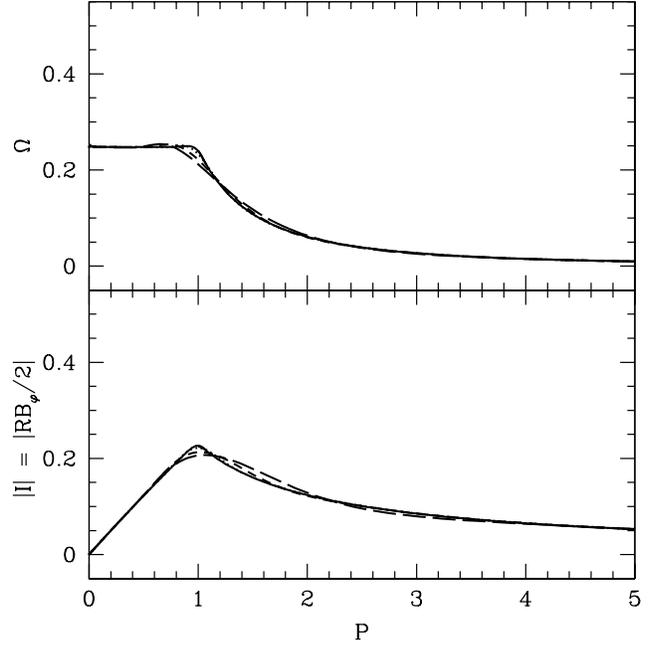,width=\columnwidth}
\end{center}
\caption{Transverse variations of the field-line invariants,
$\Omega$ and $\abs{I}$, for model~A
as a function of the magnetic field stream function $P$.  In each panel
four curves are shown: the star-disk surface (solid line),
$r = 10^2$ (dotted), $r=10^4$ (short-dashed), and $r= 10^6$
(long-dashed).  Over a range of six orders of
magnitude in distance, the values of  $\Omega$ and $\abs{I}$ are
conserved to better than $12$\%.
}
\label{fig_omegaitransversal}
\end{figure}

The model A simulation described here is well-converged: the angular
frequency of field line rotation $\Omega$ and the enclosed poloidal
current $I$ are accurately preserved along each field line, as they
should be in a stationary axisymmetric force-free
solution~\citep{mes61,okamoto1978,tpm86,bes97,nar07}.  Even though the
simulation domain extends over more than six decades in radius, these
field-line invariants are conserved to better than~$12$\%, see Fig.~\ref{fig_omegaitransversal}.

\begin{figure*}
\begin{center}
\subfigure{
\epsfig{figure=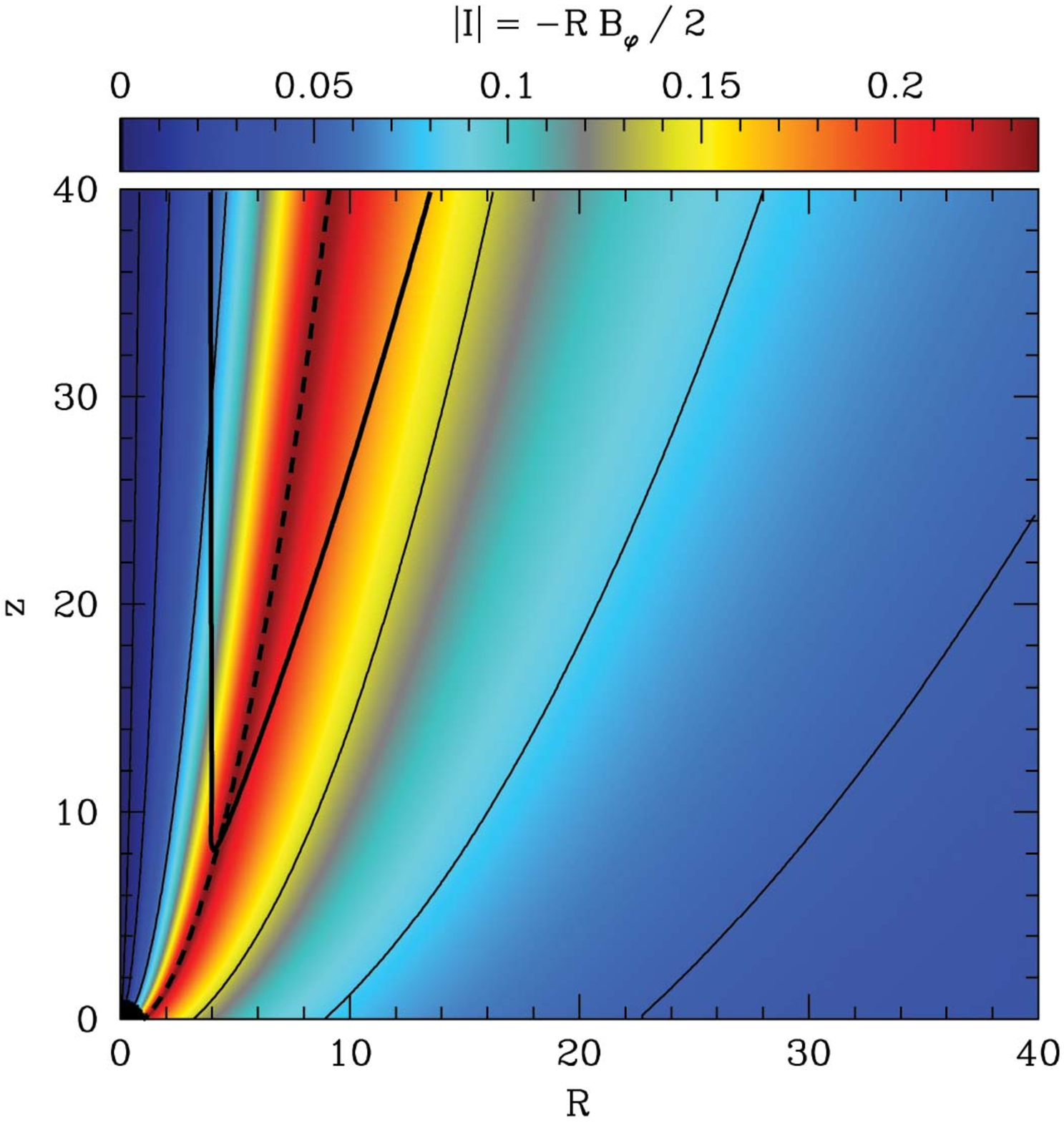,height=0.9 \columnwidth}
}
\subfigure{
\epsfig{figure=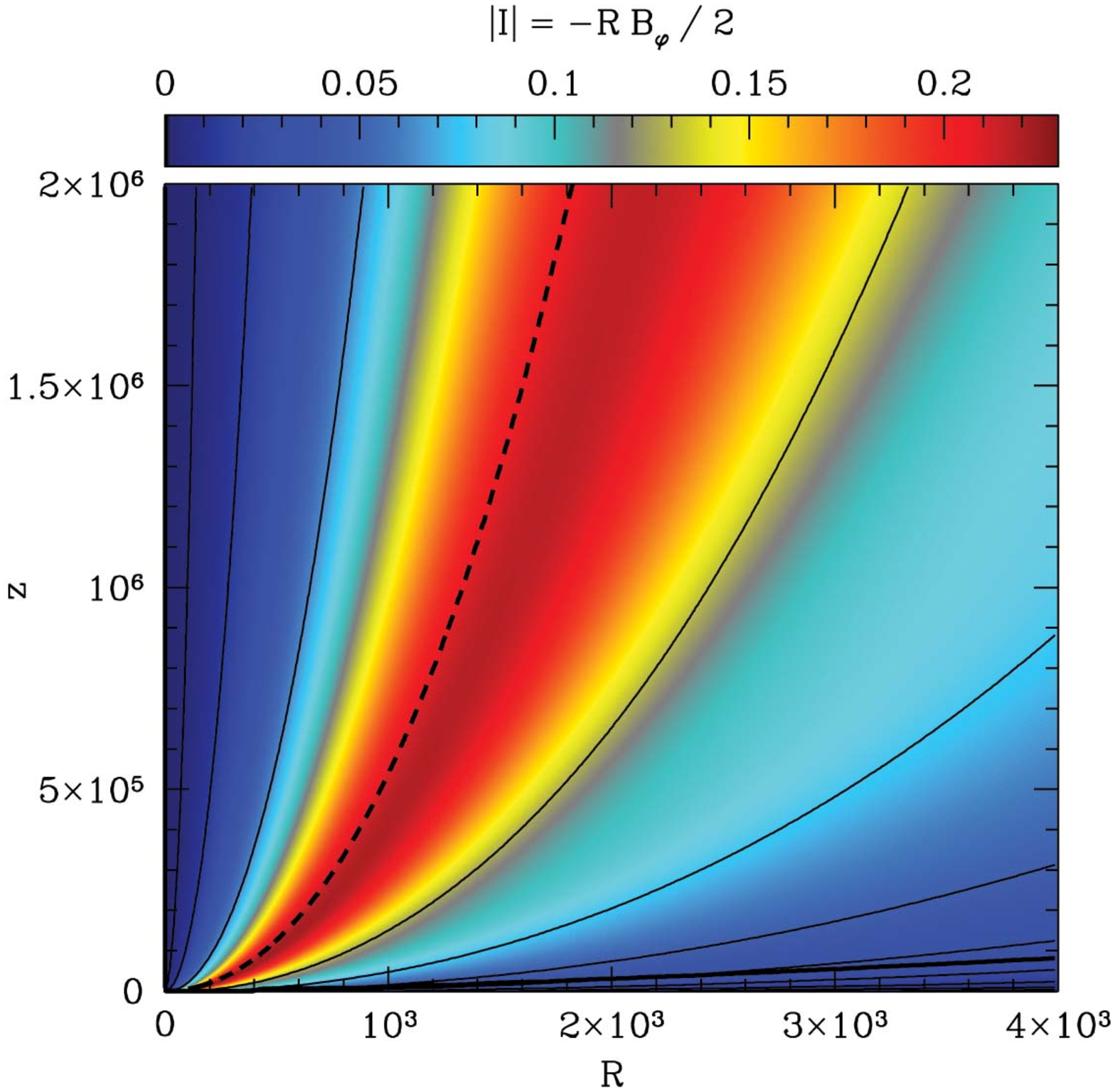,height=0.9 \columnwidth}
}\\
\subfigure{
\epsfig{figure=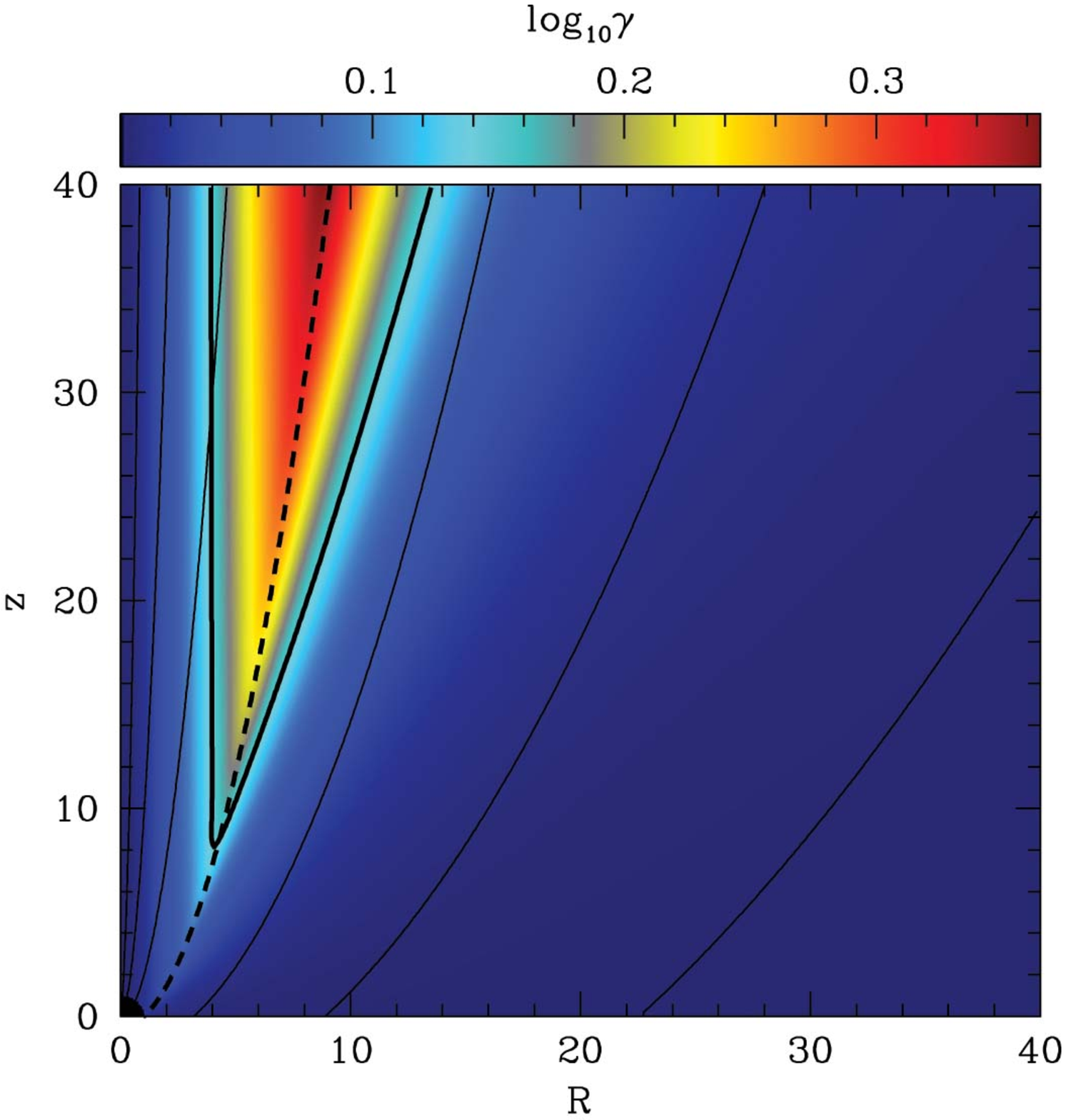,height=0.9 \columnwidth}
}
\subfigure{
\epsfig{figure=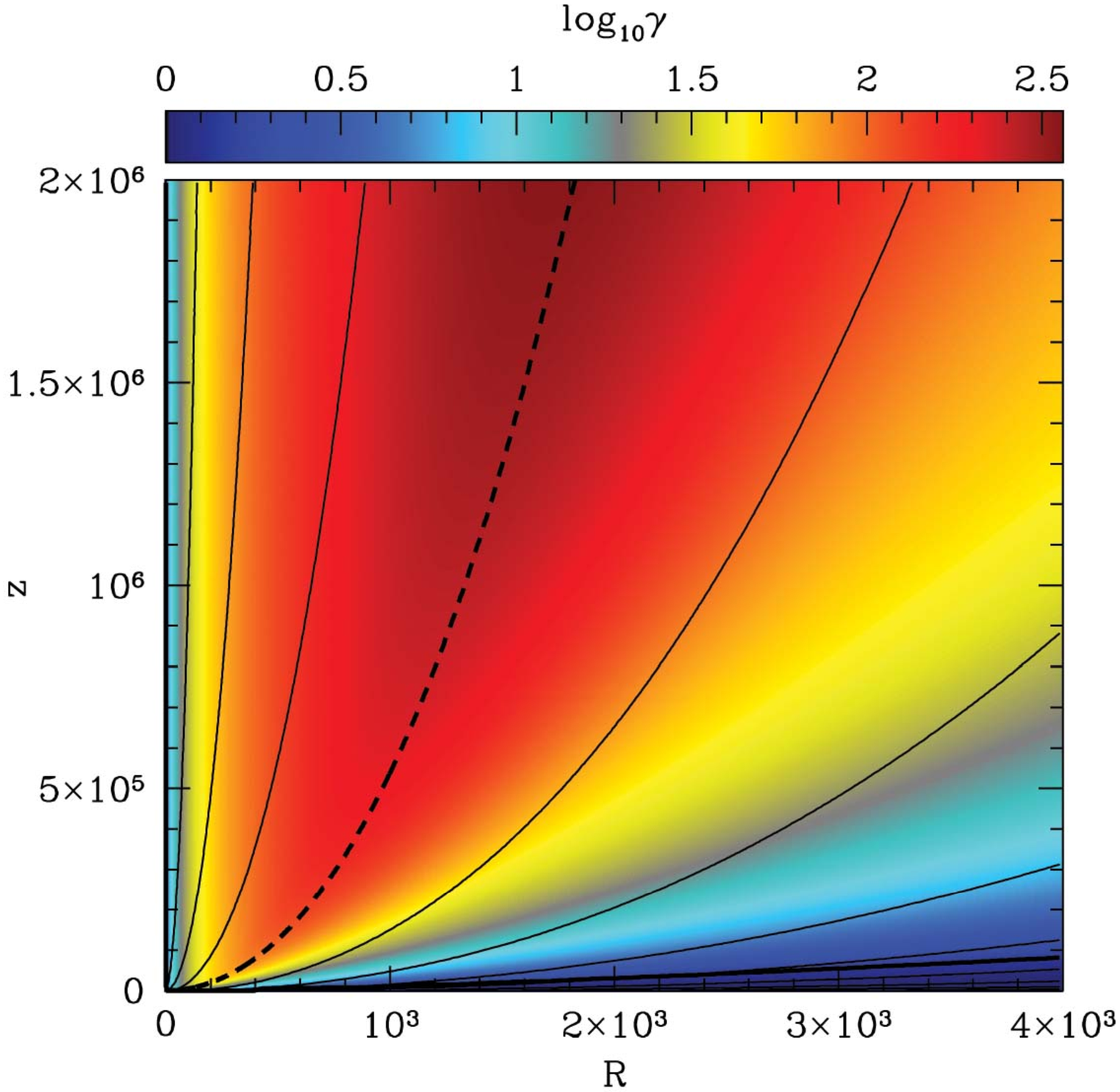,height=0.9 \columnwidth}
}
\end{center}
\caption{
Similar to Fig.~\ref{fig_nu075m025a15field}, but for model~B, which
corresponds to $\nu = 1$, $\Omega_\co = 0.25$, $\beta = 1.5$.  Field
lines in this model show faster collimation than in our fiducial
model~A.  This is because the confining magnetic pressure of the disc
wind falls off more slowly with distance -- as $r^{-2}$ here
instead of as $r^{-2.5}$ in model~A.  The difference in
collimation is not due to hoop-stress or rotation.  Note that
the maximum Lorentz factor at any distance coincides with the jet-wind
boundary, in contrast to what is seen in model~A. }
\label{fig_nu10m025a15field}
\end{figure*}

\subsection{Comparison to Analytical Results}
\label{sec_analyticalanalysis}

We now interpret the numerical results described above in terms of
simple analytic formulae.  Details may be found in the
Appendix~\ref{sec_approximate_analytic_jet_solution}.  Here we merely summarize the
relevant results.

In an axisymmetric force-free electromagnetic configuration, the drift
Lorentz factor~$\gamma$ can be described quite well by the following
analytic formula (see Appendix~\ref{sec_lorentz_factor_general}),
\begin{equation}
\frac{1}{\gamma^2} = \frac{1}{\gamma_1^2} + \frac{1}{\gamma_2^2},
\label{eq_asymalltext}
\end{equation}
where $\gamma_1$ and $\gamma_2$ are given by
\begin{align}
\gamma_1 &\approx \left[1+(\OmegaF R)^2\right]^{1/2} \approx \Omega R, \label{eq_asym1text} \\
\gamma_2 &\approx C \; \left(\frac{R_c}{R}\right)^{1/2},
\label{eq_asym2text}
\end{align}
where the last equality in~\eqref{eq_asym1text} holds for $\Omega R\gg1$.
Here, $R = r\sin\theta$ is the cylindrical radius, $\OmegaF$ is the
rotation frequency of the local field line, $R_c$ is the poloidal
radius of curvature of the field line, and $C$ is a numerical factor
of order unity that depends on the field line rotation profile
(see Appendix~\ref{sec_transversal_force_balance}),
\begin{equation}
C \approx \left(3+\frac{\p\log\OmegaF}{\p\log R}\right)^{1/2}.
\label{eq_c_coef_text}
\end{equation}
In the jet region
we have $\OmegaF = \const$ and, therefore, we expect
\begin{equation}
C \approx \sqrt3 \approx 1.7. \label{eq_bnubestfit}
\end{equation}
As we will see in~\S\ref{sec_othermodels}, in the simulations we
find values of $C$ slightly below this value because
$\OmegaF$ slightly decreases with increasing $R$ toward the edge of the jet (due
to numerical diffusion, see Fig.~\ref{fig_omegaitransversal}).

Equation~\eqref{eq_asymalltext}
gives the drift speed of an infinitely
magnetized magnetodynamic, or force-free, flow.  One might
worry that the velocity of a fluid carried along with such a flow will be
very different.  In Appendix~\ref{sec_fluid_speed}
we show that any such fluid has only a slightly
modified Lorentz factor relative to the drift speed, in the limit when the fluid
is massless, \ie, $\sigma\to\infty$. Therefore,
for all practical purposes we can assume that the fluid Lorentz factor
is given by eq.~\eqref{eq_asymalltext}.

Since $\gamma^2$ is the harmonic sum of two terms, the value of
$\gamma$ is determined by whichever of the two quantities, $\gamma_1$
and~$\gamma_2$, is {\it smaller}.  Close to the central compact star,
$\gamma_1$ is smaller, and the first term in equation (\ref{eq_asymalltext})
dominates.  Thus, for a given rotation frequency $\Omega_0$ of the
field lines in the jet (determined by the spin of the compact object),
the Lorentz factor increases linearly with distance from the jet
rotation axis~\citep{bz77,bes97}. In this well-known regime,
which we refer to as the \emph{first acceleration regime}, a faster
compact object spin leads to faster acceleration along the jet.  Also,
for a given rotation, the outermost field lines in the jet, which
emerge from the equator of the star, have the largest acceleration and
largest~$\gamma$ at any given~$z$.

The second term in equation~\eqref{eq_asymalltext} represents a slower
regime of acceleration, which we refer to as the \emph{second
  acceleration regime}.  It is present only for certain field
geometries and is generally realized only at large distances from the
compact object. For the self-similar solutions described in
\citet{nar07}, this acceleration regime is important only if the
self-similar index $\nu<1$.  Since model A has $\nu=0.75$, this term
is important for our simulation.  Note that models with $\nu\le 1$ are
astrophysically the most interesting and relevant (\S\ref{sec_application_lgrbs}),
so it is important to understand the
second acceleration regime.
A feature of the second acceleration regime is that the Lorentz factor
does not depend explicitly on the field line rotation frequency, but
is determined purely by the local poloidal curvature of the field
line~\citep{beszak04}.  Moreover, as we saw earlier, the poloidal structure of the
field line is itself largely independent of rotation.

Let us ignore the small distortion of the field line shape caused by
rotation (Fig.~\ref{fig_nu0.75m0.25all}), and take the shape to be
given by the initial nonrotating solution (see
Appendix~\ref{appendix_fieldlineshape}):
\begin{equation}
z \propto R^{2/(2-\nu)}, \qquad \theta \propto r^{-\nu/2}.
   \label{eq_fieldlineshapecylzRtext}
\end{equation}
The latter scaling, shown by the dashed line in
Fig.~\ref{fig_nu0.75m0.25all}, provides a good description of the
field line shape.  Using this scaling, we can evaluate $\gamma_1$ and
$\gamma_2$ in the jet using equations~\eqref{eq_asym1text} and
\eqref{eq_asym2text} (see Appendix~\ref{sec_lorentz_factor_jet_and_wind}),
\begin{align}
\gamma_1 &\approx \Omega_\co \,  r \, \sin\theta,
\label{eq_asym1thetascalingtext} \\
\gamma_2 &\approx \kappa/\theta,
         \label{eq_asym2thetascalingtext}
\end{align}
where $\kappa = {2C}/{\sqrt{(2-\nu)\nu}}$ does not have any explicit
dependence on $\Omega$ or position.  This gives the following
scaling of the Lorentz factor along field lines,
\begin{align}
\gamma_1 &\propto r^{1-\nu/2},
\label{eq_asym1thetascalingtextalongfieldline} \\
\gamma_2 &\propto r^{\nu/2}.
         \label{eq_asym2thetascalingtextalongfieldline}
\end{align}

Close to the central star, $\gamma_2$ is
always larger than $\gamma_1$, and thus the jet $\gamma$ is determined by
$\gamma_1$.  With increasing distance along a field line,
$\gamma_1$ and $\gamma_2$ grow at different rates.  If $\nu>1$,
$\gamma_2$ rises more rapidly than $\gamma_1$ and the Lorentz
factor of the jet is always determined only by $\gamma_1$ (e.g., model
B below, which has $\nu=1$).  However, for $\nu<1$ (model A and model
C), $\gamma_2$ rises more slowly than $\gamma_1$ and takes over at a
certain distance from the star.  This corresponds to the second slower
acceleration regime seen in Fig.~\ref{fig_nu0.75m0.25gamma}.

Figure~\ref{fig_nu0.75m0.25gamma} shows a comparison of the actual
$\gamma$ measured in the model A simulation with the prediction from
the analytic formula~\eqref{eq_asymalltext}.  We set $C = \sqrt{3}\approx1.7$ (see eq.~\eqref{eq_bnubestfit} and
the next subsection).  We find that the analytic formula for $\gamma$
agrees remarkably well with the numerical results.
The formula gives the correct slopes and reproduces the
distance at which the break between the two acceleration regimes
occurs.

\begin{figure}
\begin{center}
\epsfig{figure=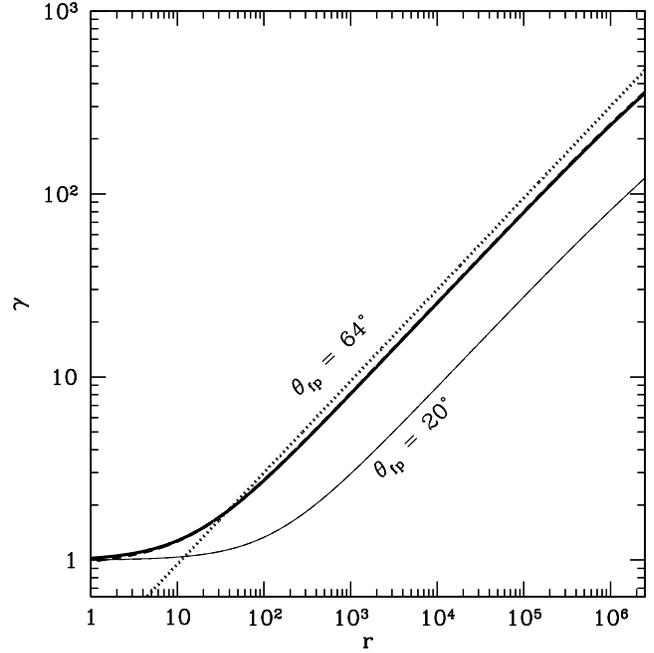,width=\columnwidth}
\end{center}
\caption{
Similar to Fig.~\ref{fig_nu0.75m0.25gamma}, but for model B.  The
analytic fit for the Lorentz factor uses the same value of $C = \sqrt3$
as before.  Compared to model~A, we see that model~B
lacks the second regime of acceleration.  Correspondingly, there is no
fast jet core present.  }
\label{fig_nu1.0m0.25gamma}
\end{figure}

\begin{figure}
\begin{center}
\epsfig{figure=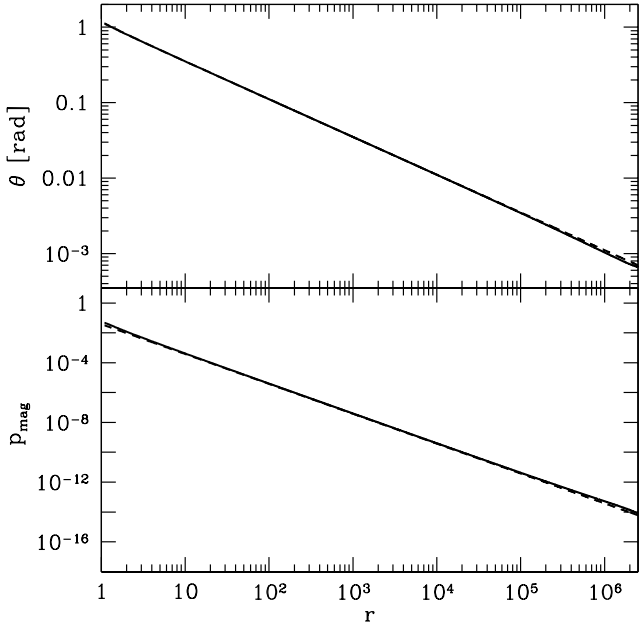,width=\columnwidth}
\end{center}
\caption{ Similar to Fig.~\ref{fig_nu0.75m0.25all}, but for model B.
  Note the excellent agreement between the numerical quantities and
  the analytical estimates.
}
\label{fig_nu1.0m0.25all}
\end{figure}

As we see from Figure~\ref{fig_nu0.75m0.25gamma}, the second regime of
Lorentz factor growth is most prominent along field lines originating
closer to the equator of the compact object. This is the reason for
the `anomalous' development of a slower-moving sheath surrounding a
faster-moving structured jet spine that we mentioned in
\S\ref{sec_fiducialmodel}. See \S\ref{sec_collimation} for
more detail.

The cause for the slight deviation of the magnetic pressure from the
$r^{-2.5}$ power-law behaviour, as seen in
Figure~\ref{fig_nu0.75m0.25all}, is discussed in
Appendix~\ref{appendix_magneticpressure}.  We show that, for $\nu<1$,
the magnetic pressure shows a broken power-law behaviour along field
lines,
\begin{equation}
p_\text{mag} \propto
  \begin{cases}
  r^{2(\nu-2)}, \quad &r \lesssim r_\text{tr}, \\
  r^{-2}, \quad &r \gtrsim r_\text{tr}.
  \end{cases}
\label{eq_pmagtext}
\end{equation}
The break radius $r_\text{tr}$ is the same as the radius where the jet
acceleration switches from the first regime~($\gamma_1$) to the
second~($\gamma_2$).  The power-law indices in~\eqref{eq_pmagtext} as
well as the predicted break radius $r_\text{tr} \approx 7\times10^3$
are consistent with the results shown in
Fig.~\ref{fig_nu0.75m0.25all}.  Note that the confining pressure
of the wind (along any field line originating in the disc sufficiently far
from the compact object) follows a single power-law, $r^{2(\nu-2)}$,
at all distances from the compact object (see \S\ref{sec_fiducialmodel}).

\subsection{Dependence of the Results on Model Details}
\label{sec_othermodels}

In order to explore which features of the results described above are
generic and which are particular to model A, we simulated a wide range
of models with $\nu$ varying from $0.5$ to $1.25$. We find that
model~A is representative of most models with $\nu < 1$.  In
particular, all these models show the two regimes of Lorentz factor
growth~\eqref{eq_asym1thetascalingtextalongfieldline}
and~\eqref{eq_asym2thetascalingtextalongfieldline}.  Similarly, we
find that model~B, which has $\nu=1$ and is described below, is
representative of models with $\nu\ge1$.

\begin{figure}
\begin{center}
\epsfig{figure=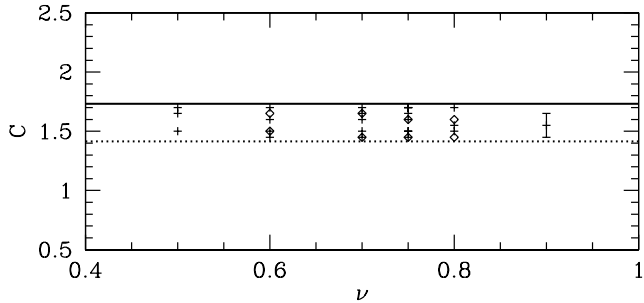,width=\columnwidth}
\end{center}
\caption{The values of the factor $C$ as determined from simulations
of jets with different choices of the field geometry threading the
star, the spin of the compact object, and the rotation profile of the
disc. Crosses and diamonds correspond to simulations with $\Omega_\co
= 0.25$ and 0.1, respectively.  The scatter is mostly due to
variations of $C$ from one field line to another (see text for more detail). For $0.5 \le
\nu\le 0.8$, the numerical results lie within the
expected analytical range~\eqref{eq_c_range}, between the solid
and dotted horizontal lines.  For $\nu \gtrsim 0.9$ we are not able to reliably
determine the value of $C$ from the simulation (an error bar
is shown).  However,
for these values of $\nu$, the second regime of acceleration does not operate for any
distance $r<10^6$, and the value of $C$ is unimportant.  Therefore the
same value of $C$ can be used here as well.}
\label{fig_bnufit}
\end{figure}

Model~B has field lines with a parabolic shape, as we expect from
equation~(\ref{eq_fieldlineshapecylzRtext}).
Figures~\ref{fig_nu10m025a15field}, \ref{fig_nu1.0m0.25gamma},
and~\ref{fig_nu1.0m0.25all} show results corresponding to this model.
The jet acceleration is always in the first regime and the
Lorentz factor of the jet is determined only by $\gamma_1$.
Consequently, the maximum acceleration always occurs for the field
line at the jet-wind boundary.  This is obvious in
Fig.~\ref{fig_nu10m025a15field}, where we see that the maximum Lorentz
factor coincides with the maximum in the enclosed current.  Also, in
Fig.~\ref{fig_nu1.0m0.25gamma}, we see that the Lorentz factor of the
field line with $\theta_{\rm fp}=20^\circ$ is always smaller than that
of the line with $\theta_{\rm fp}=64^\circ$. Our model~B simulation is well-converged,
with $\Omega$ and $I$ preserved along the field lines
to better than $15$\%, even though the simulation domain extends
over six orders of magnitude in distance.

In model~A, it was the presence of the second regime of Lorentz factor
growth that was responsible for the development of a faster jet core.
This regime is absent in model~B (compare
Fig.~\ref{fig_nu0.75m0.25all} and Fig.~\ref{fig_nu1.0m0.25gamma}).
Indeed, everything is much simpler in model~B.  For instance,
Fig.~\ref{fig_nu1.0m0.25gamma} shows that the analytic formula for the
Lorentz factor accurately reproduces the numerical profile, and
Fig.~\ref{fig_nu1.0m0.25all} shows that both the field line shape and
the comoving pressure accurately follow the predicted dependencies,
$\theta \propto r^{-1/2}$, $p_{\rm mag}\propto r^{-2}$.  We obtain
this kind of close agreement for all models with $\nu\ge1$.

We have investigated the sensitivity of the models to the rotation
profile in the disc (the value of $\beta$), the magnitude of the
stellar spin (the value of $\Omega_0$), and the geometry of the field
threading the star.  For $\nu$ ranging from $0.5$ to $1.25$ we tried
different values of these parameters. In particular, we have done
simulations with a uniform rotation velocity in the disc, i.e.,
$\beta=1$, which corresponds to the self-similar model of
\citet{nar07}, and we have tried both a monopole field and
a uniform vertical field threading the
star. We find that these changes do not noticeably affect the jet; in
particular, the field line shape changes negligibly. We have also
investigated the effect of a slower stellar spin: $\Omega_\co =
0.1$. We find that the field line shape stays very close to that of
the nonrotating solution so long as $\nu \ge 1$, but changes
logarithmically for $\nu < 1$, as in model~A.

We were particularly interested to see how well the general formula
for the Lorentz factor~\eqref{eq_asymalltext} performs for the range
of models we considered.  Since $\OmegaF$ in the jet region
is not perfectly constant due
to inevitably present numerical diffusion (Fig.~\ref{fig_omegaitransversal}),
\begin{equation}
-1\lesssim\frac{\p\log\OmegaF}{\p\log R} \lesssim 0,
\end{equation}
we expect a range of values for the factor $C$~(eq.~\ref{eq_c_coef_text}),
\begin{equation}
\sqrt2 \lesssim C \lesssim \sqrt{3}. \label{eq_c_range}
\end{equation}
The upper bound $C=\sqrt3$~(eq.~\ref{eq_bnubestfit})
is the analytical value for the case $\OmegaF = \const$.
Figure~\ref{fig_bnufit} shows that for $0.5\le\nu\le0.8$ the best-fit
values of the factor $C$ for various field lines threading
the star are within the expected range~\eqref{eq_c_range},
for all models we considered.
For $\nu \gtrsim 0.9$ the second regime of the Lorentz
factor~\eqref{eq_asym2text} is unimportant (it is realized if at all
only at greater distances than are of interest to us), and so the
value of $C$ is irrelevant.  Thus, we can use the analytical value of~$C$~\eqref{eq_bnubestfit}
with equation~\eqref{eq_asymalltext} for all values of $\nu$ in the range
$0.5 \le \nu \le 1.25$, for all physically relevant values of
$\Omega_0$, and for any value of $\beta$ between $1$ and $1.5$ (we have
not explored other values).  In all cases, for most of the jet (for
field lines with $0\le\theta_\fpt \le 80^\circ$),
$\OmegaF$~is constant to within a few percent and the Lorentz
factor predicted by equation~\eqref{eq_asymalltext} agrees with
the numerical results to better than $10$\%
(see, \eg, Figs.~\ref{fig_nu0.75m0.25gamma} and~\ref{fig_nu1.0m0.25gamma}).

\subsection{Collimation and Transverse Structure of Jets}
\label{sec_collimation}

\begin{figure}
\begin{center}
\epsfig{figure=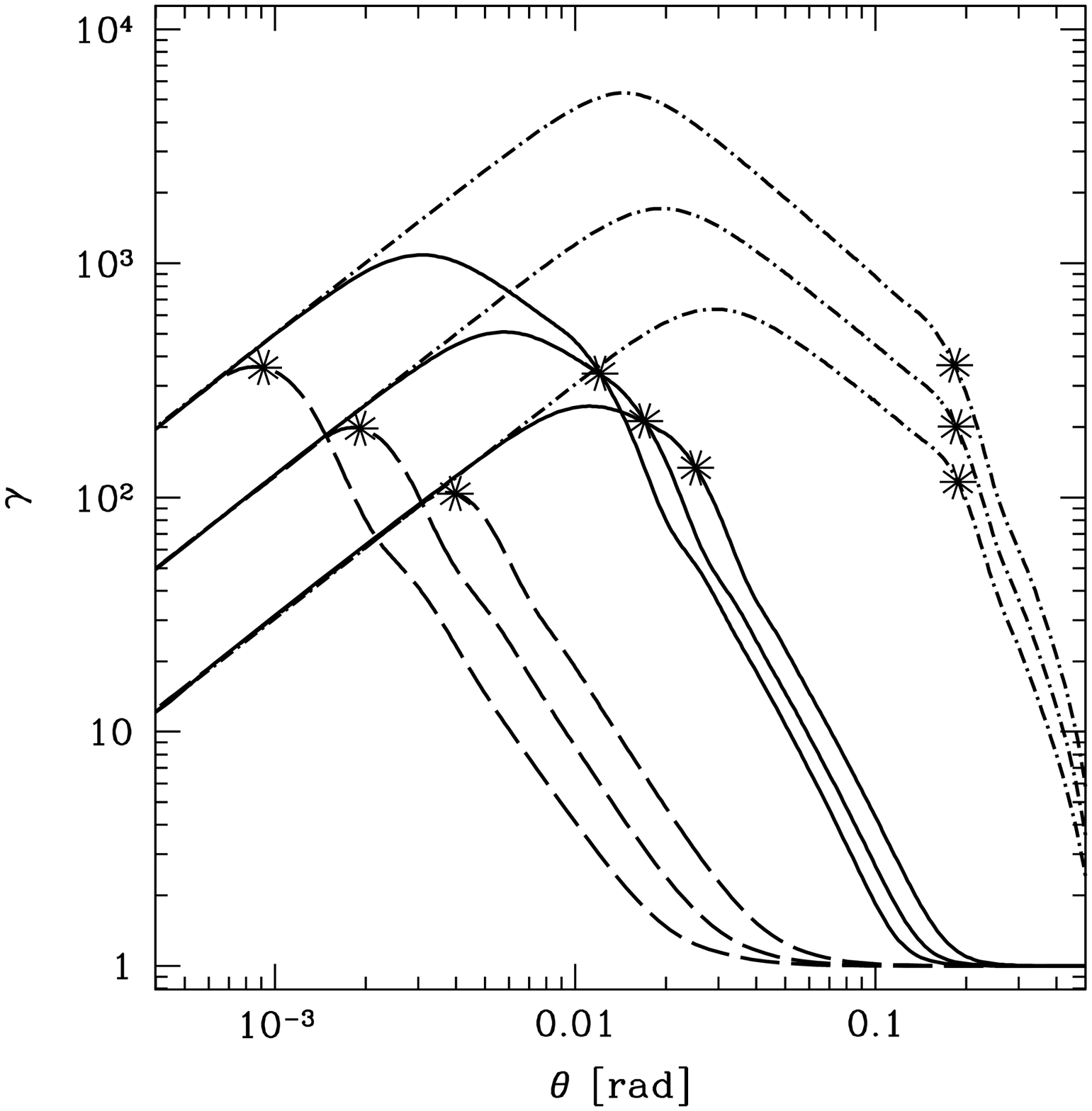,width=\columnwidth}\\
\epsfig{figure=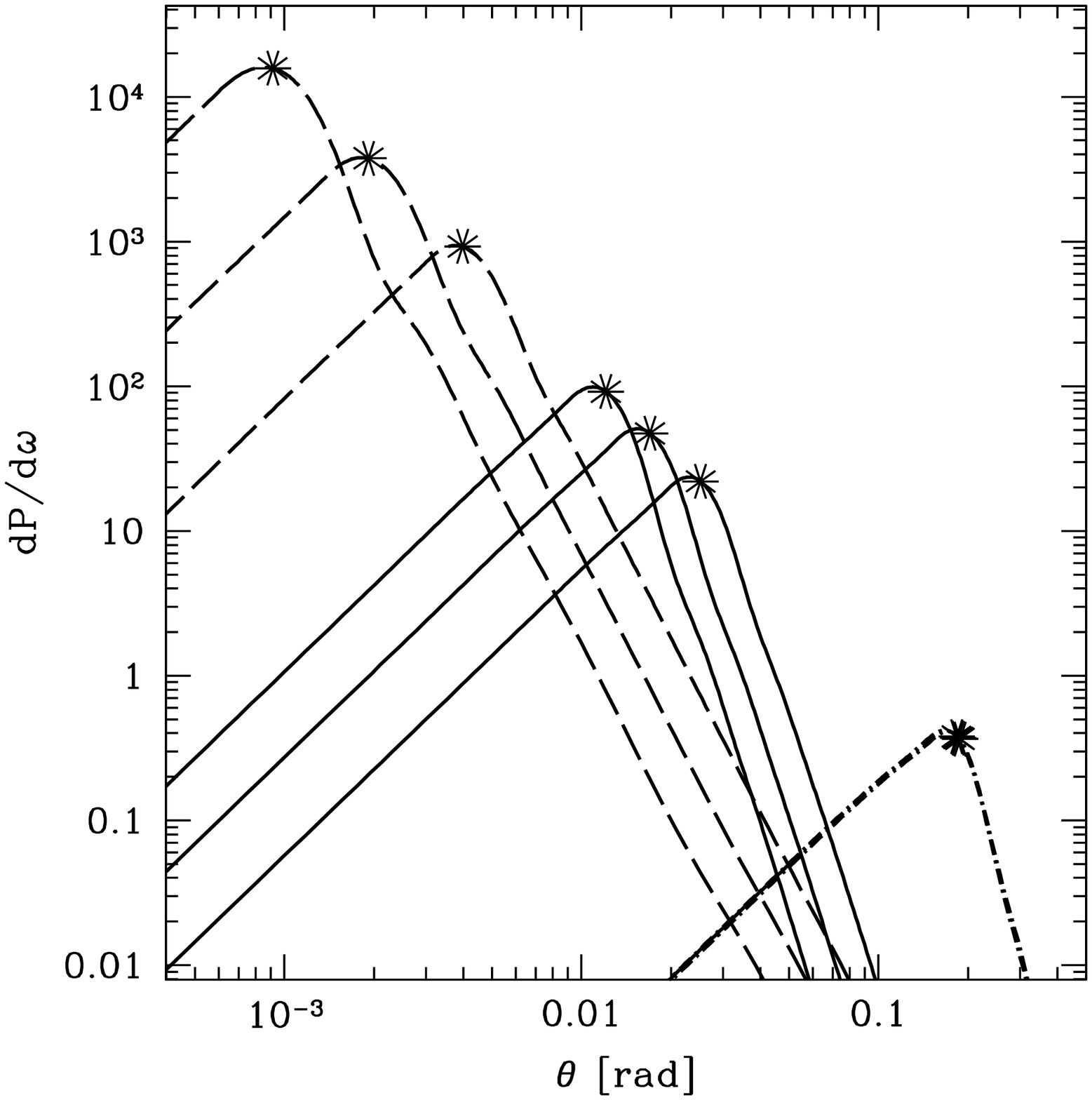,width=\columnwidth}
\end{center}
\caption{ Upper Panel: Angular dependence of the Lorentz factor
$\gamma$ for the fiducial model~A ({}$\nu=0.75$, solid lines),
model~B ({}$\nu=1$, dashed lines), and model~C ({}$\nu=0.6$, dot-dashed
lines).  For each model, three distances from the compact object are
shown.  From bottom to top, they are $r = 1.25\times10^5, ~5\times10^5, ~2
\times 10^6$. For a maximally spinning BH of mass $M = 3 M_\odot$, these distances
correspond to $r = 6 \times 10^{10}$, $2 \times 10^{11}$ and $9 \times
10^{11}$~cm; for a neutron star of radius $10$~km, the distances
in centimeters are approximately half these values.
Asterisks indicate the boundary between the jet and the
wind.  Note that for all models and distances shown, $\gamma$ is
$\sim100$ or larger.  Note also that, for models A and C, the maximum
value of $\gamma$ occurs inside the jet, whereas for model B, it
occurs at the boundary between the jet and the wind.
Lower Panel: Angular dependence of the jet energy
flux per unit solid angle, $dP/d\omega$, for the same models and
distances.  The energy flux is maximum at the jet-wind boundary
for all three models.  }
\label{fig_gammathetaprofile}
\end{figure}

Figures~\ref{fig_nu0.75m0.25gamma}, \ref{fig_nu0.75m0.25all},
\ref{fig_nu1.0m0.25gamma}, and~\ref{fig_nu1.0m0.25all} show the
behaviour of various quantities along field lines in models A and B.
We now consider how these quantities vary across the jet at a given
distance from the compact object.  The results are shown in
Fig.~\ref{fig_gammathetaprofile}.

The upper panel of Fig.~\ref{fig_gammathetaprofile} shows the angular
profile of the Lorentz factor at various distances from the central
object for each of the three models, A, B, C.  Consider the simplest
of the three models, model B, which has $\nu=1$.  As
Fig.~\ref{fig_nu1.0m0.25gamma} shows, in this case the Lorentz factor
is simply equal to $\Omega R$ at large distances.  Since the field
lines in the `jet' region of the outflow are all connected to the
rigidly rotating compact object at the center (see
Fig.~\ref{fig_bhcartoon}), all of them have the same $\Omega=\Omega_\co$.
Therefore, we expect $\gamma=\Omega_\co R\approx\Omega_\co r\theta$.
This linear increase of $\gamma$ with $\theta$ at a fixed $r$
terminates at the edge of the jet, $\theta=\theta_j$.  At the jet
boundary we have
\begin{equation}
\theta = \theta_j\approx\sqrt{\frac{2}{r}}, \quad
\gamma\approx {2\frac{\Omega_\co}{\theta_j}}={\frac{0.5}{\theta_j}}
\approx 0.35 \sqrt{r}.
\label{Bgammatheta}
\end{equation}

Beyond the edge of the jet, the field lines are attached to the disc,
where $\Omega$ falls rapidly with increasing radius.  Therefore, the
Lorentz factor decreases quickly.
Appendix~\ref{sec_lorentz_factor_jet_and_wind} shows that the expected
dependence is $\gamma\propto \theta^{-2}$.  The dashed lines in
Fig.~\ref{fig_gammathetaprofile} confirm the scalings of $\gamma$
with $\theta$ both inside and outside the jet.  They also show how
$\theta_j$ and $\gamma$ vary with distance from the compact object.

Consider next model A, with $\nu=0.75$.  Now we expect the jet angle
to scale as
\begin{equation}
\theta_j \approx{\frac{\sqrt{2}}{r^{\nu/2}}}=
{\frac{\sqrt{2}}{r^{3/8}}}.
\label{eq_jet_opening_angle}
\end{equation}
This is approximately verified in Figs.~\ref{fig_nu0.75m0.25all} and
\ref{fig_gammathetaprofile}. However, the agreement is not perfect
because field lines open out slightly at large distances relative to
the analytical approximation of the field line shape.

The Lorentz factor profile has a more complicated behaviour in this
model because of the presence of two different regimes of acceleration
(Fig.~\ref{fig_nu0.75m0.25gamma}).  Close to the axis, the field lines
are in the first acceleration regime, where $\gamma\approx\Omega_0R$
and the behaviour is the same as in model B, \ie\ $\gamma \propto \theta$.
However, at an angle
$\theta=\theta_m$ (`m' for maximum), we begin to see field lines that
have switched to the second acceleration regime, and beyond this point
the Lorentz factor decreases with increasing~$\theta$.  Thus we have
\begin{equation}
\gamma \approx
  \begin{cases}
  \Omega_0r\theta, \quad &0<\theta<\theta_m, \\
  3.8/\theta, \quad &\theta_m<\theta<\theta_j.
  \end{cases}
\end{equation}
The coefficient 3.8 is obtained from the simulations, but it is close
to the analytical value of $\kappa=3.6$.  (The small difference is
because of the slight opening up of the field lines in the second
acceleration regime).  The angle corresponding to the maximum
Lorentz factor is
\begin{equation}
\theta_m \approx \sqrt{\frac{3.8}{\Omega_0r}}=\frac{3.9}{\sqrt{r}},
\end{equation}
and the value of the maximum Lorentz factor is (eq.~\ref{eq_asymalltext})
\begin{equation}
\gamma_m \approx \frac{1}{\sqrt{2}} \Omega_\co r \theta_m
         \approx \sqrt{1.9 \Omega_\co r} \approx 0.7 \sqrt{r}.
         \label{eq_gammam_modela}
\end{equation}

Interestingly, over the entire range of angles from $\theta_m$ to
$\theta_j$, we have the simple scaling
\begin{equation}
\gamma=\frac{3.8}{\theta}, \quad \theta_m < \theta < \theta_j.
\end{equation}
Note also that the coefficient in this relation is larger than the one
corresponding to model B (eq.~\ref{Bgammatheta}).  Thus, for the same
Lorentz factor, the jet in model A has at least six times larger
opening angle than the jet in model~B.
Beyond the edge of the jet, in the wind region, the Lorentz factor
drops rapidly as $\theta^{-2.5}$.

We can now make a general prediction for how the peak Lorentz factor $\gamma_m$ scales
with distance from the compact object.  For a maximally
spinning BH, $\gamma_m$ attained at a distance~$r$ is likely
to lie in the range bounded by models $A$ and $B$ (eqs.~\ref{Bgammatheta}
and~\ref{eq_gammam_modela}):
\begin{equation}
\gamma_m(r) = (0.3-0.7) \sqrt{r}.
\label{eq_gamma_sqrt_scaling}
\end{equation}
Note that this formula gives the maximum value of the
Lorentz factor over all field lines, as opposed to
a value of the Lorentz factor along a single field line.

As we see from Fig.~\ref{fig_gammathetaprofile}, model~C ($\nu=0.6$)
is a more extreme version of model~A. The jet is significantly wider
({}$\theta_j\sim0.2$ radians), the maximum Lorentz factor is much
larger ($\sim5000$ at $r=2\times10^6$), and the maximum of~$\gamma$
occurs at $\theta \ll \theta_j$. Unfortunately, our numerical results
for this model show large deviations from the analytical model
described in the Appendix. In particular, the poloidal field line
shape is nearly monopolar at large distance, i.e., $\theta$ is nearly
independent of $r$ along each field line, instead of behaving as
$\theta \sim r^{-0.3}$ (see eq.~\ref{eq_thetaj}).\footnote{We note
that the monopole has a low acceleration efficiency for
a finite value of magnetization~$\sigma$ \citep{bes98,bog99}.
Therefore, for a finite magnetization, the
acceleration efficiency might be also low for model~C. } Qualitatively,
however, model~C is similar to model~A.  Model~C is well-converged
with $\Omega$ and $I$ preserved along the field
lines to better than~$15$\% throughout the simulation domain.

We consider next the power output of the jet.  We define the angular
density of electromagnetic energy flux as
\begin{equation}
\frac{dP}{d\omega} = r^2 \Abs{\myvec{S}}
                   = \frac{r^2}{4\pi} \Abs{\myvec E \times \myvec B},
\end{equation}
where $\myvec S$ is the Poynting vector and $d\omega =
d\varphi\,d\theta\sin\theta$ is the solid angle.  As we show in
Appendix~\ref{appendix_poweroutput} and verify in
Fig.~\ref{fig_gammathetaprofile}, the power output grows quadratically
with distance from the jet axis for all models,
\begin{equation}
\frac{dP}{d\omega} \approx \frac{1}{4\pi} \Omega_\co^2 r^{2\nu} \theta^2,
                   \quad \theta \le \theta_j,
                   \label{eq_dpdomega}
\end{equation}
reaches its maximum at the jet boundary, indicated by asterisks in
Fig.~\ref{fig_gammathetaprofile},
\begin{equation}
\frac{dP}{d\omega}\biggr|_{\theta=\theta_j}
   \approx \frac{\Omega_\co^2}{\pi\theta_j^2}
   \approx \frac{0.02}{\theta_j^2},
   \label{eq_dpdomegamax}
\end{equation}
and falls off rapidly in the wind region as~$\theta^{2-4\beta/\nu}$.
Equation~\eqref{eq_dpdomegamax} illustrates that jets with a smaller
opening angle have a larger power output per unit solid angle. The
steepest decline of angular power in the wind region
is~\hbox{$\theta^{-10}$} in model~C, followed by less steep declines
of~\hbox{$\theta^{-6}$} in model~A and~\hbox{$\theta^{-4}$} in
model~B.  This behaviour can be seen in the lower panel of
Fig.~\ref{fig_gammathetaprofile}.  The angular power output
profile in model~C does not evolve with distance
since the opening angle of the jet is nearly
independent of distance.

The total power output in the jet and the wind is~(see
Appendix~\ref{appendix_poweroutput})
\begin{align}
P^\text{jet}  &\approx \frac{\Omega_\co^2B_r^2}{2} = \frac{B_r^2}{32}, \\
P^\text{wind} &\approx \frac{\Omega_\co^2B_r^2}{2(\beta/\nu-1)}
              = \frac{B_r^2}{32(\beta/\nu-1)},
\end{align}
where $B_r$ is the radial magnetic field strength near the BH.
The total power is independent of the distance at which it is evaluated, which is
a manifestation of the fact that energy flows along poloidal field
lines.  The total power output in the jet is the same for all models,
and the total power in the wind varies from model to model.  The most
energetic wind is in model~B and carries twice as much power as the
jet.  In model~A, the wind and the jet have equal power outputs, and
in model~C, the power output in the wind is two thirds of the power in the
jet.

To obtain the total power of the jet in physical (cgs) units, we
specialize to an astrophysical system with a BH of mass $M$, radial
field strength near the BH $B_r$, and angular rotation
frequency of field lines $\Omega_\co$.  We obtain
\begin{align}
P^\text{jet}
  &= \frac{1}{2} \Omega_\co^2 \, B_r^2 \, r_0^2 \, c  \notag\\
  &\approx 1.8\times10^{50} \left(\frac{\Omega_\co}{\Omega_\text{max}}\right)^2
                     \left(\frac{B_r}{10^{15}\text{G}}\right)^2
                     \left(\frac{M}{3M_\odot}\right)^2
                     \left[\frac{\text{erg}}{\text{s}}\right],
                     \label{eq_totaljetpower}
\end{align}
where $r_0 \approx r_g = GM/c^2$ for a rapidly spinning BH.  For
characteristic values, $\Omega_\co\sim\Omega_\text{max}=0.25c/r_g$, $B_r \sim 10^{15}$~G, $M
\sim 3M_\odot$, and taking the typical duration of a long GRB $\sim 10
- 100$~s, the model predicts a total energy output of
$10^{51}-10^{52}$ erg, which is comparable to the energy output
inferred for GRB jets~\citep{pir05}.  We note that the actual
value of the magnetic field $B_r$ might be higher since the observations
only account for a fraction of the electromagnetic energy flux.
Recent GRMHD simulations suggest a value of $10^{16}$~G~\citep{mck05}.

\section{Comparison to Other Work}
\label{sec_comparison_to_other_work}

Since the observed energy output of long GRBs is
$10^{51}-10^{52}$~erg \citep{pir05}, any model of long GRBs requires a
central engine capable of supplying this copious amount of energy. In
the absence of magnetic fields, a possible energy source is
annihilation of neutrinos from the accretion
disc~\citep{kohri_neutrino_dominated_2005, chen_neutrino_cooled_2007,
kawanaka_neutrino_cooled_2007}. Attempts to include the neutrino
physics self-consistently have so far not succeeded in producing
relativistic jets~\citep{nagataki_neutrino_jets_2007,
takiwaki_special_2007}. However, an ad-hoc quasi-isotropic energy
deposition of $10^{50} - 10^{52}$ ergs into the polar regions seems to lead
to jets with Lorentz factors~$\sim100$~\citep{alo00,
zhang_relativistic_jets_2003, zwh04,mor06, wang_relativistic_2007}. The jets
so formed are found to be stable in two and three dimensions, to accelerate via
the conversion of internal energy into bulk kinetic energy, and to
collimate to angles $\lesssim 5^\circ$ as a result of interacting with the
dense stellar envelope. We note that similar hydrodynamic simulations of
short GRB jets have attained much higher Lorentz factors $\sim700$
due to the absence of a stellar envelope that the jet has to
penetrate~\citep{aloy_shortgrb_2005}.

The inclusion of magnetic fields enables the extraction of energy from
spinning BHs via the Blandford-Znajek mechanism~\citep{bz77,km07}
and from accretion discs via the action of magnetic torques~\citep{bp82}.
The paradigm of electromagnetically powered jets
is particularly appealing since it is able
to self-consistently reproduce the observed jet energetics of
long GRBs~\citep[see \S\ref{sec_collimation} and][]{mck05}, without
any need for ad-hoc energy deposition.
Numerical simulations, within the framework of general relativistic
magnetohydrodynamics (GRMHD), of magnetized accretion discs around spinning
BHs naturally produce mildly relativistic
jets~\citep{mck04,mck05,dhkh05,hk06,bhk07,barkov_stellar_2007} as well as relativistic
jets with Lorentz factors $\sim 10$~\citep{mck06jf}.

Our results are consistent with those of~\citet{mck06jf} of a highly magnetized jet
that is supported by the corona and disc wind up to a radius $r\sim\text{few}\times10^2$ by which
the Lorentz factor is $\sim 10$ similar to expected by our $\nu=3/4$ model.
They did not include a dense stellar envelope.  Beyond the distance $r\sim\text{few}\times10^2$ the disk wind
no longer confines the jet, which proceeds to open up and
become conical as it passes the fast magnetosonic surface.
Beyond the fast magnetosonic surface they find the jet is no longer
efficiently accelerated, as consistent with expectations of
unconfined, conical MHD solutions~\citep{bes98,bog99}.  The jet simulated by \citet{mck06jf}
shows a mild hollow core in $\gamma$ but shows no hollow core in energy flux,
although this may be a result of numerical diffusion
or significant time-dependence within the jet.

\citet{buc06,buc07,buc08,kb07} studied the formation and propagation of relativistic jets from neutron
stars as a model for core-collapse-driven GRB jets.  They were unable to study the formation and
propagation of ultrarelativistic jets likely due to computational constraints and their model setup.  While our model is somewhat idealized
compared to those studies, we demonstrate the generic process of magnetic-driven acceleration to
ultrarelativistic speeds and thus extend their simulations of magnetar-driven GRB jets.

\citet{mck07a,mck07b} found that the jets in their simulations are
collimated by the pressure support from a surrounding ambient medium, such as
a disc corona/wind or stellar envelope, rather than being
self-collimated.  Thus the treatment of the medium that confines the jet appears to be crucial.
\citet{kom07,barkov_ultrarel_proceeding_2008} modelled the action of such an ambient medium by
introducing a rigid wall at the outer boundary of the jet and studied the magnetic acceleration.
They obtained solutions with Lorentz factors of~$\sim 300$ with an efficient conversion of magnetic to
kinetic energy.
In the current
work, instead of keeping the shape of the jet boundary fixed via a rigid wall, we
prescribe the ambient pressure profile, so that the shape of the jet
boundary can self-adjust in response to pressure changes inside the
jet~\citep{mck07a}.  We believe that this is a more natural boundary condition
for modeling GRB jets confined by the pressure of
the stellar envelope.

The magnetodynamical, or force-free, approximation may provide a good approximation to the
field geometry even in the mass-loaded MHD regime as long as the flow
is far from the singular monopole case of an unconfined flow.  \citet{fo04} used this fact to study
ultrarelativistic jets in the GRB context.  As in our work, they determine the field line
geometry in the magnetodynamical regime, but they use energy conservation for the particles to determine
an approximate MHD solution and approximate efficiency of conversion from magnetic to kinetic energy.

The conversion efficiency for fully self-consistent highly relativistic MHD solutions
has been studied only for a limited number of field geometries.
High conversion efficiency
was found for a parabolic $\nu = 1$ solution and an intermediate $\nu = 2/3$ solution,
but not for the singular monopole $\nu = 0$ solution~\citep{bes98,bes06,barkov_ultrarel_proceeding_2008}.
Following this work we plan to
include particle rest-mass and systematically study the efficiency of particle
acceleration in highly relativistic self-consistent MHD solutions.  For
this we will use the same numerical scheme as in the present numerical work but optimized for the ultrarelativistic MHD
regime~\citep{gam03,migmck07,tch_wham07}.

Since we assume axial symmetry in this study, our simulations cannot address the question of jet
stability to the $3$D kink (screw) mode. Unlike the hydrodynamic case, we are not aware of
any studies of relativistically magnetized jets in $3$D that address this issue.
However, all jets in this paper have $B_\varphi \approx - \Omega R
B_r$ (see eq.~\ref{eq_bphiofbp}). Thus, they marginally satisfy the
stability criterion of~\citet{tom01}, suggesting that our jets
are marginally stable to kink instability.  Also, the spontaneous development of the spine-sheath
structure in models with $\nu<1$ may be naturally stabilizing \citep{mhn07,hardee07,hmn07}.

\section{Astrophysical Application}
\label{sec_astrophysical_applications}

Since we consider magnetodynamic, or force-free, jets in this paper, we
are not able to study the effects of mass-loading of the jet. However, we expect
the main properties of our infinitely magnetized jets to carry over to
mass-loaded jets, provided the latter are electromagnetically
dominated.  Mass-loaded jets stay electromagnetically dominated as
long as the Lorentz factor is well below the
initial magnetization~$\sigma$ of the jet~\citep{bes06}, where~$\sigma$ is the ratio of
electromagnetic energy flux to mass energy flux at the base of the
jet.  We assume that this condition is satisfied and proceed to apply
our results to GRBs and other astrophysical systems with relativistic
jets.

\subsection{Application to Long GRBs}
\label{sec_application_lgrbs}

The first question of interest is what sets the terminal Lorentz
factor of a relativistic jet.  We have shown in this paper that
$\gamma$ increases with distance from the compact object as $\gamma
\sim 1/\theta_j \sim r^{\nu/2}$.  The value of $\nu$ is determined by
the radial dependence of the confining pressure: if pressure varies as
$r^{-\alpha} \sim r^{-2.5}$, then $\nu=2-(\alpha/2) \sim 0.75$.

In the context of the collapsar or magnetar model of GRBs, the
confining pressure is primarily due to the stellar envelope, and hence
acceleration is expected to continue only until the jet leaves the
star.  Once outside, the jet loses pressure support and the magnetic
field configuration will probably become conical (monopolar).
This geometry is inefficient
for accelerating particles to Lorentz factors larger
than~$\gamma_\text{max}\sim\sigma^{1/3}$~\citep{bes98,bog99}.
We note that mildly relativistic edges of the jet will
expand quasi-spherically after the jet loses pressure support,
while the ultrarelativistic jet core will open up at most
conically~\citep[][]{lyutikov_grbs_2003}.  In fact,
if the jet loses pressure support at a distance $r\sim r_\text{s}$
(which corresponds to the radius of the progenitor star),
the opening angle of the jet core will stay approximately constant
until the jet reaches a much larger distance
$r\sim(\gamma\theta_j)^2_{r=r_\text{s}} r_\text{s}\gtrsim10^2 r_\text{s}$
(for our models A and~C).
We note also that current-driven instabilities may set in when the jet loses pressure
support, and much of the electromagnetic energy may be converted into
thermal energy \citep{mck06jf,lyutikov_em_model_grbs_2006}.
These additional topics are beyond the scope of the present paper
and require simulations that include a loss of pressure support at
large radius to model the stellar surface.

From the above discussion, we expect that \emph{the terminal Lorentz
factor of the jet in a long GRB is determined primarily by the size of the
progenitor star}. Figure~\ref{fig_gammathetaprofile} shows that, for
a BH of a few solar masses and
stellar radii in the range ${\rm few}\times10^{10}$ cm to $10^{12}$
cm, the Lorentz factor of the jet is expected to be between about $100$
and $5000$.  Such masses and stellar radii are just what we expect for GRB
progenitors, and the calculated Lorentz factors are perfectly
consistent with the values of $\gamma$ inferred from GRB observations.

For our fiducial model, the size of the star determines the terminal
Lorentz factor as long as the initial magnetization of the jet
is sufficiently high, $\sigma\gtrsim10^3$, but not infinite, $\sigma\lesssim10^9$.
The first condition ensures that the jet inside the star is
well-described by the force-free approximation. However, once the jet
breaks out of the star and its geometry becomes monopolar, the effects
of finite magnetization will set in, provided the second condition is
satisfied.  Acceleration in such monopolar field is ineffective,
therefore the growth of the Lorentz factor stops.
Estimates of baryon loading of magnetized GRB jets from black hole
accretion systems are
somewhat uncertain. We can estimate the initial magnetization of a GRB jet
as the ratio of electromagnetic jet power output~\eqref{eq_totaljetpower} to
the rate of baryon mass-loading of the
jet~\citep[their eq.~A10]{levinson_eichler_grb_baryon_loading_2003,mckinney2005},
which gives a value of initial jet magnetization
$\sigma \sim 10^3$ for characteristic values
of the accretion system parameters.
However, given the uncertainty in these parameters, the actual value
of magnetization could be an order of magnitude lower or higher.
Note that magnetically accelerated jets behave very differently from
relativistic fireballs.  
Beyond the star fireballs adiabatically expand leading to $\gamma\propto r$
until nearly all thermal energy is exhausted, and so there is no spatial
scale that sets a terminal Lorentz factor.  Fireball models require the
baryon-loading of the jet be fine-tuned to obtain the observed Lorentz
factor and opening angle.  However, in the MHD case, acceleration
essentially ceases at the stellar surface for a wide range of values of
$\sigma$. 

\begin{figure*}
{\hfill\epsfig{figure=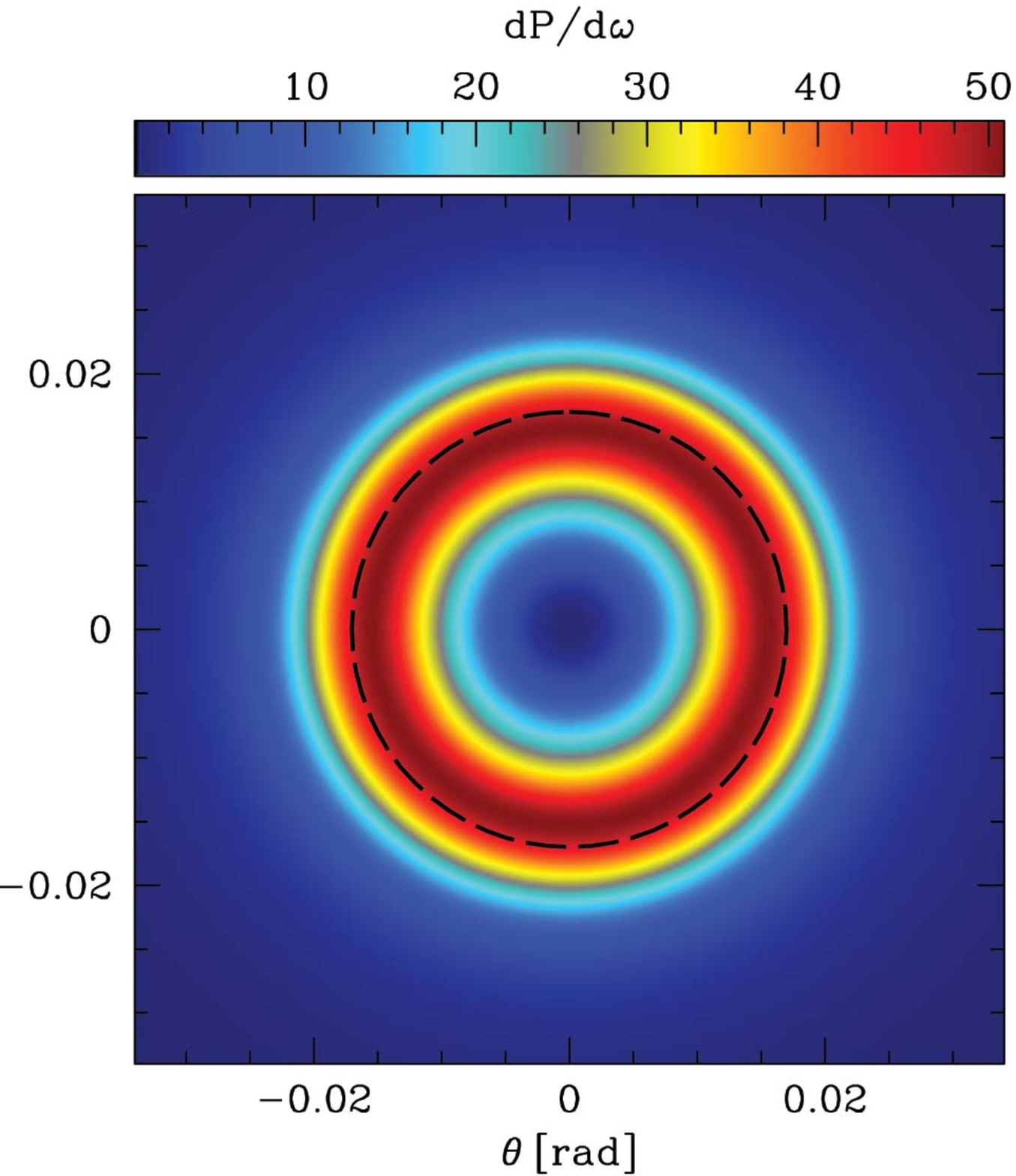,width=0.45\textwidth}\hfill\epsfig{figure=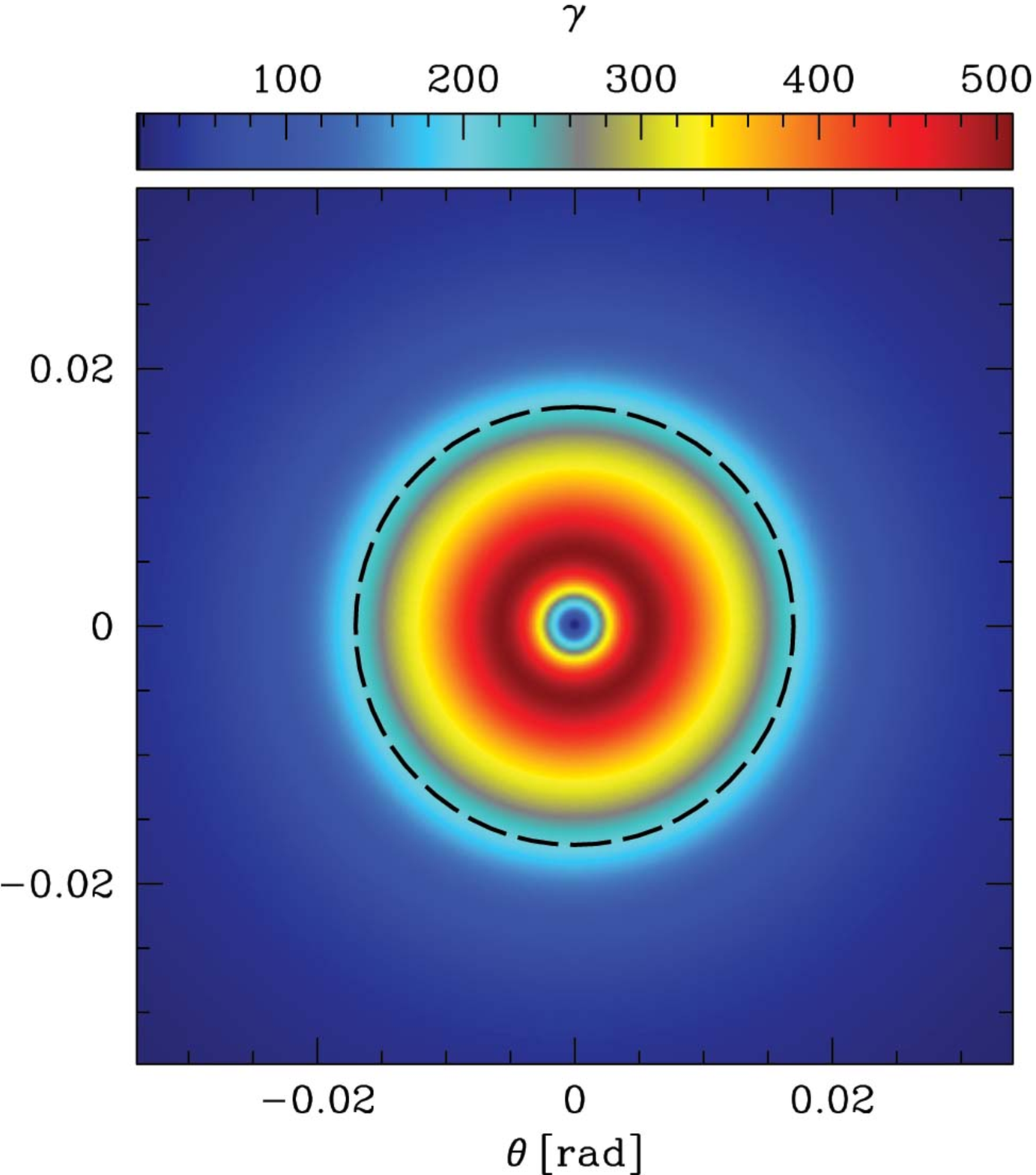,width=0.45\textwidth}\hfill{}}
\caption{Face-on view of the lateral structure of the jet in
model~A at a distance of $r = 5\times 10^5$ ({}$2\times10^{11}$~cm for
a maximally spinning BH of mass \hbox{$M = 3M_\odot$}).  The left panel shows the
energy flux of the jet per unit solid angle, $dP/d\omega$, and the right panel
shows the Lorentz factor, $\gamma$.  Red colour indicates large and
blue colour small values of the respective quantities.
The dashed line shows the jet-wind boundary. Note that the maximum
value of the energy flux, $dP/d\omega$, occurs at the jet-wind boundary,
while $\gamma$ is maximal inside the jet.
}
\label{fig_faceon_jetview}
\end{figure*}

We discuss next the degree of collimation of the jet.  The three
models we have described in this paper form a one-parameter sequence.
Model~B has the most highly collimated jet and model~C has the least
collimated jet, while our fiducial model~A lies in-between.  From
Figure~\ref{fig_gammathetaprofile} we see that the models produce jet
opening angles in the range $10^{-3} - 0.2$ radians, which is
consistent with the typical opening angles observed in long
GRBs $\sim0.03 - 1$~rad \citep{frail01,zeh_long_grb_angles_2006}.  In making this comparison,
we are assuming that the opening angle of the jet is set by its value
when the jet emerges from the star.

Observations of GRB afterglows often indicate an achromatic `jet
break' in the light curve a day or two after the initial burst.
These breaks suggest that the opening angle $\theta_j$ of the jet is
substantially larger than the initial beaming angle $1/\gamma$ of the
GRB.  Roughly, it seems that GRBs have $\gamma\theta_j \gtrsim 10$.
In this regard, the three models described in this paper have very
different properties.  Model~B ($\nu=1$) has $\gamma \theta_j<1$.
This model would predict a jet break immediately at the
end of the prompt GRB, which is very different from what is seen.
Model~A ($\nu=0.75$) has $\gamma\theta_j\sim 4$ for the plasma at the
boundary of the jet.  This model has larger values of
$\gamma$ in the interior, so it predicts a range of values of
$\gamma\theta_j$ from $4$ up to about $15$.  Model~C ($\nu=0.6$) is even
more extreme, giving values of $\gamma\theta_j \sim 25-950$.  These
models span a range that is more than wide enough to match
observations.  Our fiducial model with $\nu=0.75$ appears to be most consistent
with the data.

Note that the different models correspond to different radial profiles
of pressure in the confining medium: model~B has pressure varying as
$r^{-2}$, model~A goes as $r^{-2.5}$, and model~C as $r^{-2.8}$.  We
selected model~A as our fiducial model because a superficial
study of GRB progenitor star models suggested that the pressure
probably varies as $r^{-2.5}$.  However, in view of the fact that
jet properties depend strongly on the value of the index, a
more detailed study of this point is warranted.

Finally, we consider the jet power.  The total energy output in a GRB
jet is given by equation~\eqref{eq_totaljetpower}.  This assumes that the power
delivered by the disc wind is deposited entirely into the stellar envelope and
so does not contribute to the jet, although the mass-loading
of field lines near the jet-wind boundary probably depends upon the amount of
turbulence in the interface \citep{mck06jf,mor06,wang_relativistic_2007}
and non-ideal diffusion physics \citep{le03,mckinney2005,rbr06}.
For characteristic values of the magnetic field strength $B_r$ near the compact object
and the angular frequency $\Omega_\co$ of the compact object, the
energy released over the typical duration of a long GRB is about
$10^{51} - 10^{52}$ erg, which is consistent with the observed
power~\citep{pir05}.  Other things being equal, the energy scales as
$\Omega_\co^2$.  This scaling~\citep{bz77} would
certainly apply to a magnetar.  In the case of an accreting black
hole, the magnetic field strength near the black hole may itself
increase rapidly with $\Omega_\co$ (for a fixed mass accretion rate),
and this may give a steeper dependence of jet power on $\Omega_\co$
\citep{mck05}.

Figure~\ref{fig_gammathetaprofile} indicates that the electromagnetic
energy flux in the jet has a substantial variation across the jet. The
power is very low near the jet axis, and most of the energy comes out
near the jet boundary.  Note in particular that, for the
observationally relevant model~A, the maximum jet power
does not coincide with the maximum Lorentz
factor~(see Fig.~\ref{fig_faceon_jetview}).  This unusual
behaviour is the result of the second acceleration regime which we
discussed in \S\ref{sec_collimation} (see Figs.~\ref{fig_nu075m025a15field},
\ref{fig_nu0.75m0.25gamma}, \ref{fig_gammathetaprofile}).
Such a ring-like shaped distribution of Lorentz factor
in a fireball naturally leads to the Amati relation of an observed correlation
between de-redshifted peak frequency in the GRB spectrum and the isotropic
equivalent luminosity of GRBs~\citep{eichler_levinson_on_amati_2004}.
Regardless, the fact that the
jet power comes out along a hollow cone, and not as a uniformly filled
cone as assumed for example by \citet{rhoads_uniform_jet_1997,rhoads_uniform_jet_1999}
and \citet{moderski_afterglow_light_curves_2000}, is
likely to have observational consequences for both the prompt emission
from a GRB and its afterglow~\citep{granot_jet_unconventional_structure_2005}.
We are assuming, of course, that the
electromagnetic power which we calculate from our model is directly
proportional to the observed radiative power (both prompt and afterglow).

\citet{rlr02} suggested the interesting possibility
that GRB jets may be `structured,' with the energy flux per unit
solid angle, $dP/d\omega$, having a flat core and the
power falling off as $\sim\theta^{-2}$ outside the core.  We have
already seen that our electromagnetic jets do not have a flat core.
In addition, we find that the power outside the jet, in the wind
region, falls off very steeply.  The variation is $\theta^{-4}$ in the
mildest case (model~B) and is as steep as $\theta^{-10}$ in the most
extreme case (model~C).  In addition,  the disc wind in
our idealized model is merely a proxy for a gaseous confining medium,
which means that the electromagnetic power in this region may be even
less than we estimate.  Entrainment and instabilities within the jet may lead to a less sharp
distribution at large angles.

\subsection{Application to AGN, X-ray Binaries, and Short GRBs}

In the case of accreting black holes in active galactic nuclei
(AGN), accreting neutron stars and black holes in X-ray binaries, or
accreting black holes for short GRBs~\citep[which presumably form
as a result of a coalescence of compact
object binary systems,][]{pir05,meszaros_grbs_2006},
there is no stellar envelope to
confine the jet.
Therefore, the only confining medium available is
the wind from the inner regions of the accretion disc.  The strength
of these winds is known to be a strong function of the radiative
efficiency of the disc.  A radiatively inefficient disc (\ie, an
advection-dominated accretion flow, ADAF), will have a strong disc
wind~\citep{nar94}, whereas a radiatively efficient disc (\ie,
a standard thin disc, \citealt{sha73}), will normally have a much weaker wind.  Thus,
jet acceleration should be more effective in systems with ADAFs, viz.,
low-luminosity AGN and
black hole binaries (BHBs) in the `hard state' and the `quiescent state'~\citep{nm08}.
Indeed, observations indicate that these systems are invariably
radio-loud, whereas higher luminosity AGN and BHBs in the `thermal state',
which are powered by thin accretion discs, are often
radio quiet.  On the other hand, some of the most energetic radio quasars
are associated with high luminosity AGN which presumably have thin
discs.  It is unknown how such jets are confined, but large-scale
force-free magnetic fields could replace the material support of the disk or wind.

The terminal Lorentz factor of the jet in an ADAF or a short GRB system will depend
on the distance out to which the disc wind is able to provide
significant pressure support.  Numerical GRMHD simulations of ADAF-like tori of extent $r\sim 40$
around rapidly spinning BHs~\citep{mck06jf} show that the disc wind
is effective to a radius $r\sim \text{few}\times10^2$, giving a terminal Lorentz factor
$\gamma\sim {\rm few}-10$ as long as $\sigma\lesssim 10^3$.  Such Lorentz factors are consistent with
values inferred for AGN~\citep{jorstad_agn_jet_2005}, BHBs~\citep{fend04a}, and
short GRB jets~\citep{nakar_short_grbs_2007}.  Under some circumstances neutron star X-ray binaries
may produce jets similar to BHBs~\citep{fend04b}.  Short GRBs can be
mass-loaded by neutron deposition and are limited to $\sigma\lesssim 10^3$ just as required.

Once the confining effect of the disc wind ceases at $r\sim100$, further acceleration will require some
other confining medium, e.g., an external interstellar medium, to take
over.  This would appear to be unlikely for a BHB or a short GRB,
but it may work for some AGN. Let us assume that the jet stops collimating
(becomes monopolar) beyond $r\sim100$.
The opening angles that we expect then lie in the range bounded by the
most collimated model~B and the least collimated model~C,
$\theta_j \sim 0.14 - 0.5$ (eq.~\ref{eq_jet_opening_angle}), \ie\
substantially larger than for long GRBs. These values are
perfectly consistent with the jet opening angles
inferred for the
systems considered in this subsection~\citep[see the above references and][]{watson_short_grb_collimation_2006}.

\section{Conclusions}
\label{sec_conclusions}

By performing axisymmetric time-dependent numerical simulations, we have
studied highly magnetized, ultrarelativistic, magnetically-driven jets.
In the context of the collapsar scenario of long GRBs, we obtain global
stationary solutions of magnetodynamic, or force-free, jets confined by an
external medium with a radial pressure profile motivated by models
of GRB progenitor stars (see \S\ref{sec_motivation}).

We find that both the size of the progenitor star
and the radial variation of pressure in the envelope of the star
determine the terminal Lorentz factor and the opening angle
of the jet for a wide range of initial magnetization of the jet.  At the radius where the jet breaks
out of the star, \ie\ at a distance $r \sim 10^{10} - 10^{12}$~cm,
the jets in our models attain bulk Lorentz factors $\gamma \sim
100 - 5000$. These are currently the largest~$\gamma$ attained
in a numerical MHD jet simulation of long duration GRBs.  The Lorentz factors
we obtain are perfectly consistent with the values of $\gamma$
inferred from observations. The simulated jets have
opening angles in the range $\theta_j \sim 10^{-3} - 0.2$~radians, in
agreement with the typical opening angles observed in long duration
GRBs \citep[$\sim0.03 - 1$~rad, ][]{frail01,zeh_long_grb_angles_2006}.

For a maximally rotating black hole or a $\sim1$-ms magnetar, a
characteristic magnetic field near the compact object of $10^{15}$~G,
and a burst duration of $10 - 100$~seconds, our simulated jets provide an
energy output of $10^{51} - 10^{52}$~erg, comparable to the
power output inferred for GRB jets~\citep{pir05}.

The angular structure of the simulated jets is not uniform. The jet
power comes out in a hollow cone, peaking at the jet boundary
(Fig.~\ref{fig_faceon_jetview}). However, the Lorentz factor peaks
neither at the jet axis nor at the jet boundary but rather in between
(Fig.~\ref{fig_faceon_jetview}). This nonuniform lateral jet structure
may have observational consequences for both the prompt
emission from GRBs and afterglows.

To fully interpret simulation results, we derive a simple approximate
analytical model (Appendix~\ref{sec_approximate_analytic_jet_solution})
which gives the scaling of~$\gamma$ and~$\theta_j$ as a function of
distance away from the compact object, as well as the variation
of~$\gamma$ and energy flux across the face of the jet. With these
scalings we are able to understand all our simulation results, and we can
predict the properties of highly magnetized jets in more general
situations.  In particular, in most of our models we find that the maximum Lorentz
factor at a distance $z$ away from the compact object is
$\gamma_m\sim (z/r_0)^{1/2}$, where $r_0$ is the central black hole/magnetar radius
(eq.~\ref{eq_gamma_sqrt_scaling}).

While the magnetodynamical regime that we have studied allows us to establish the ability of highly magnetized
flows to accelerate and collimate into ultrarelativistic jets, this approximation cannot be used to establish the efficiency of
conversion from magnetic to kinetic energy.  Following this work we plan to include
particle rest-mass to study the properties of jets in the MHD regime and to determine the efficiency
with which plasma is accelerated in the ultrarelativistic MHD regime.
Future simulations will focus on the time-dependent formation of ultrarelativistic magnetized
jets from the central engine and propagation of the
jet through a realistic stellar envelope (see, e.g., \citealt{takiwaki_special_2007,buc08}).
Our present magnetodynamical and future cold MHD results should be a useful theoretical guide for
understanding these more realistic and complicated simulations.

\section*{Acknowledgements}
We thank Vasily Beskin, Pawan Kumar, Matthew McQuinn, Shin Mineshige, Ehud Nakar, Tsvi Piran, and Dmitri Uzdensky for useful discussions
and the referee, Serguei Komissarov, for various suggestions that helped to
improve the paper.
JCM has been supported by a Harvard Institute for Theory and Computation Fellowship and
by NASA through Chandra Postdoctoral Fellowship PF7-80048 awarded by the Chandra X-Ray Observatory Center.
The simulations described in this paper were run on the BlueGene/L system
at the Harvard SEAS CyberInfrastructures Lab.


\begin{thebibliography}{}

\bibitem[\protect\citeauthoryear{{Aloy}, {Janka} \& {M{\"u}ller}}{{Aloy}
  et~al.}{2005}]{aloy_shortgrb_2005}
{Aloy} M.~A.,  {Janka} H.-T.,    {M{\"u}ller} E.,  2005, \aap, 436, 273

\bibitem[\protect\citeauthoryear{{Aloy}, {M{\"u}ller}, {Ib{\'a}{\~n}ez},
  {Mart{\'{\i}}} \& {MacFadyen}}{{Aloy} et~al.}{2000}]{alo00}
{Aloy} M.~A.,  {M{\"u}ller} E.,  {Ib{\'a}{\~n}ez} J.~M.,  {Mart{\'{\i}}} J.~M.,
     {MacFadyen} A.,  2000, \apjl, 531, L119

\bibitem[\protect\citeauthoryear{{Aloy} \& {Obergaulinger}}{{Aloy} \&
  {Obergaulinger}}{2007}]{aloy_relativistic_outflows_2007}
{Aloy} M.~A.,  {Obergaulinger} M.,  2007, in Revista Mexicana de Astronomia y
  Astrofisica Conference Series Vol.~30 of Revista Mexicana de Astronomia y
  Astrofisica Conference Series, {Relativistic Outflows in Gamma-Ray Bursts}.
pp 96--103

\bibitem[\protect\citeauthoryear{{Appl} \& {Camenzind}}{{Appl} \&
  {Camenzind}}{1993}]{appl_camenzind_asymptotic_1993}
{Appl} S.,  {Camenzind} M.,  1993, \aap, 274, 699

\bibitem[\protect\citeauthoryear{Barkov \& Komissarov}{Barkov \&
  Komissarov}{2008}]{barkov_ultrarel_proceeding_2008}
Barkov M.,  Komissarov S.,  2008, preprint (arXiv:0801.4861)

\bibitem[\protect\citeauthoryear{Barkov \& Komissarov}{Barkov \&
  Komissarov}{2007}]{barkov_stellar_2007}
Barkov M.~V.,  Komissarov S.~S.,  2007, preprint (arXiv:0710.2654)

\bibitem[\protect\citeauthoryear{{Beckwith}, {Hawley} \& {Krolik}}{{Beckwith}
  et~al.}{2007}]{bhk07}
{Beckwith} K.,  {Hawley} J.~F.,    {Krolik} J.~H.,  2007, preprint
  (arXiv:0709.3833)

\bibitem[\protect\citeauthoryear{Begelman \& Li}{Begelman \&
  Li}{1994}]{begelman_asymptotic_1994}
Begelman M.~C.,  Li Z.-Y.,  1994, \apj, 426, 269

\bibitem[\protect\citeauthoryear{{Bekenstein} \& {Oron}}{{Bekenstein} \&
  {Oron}}{1978}]{bekenstein_new_conservation_1978}
{Bekenstein} J.~D.,  {Oron} E.,  1978, \prd, 18, 1809

\bibitem[\protect\citeauthoryear{{Beskin}}{{Beskin}}{1997}]{bes97}
{Beskin} V.~S.,  1997, Soviet Physics Uspekhi, 40, 659

\bibitem[\protect\citeauthoryear{{Beskin}, {Kuznetsova} \& {Rafikov}}{{Beskin}
  et~al.}{1998}]{bes98}
{Beskin} V.~S.,  {Kuznetsova} I.~V.,    {Rafikov} R.~R.,  1998, \mnras, 299,
  341

\bibitem[\protect\citeauthoryear{{Beskin} \& {Nokhrina}}{{Beskin} \&
  {Nokhrina}}{2006}]{bes06}
{Beskin} V.~S.,  {Nokhrina} E.~E.,  2006, \mnras, 367, 375

\bibitem[\protect\citeauthoryear{{Beskin}, {Zakamska} \& {Sol}}{{Beskin}
  et~al.}{2004}]{beszak04}
{Beskin} V.~S.,  {Zakamska} N.~L.,    {Sol} H.,  2004, \mnras, 347, 587

\bibitem[\protect\citeauthoryear{{Bethe}}{{Bethe}}{1990}]{bethe_supernova_mech%
anisms_1990}
{Bethe} H.~A.,  1990, Reviews of Modern Physics, 62, 801

\bibitem[\protect\citeauthoryear{{Blandford}}{{Blandford}}{1976}]{blandford_ac%
cretion_disk_electrodynamics_1976}
{Blandford} R.~D.,  1976, \mnras, 176, 465

\bibitem[\protect\citeauthoryear{{Blandford} \& {Payne}}{{Blandford} \&
  {Payne}}{1982}]{bp82}
{Blandford} R.~D.,  {Payne} D.~G.,  1982, \mnras, 199, 883

\bibitem[\protect\citeauthoryear{{Blandford} \& {Znajek}}{{Blandford} \&
  {Znajek}}{1977}]{bz77}
{Blandford} R.~D.,  {Znajek} R.~L.,  1977, \mnras, 179, 433

\bibitem[\protect\citeauthoryear{{Bogovalov} \& {Tsinganos}}{{Bogovalov} \&
  {Tsinganos}}{1999}]{bog99}
{Bogovalov} S.,  {Tsinganos} K.,  1999, \mnras, 305, 211

\bibitem[\protect\citeauthoryear{{Bogovalov}}{{Bogovalov}}{1997}]{bog97}
{Bogovalov} S.~V.,  1997, \aap, 323, 634

\bibitem[\protect\citeauthoryear{{Bucciantini}, {Quataert}, {Arons}, {Metzger}
  \& {Thompson}}{{Bucciantini} et~al.}{2007}]{buc07}
{Bucciantini} N.,  {Quataert} E.,  {Arons} J.,  {Metzger} B.~D.,    {Thompson}
  T.~A.,  2007, \mnras, 380, 1541

\bibitem[\protect\citeauthoryear{{Bucciantini}, {Quataert}, {Arons}, {Metzger}
  \& {Thompson}}{{Bucciantini} et~al.}{2008}]{buc08}
{Bucciantini} N.,  {Quataert} E.,  {Arons} J.,  {Metzger} B.~D.,    {Thompson}
  T.~A.,  2008, \mnras, 383, L25

\bibitem[\protect\citeauthoryear{{Bucciantini}, {Thompson}, {Arons}, {Quataert}
  \& {Del Zanna}}{{Bucciantini} et~al.}{2006}]{buc06}
{Bucciantini} N.,  {Thompson} T.~A.,  {Arons} J.,  {Quataert} E.,    {Del
  Zanna} L.,  2006, \mnras, 368, 1717

\bibitem[\protect\citeauthoryear{{Burrows}, {Dessart}, {Livne}, {Ott} \&
  {Murphy}}{{Burrows} et~al.}{2007}]{burrows_snjets_2007}
{Burrows} A.,  {Dessart} L.,  {Livne} E.,  {Ott} C.~D.,    {Murphy} J.,  2007,
  \apj, 664, 416

\bibitem[\protect\citeauthoryear{{Camenzind}}{{Camenzind}}{1987}]{camenzind_fi%
nite_element_1987}
{Camenzind} M.,  1987, \aap, 184, 341

\bibitem[\protect\citeauthoryear{Chen \& Beloborodov}{Chen \&
  Beloborodov}{2007}]{chen_neutrino_cooled_2007}
Chen W.-X.,  Beloborodov A.~M.,  2007, \apj, 657, 383

\bibitem[\protect\citeauthoryear{{Contopoulos}, {Kazanas} \&
  {Fendt}}{{Contopoulos} et~al.}{1999}]{ckf99}
{Contopoulos} I.,  {Kazanas} D.,    {Fendt} C.,  1999, \apj, 511, 351

\bibitem[\protect\citeauthoryear{{Contopoulos}}{{Contopoulos}}{1995}]{con95}
{Contopoulos} J.,  1995, \apj, 446, 67

\bibitem[\protect\citeauthoryear{{Contopoulos} \& {Lovelace}}{{Contopoulos} \&
  {Lovelace}}{1994}]{cl94}
{Contopoulos} J.,  {Lovelace} R.~V.~E.,  1994, \apj, 429, 139

\bibitem[\protect\citeauthoryear{{De Villiers}, {Hawley} \& {Krolik}}{{De
  Villiers} et~al.}{2003}]{dev03}
{De Villiers} J.-P.,  {Hawley} J.~F.,    {Krolik} J.~H.,  2003, \apj, 599, 1238

\bibitem[\protect\citeauthoryear{{De Villiers}, {Hawley}, {Krolik} \&
  {Hirose}}{{De Villiers} et~al.}{2005}]{dhkh05}
{De Villiers} J.-P.,  {Hawley} J.~F.,  {Krolik} J.~H.,    {Hirose} S.,  2005,
  \apj, 620, 878

\bibitem[\protect\citeauthoryear{{Di Matteo}, {Perna} \& {Narayan}}{{Di Matteo}
  et~al.}{2002}]{dm02}
{Di Matteo} T.,  {Perna} R.,    {Narayan} R.,  2002, \apj, 579, 706

\bibitem[\protect\citeauthoryear{{Eichler} \& {Levinson}}{{Eichler} \&
  {Levinson}}{2004}]{eichler_levinson_on_amati_2004}
{Eichler} D.,  {Levinson} A.,  2004, \apjl, 614, L13

\bibitem[\protect\citeauthoryear{{Fender}, {Wu}, {Johnston}, {Tzioumis},
  {Jonker}, {Spencer} \& {van der Klis}}{{Fender} et~al.}{2004}]{fend04b}
{Fender} R.,  {Wu} K.,  {Johnston} H.,  {Tzioumis} T.,  {Jonker} P.,  {Spencer}
  R.,    {van der Klis} M.,  2004, \nat, 427, 222

\bibitem[\protect\citeauthoryear{{Fender}, {Belloni} \& {Gallo}}{{Fender}
  et~al.}{2004}]{fend04a}
{Fender} R.~P.,  {Belloni} T.~M.,    {Gallo} E.,  2004, \mnras, 355, 1105

\bibitem[\protect\citeauthoryear{{Fendt}}{{Fendt}}{1997}]{fendt97}
{Fendt} C.,  1997, \aap, 319, 1025

\bibitem[\protect\citeauthoryear{{Fendt}, {Camenzind} \& {Appl}}{{Fendt}
  et~al.}{1995}]{fendt_collimation1_1995}
{Fendt} C.,  {Camenzind} M.,    {Appl} S.,  1995, \aap, 300, 791

\bibitem[\protect\citeauthoryear{{Fendt} \& {Ouyed}}{{Fendt} \&
  {Ouyed}}{2004}]{fo04}
{Fendt} C.,  {Ouyed} R.,  2004, \apj, 608, 378

\bibitem[\protect\citeauthoryear{{Frail}, {Kulkarni}, {Sari}, {Djorgovski},
  {Bloom}, {Galama}, {Reichart}, {Berger}, {Harrison}, {Price}, {Yost},
  {Diercks}, {Goodrich} \& {Chaffee}}{{Frail} et~al.}{2001}]{frail01}
{Frail} D.~A.,  {Kulkarni} S.~R.,  {Sari} R.,  {Djorgovski} S.~G.,  {Bloom}
  J.~S.,  {Galama} T.~J.,  {Reichart} D.~E.,  {Berger} E.,  {Harrison} F.~A.,
  {Price} P.~A.,  {Yost} S.~A.,  {Diercks} A.,  {Goodrich} R.~W.,    {Chaffee}
  F.,  2001, \apjl, 562, L55

\bibitem[\protect\citeauthoryear{{Gammie}, {McKinney} \& {T{\'o}th}}{{Gammie}
  et~al.}{2003}]{gam03}
{Gammie} C.~F.,  {McKinney} J.~C.,    {T{\'o}th} G.,  2003, \apj, 589, 444

\bibitem[\protect\citeauthoryear{{Gammie}, {Shapiro} \& {McKinney}}{{Gammie}
  et~al.}{2004}]{gammie_bh_spin_evolution_2004}
{Gammie} C.~F.,  {Shapiro} S.~L.,    {McKinney} J.~C.,  2004, \apj, 602, 312

\bibitem[\protect\citeauthoryear{{Goldreich} \& {Julian}}{{Goldreich} \&
  {Julian}}{1969}]{gol69}
{Goldreich} P.,  {Julian} W.~H.,  1969, \apj, 157, 869

\bibitem[\protect\citeauthoryear{{Goldreich} \& {Julian}}{{Goldreich} \&
  {Julian}}{1970}]{goldreich_julian_stellar_winds_1970}
{Goldreich} P.,  {Julian} W.~H.,  1970, \apj, 160, 971

\bibitem[\protect\citeauthoryear{{Granot}}{{Granot}}{2005}]{granot_jet_unconve%
ntional_structure_2005}
{Granot} J.,  2005, \apj, 631, 1022

\bibitem[\protect\citeauthoryear{{Hardee}, {Mizuno} \& {Nishikawa}}{{Hardee}
  et~al.}{2007}]{hmn07}
{Hardee} P.,  {Mizuno} Y.,    {Nishikawa} K.-I.,  2007, \apss, 311, 281

\bibitem[\protect\citeauthoryear{{Hardee}}{{Hardee}}{2007}]{hardee07}
{Hardee} P.~E.,  2007, arXiv:astro-ph/0704.1621, 704

\bibitem[\protect\citeauthoryear{{Hawley} \& {Krolik}}{{Hawley} \&
  {Krolik}}{2006}]{hk06}
{Hawley} J.~F.,  {Krolik} J.~H.,  2006, \apj, 641, 103

\bibitem[\protect\citeauthoryear{Heger, Woosley \& Spruit}{Heger
  et~al.}{2005}]{heger_presupernova_2005}
Heger A.,  Woosley S.~E.,    Spruit H.~C.,  2005, \apj, 626, 350

\bibitem[\protect\citeauthoryear{{Jorstad}, {Marscher}, {Lister}, {Stirling},
  {Cawthorne}, {Gear}, {G{\'o}mez}, {Stevens}, {Smith}, {Forster} \&
  {Robson}}{{Jorstad} et~al.}{2005}]{jorstad_agn_jet_2005}
{Jorstad} S.~G.,  {Marscher} A.~P.,  {Lister} M.~L.,  {Stirling} A.~M.,
  {Cawthorne} T.~V.,  {Gear} W.~K.,  {G{\'o}mez} J.~L.,  {Stevens} J.~A.,
  {Smith} P.~S.,  {Forster} J.~R.,    {Robson} E.~I.,  2005, \aj, 130, 1418

\bibitem[\protect\citeauthoryear{Kawanaka \& Mineshige}{Kawanaka \&
  Mineshige}{2007}]{kawanaka_neutrino_cooled_2007}
Kawanaka N.,  Mineshige S.,  2007, \apj, 662, 1156

\bibitem[\protect\citeauthoryear{Kohri, Narayan \& Piran}{Kohri
  et~al.}{2005}]{kohri_neutrino_dominated_2005}
Kohri K.,  Narayan R.,    Piran T.,  2005, \apj, 629, 341

\bibitem[\protect\citeauthoryear{{Komissarov}}{{Komissarov}}{2001}]{kom01}
{Komissarov} S.~S.,  2001, \mnras, 326, L41

\bibitem[\protect\citeauthoryear{{Komissarov}}{{Komissarov}}{2002}]{kom02}
{Komissarov} S.~S.,  2002, \mnras, 336, 759

\bibitem[\protect\citeauthoryear{{Komissarov}}{{Komissarov}}{2005}]{kom05}
{Komissarov} S.~S.,  2005, \mnras, 359, 801

\bibitem[\protect\citeauthoryear{{Komissarov} \& {Barkov}}{{Komissarov} \&
  {Barkov}}{2007}]{kb07}
{Komissarov} S.~S.,  {Barkov} M.~V.,  2007, \mnras, 382, 1029

\bibitem[\protect\citeauthoryear{{Komissarov}, {Barkov}, {Vlahakis} \&
  {K{\"o}nigl}}{{Komissarov} et~al.}{2007}]{kom07}
{Komissarov} S.~S.,  {Barkov} M.~V.,  {Vlahakis} N.,    {K{\"o}nigl} A.,  2007,
  \mnras, 380, 51

\bibitem[\protect\citeauthoryear{{Komissarov} \& {McKinney}}{{Komissarov} \&
  {McKinney}}{2007}]{km07}
{Komissarov} S.~S.,  {McKinney} J.~C.,  2007, \mnras, 377, L49

\bibitem[\protect\citeauthoryear{{Levinson} \& {Eichler}}{{Levinson} \&
  {Eichler}}{1993}]{le93}
{Levinson} A.,  {Eichler} D.,  1993, \apj, 418, 386

\bibitem[\protect\citeauthoryear{{Levinson} \& {Eichler}}{{Levinson} \&
  {Eichler}}{2003a}]{levinson_eichler_grb_baryon_loading_2003}
{Levinson} A.,  {Eichler} D.,  2003a, \apjl, 594, L19

\bibitem[\protect\citeauthoryear{{Levinson} \& {Eichler}}{{Levinson} \&
  {Eichler}}{2003b}]{le03}
{Levinson} A.,  {Eichler} D.,  2003b, \apjl, 594, L19

\bibitem[\protect\citeauthoryear{{Lithwick} \& {Sari}}{{Lithwick} \&
  {Sari}}{2001}]{lithwick_lower_limits_2001}
{Lithwick} Y.,  {Sari} R.,  2001, \apj, 555, 540

\bibitem[\protect\citeauthoryear{{Liu}, {Shapiro} \& {Stephens}}{{Liu}
  et~al.}{2007}]{liu_shapiro_collapse_2007}
{Liu} Y.~T.,  {Shapiro} S.~L.,    {Stephens} B.~C.,  2007, \prd, 76, 084017

\bibitem[\protect\citeauthoryear{{Lovelace}}{{Lovelace}}{1976}]{lov76}
{Lovelace} R.~V.~E.,  1976, \nat, 262, 649

\bibitem[\protect\citeauthoryear{{Lovelace} \& {Romanova}}{{Lovelace} \&
  {Romanova}}{2003}]{lovelace_poynting_jets_2003}
{Lovelace} R.~V.~E.,  {Romanova} M.~M.,  2003, \apjl, 596, L159

\bibitem[\protect\citeauthoryear{{Lovelace}, {Turner} \& {Romanova}}{{Lovelace}
  et~al.}{2006}]{lovelace_jet_pulsar_2006}
{Lovelace} R.~V.~E.,  {Turner} L.,    {Romanova} M.~M.,  2006, \apj, 652, 1494

\bibitem[\protect\citeauthoryear{{Lyutikov}}{{Lyutikov}}{2006}]{lyutikov_em_mo%
del_grbs_2006}
{Lyutikov} M.,  2006, New Journal of Physics, 8, 119

\bibitem[\protect\citeauthoryear{{Lyutikov} \& {Blandford}}{{Lyutikov} \&
  {Blandford}}{2003}]{lyutikov_grbs_2003}
{Lyutikov} M.,  {Blandford} R.,  2003, preprint (arXiv:astro-ph/0312347)

\bibitem[\protect\citeauthoryear{{MacDonald} \& {Thorne}}{{MacDonald} \&
  {Thorne}}{1982}]{macdonald_thorne_bh_forcefree_1982}
{MacDonald} D.,  {Thorne} K.~S.,  1982, \mnras, 198, 345

\bibitem[\protect\citeauthoryear{{MacFadyen} \& {Woosley}}{{MacFadyen} \&
  {Woosley}}{1999}]{mac99}
{MacFadyen} A.~I.,  {Woosley} S.~E.,  1999, \apj, 524, 262

\bibitem[\protect\citeauthoryear{{McClintock}, {Shafee}, {Narayan},
  {Remillard}, {Davis} \& {Li}}{{McClintock}
  et~al.}{2006}]{mcclintock_grs1915_2006}
{McClintock} J.~E.,  {Shafee} R.,  {Narayan} R.,  {Remillard} R.~A.,  {Davis}
  S.~W.,    {Li} L.-X.,  2006, \apj, 652, 518

\bibitem[\protect\citeauthoryear{{McKinney}}{{McKinney}}{2004}]{mckinney2004}
{McKinney} J.~C.,  2004, PhD thesis, PhD Thesis, University of Illinois at
  Urbana-Champaign, 255 pages, DAI-B 65/11, p.~5779

\bibitem[\protect\citeauthoryear{{McKinney}}{{McKinney}}{2005a}]{mckinney2005}
{McKinney} J.~C.,  2005a, preprint (arXiv:astro-ph/0506368)

\bibitem[\protect\citeauthoryear{{McKinney}}{{McKinney}}{2005b}]{mck05}
{McKinney} J.~C.,  2005b, \apjl, 630, L5

\bibitem[\protect\citeauthoryear{{McKinney}}{{McKinney}}{2006a}]{mck06ffcode}
{McKinney} J.~C.,  2006a, \mnras, 367, 1797

\bibitem[\protect\citeauthoryear{{McKinney}}{{McKinney}}{2006b}]{mck06jf}
{McKinney} J.~C.,  2006b, \mnras, 368, 1561

\bibitem[\protect\citeauthoryear{{McKinney}}{{McKinney}}{2006c}]{mck06pulff}
{McKinney} J.~C.,  2006c, \mnras, 368, L30

\bibitem[\protect\citeauthoryear{{McKinney} \& {Gammie}}{{McKinney} \&
  {Gammie}}{2004}]{mck04}
{McKinney} J.~C.,  {Gammie} C.~F.,  2004, \apj, 611, 977

\bibitem[\protect\citeauthoryear{{McKinney} \& {Narayan}}{{McKinney} \&
  {Narayan}}{2007a}]{mck07a}
{McKinney} J.~C.,  {Narayan} R.,  2007a, \mnras, 375, 513

\bibitem[\protect\citeauthoryear{{McKinney} \& {Narayan}}{{McKinney} \&
  {Narayan}}{2007b}]{mck07b}
{McKinney} J.~C.,  {Narayan} R.,  2007b, \mnras, 375, 531

\bibitem[\protect\citeauthoryear{{Mestel}}{{Mestel}}{1961}]{mes61}
{Mestel} L.,  1961, \mnras, 122, 473

\bibitem[\protect\citeauthoryear{{Mestel} \& {Shibata}}{{Mestel} \&
  {Shibata}}{1994}]{mestel_shibata_pulsar_1994}
{Mestel} L.,  {Shibata} S.,  1994, \mnras, 271, 621

\bibitem[\protect\citeauthoryear{{Meszaros}}{{Meszaros}}{2006}]{meszaros_grbs_%
2006}
{Meszaros} P.,  2006, Reports of Progress in Physics, 69, 2259

\bibitem[\protect\citeauthoryear{{Meszaros} \& {Rees}}{{Meszaros} \&
  {Rees}}{1997}]{mr97}
{Meszaros} P.,  {Rees} M.~J.,  1997, \apjl, 482, L29

\bibitem[\protect\citeauthoryear{{Michel}}{{Michel}}{1969}]{mic69}
{Michel} F.~C.,  1969, \apj, 158, 727

\bibitem[\protect\citeauthoryear{{Mignone} \& {McKinney}}{{Mignone} \&
  {McKinney}}{2007}]{migmck07}
{Mignone} A.,  {McKinney} J.~C.,  2007, \mnras, 378, 1118

\bibitem[\protect\citeauthoryear{{Mizuno}, {Hardee} \& {Nishikawa}}{{Mizuno}
  et~al.}{2007}]{mhn07}
{Mizuno} Y.,  {Hardee} P.,    {Nishikawa} K.-I.,  2007, \apj, 662, 835

\bibitem[\protect\citeauthoryear{{Mizuno}, {Yamada}, {Koide} \&
  {Shibata}}{{Mizuno} et~al.}{2004}]{mizuno04}
{Mizuno} Y.,  {Yamada} S.,  {Koide} S.,    {Shibata} K.,  2004, \apj, 615, 389

\bibitem[\protect\citeauthoryear{{Moderski}, {Sikora} \& {Bulik}}{{Moderski}
  et~al.}{2000}]{moderski_afterglow_light_curves_2000}
{Moderski} R.,  {Sikora} M.,    {Bulik} T.,  2000, \apj, 529, 151

\bibitem[\protect\citeauthoryear{{Morsony}, {Lazzati} \& {Begelman}}{{Morsony}
  et~al.}{2007}]{mor06}
{Morsony} B.~J.,  {Lazzati} D.,    {Begelman} M.~C.,  2007, \apj, 665, 569

\bibitem[\protect\citeauthoryear{{Nagataki}, {Takahashi}, {Mizuta} \&
  {Takiwaki}}{{Nagataki} et~al.}{2007}]{nagataki_neutrino_jets_2007}
{Nagataki} S.,  {Takahashi} R.,  {Mizuta} A.,    {Takiwaki} T.,  2007, \apj,
  659, 512

\bibitem[\protect\citeauthoryear{{Nakar}}{{Nakar}}{2007}]{nakar_short_grbs_200%
7}
{Nakar} E.,  2007, \physrep, 442, 166

\bibitem[\protect\citeauthoryear{{Narayan} \& {McClintock}}{{Narayan} \&
  {McClintock}}{2008}]{nm08}
{Narayan} R.,  {McClintock} J.~E.,  2008, New Astron. Rev., in press
  (arXiv:0803.0322)

\bibitem[\protect\citeauthoryear{{Narayan}, {McKinney} \& {Farmer}}{{Narayan}
  et~al.}{2007}]{nar07}
{Narayan} R.,  {McKinney} J.~C.,    {Farmer} A.~J.,  2007, \mnras, 375, 548

\bibitem[\protect\citeauthoryear{{Narayan}, {Paczynski} \& {Piran}}{{Narayan}
  et~al.}{1992}]{nar92}
{Narayan} R.,  {Paczynski} B.,    {Piran} T.,  1992, \apjl, 395, L83

\bibitem[\protect\citeauthoryear{{Narayan}, {Piran} \& {Kumar}}{{Narayan}
  et~al.}{2001}]{narpirankumar01}
{Narayan} R.,  {Piran} T.,    {Kumar} P.,  2001, \apj, 557, 949

\bibitem[\protect\citeauthoryear{{Narayan} \& {Yi}}{{Narayan} \&
  {Yi}}{1994}]{nar94}
{Narayan} R.,  {Yi} I.,  1994, \apjl, 428, L13

\bibitem[\protect\citeauthoryear{{Okamoto}}{{Okamoto}}{1974}]{okamoto_magnetic%
_braking4_1974}
{Okamoto} I.,  1974, \mnras, 166, 683

\bibitem[\protect\citeauthoryear{{Okamoto}}{{Okamoto}}{1978}]{okamoto1978}
{Okamoto} I.,  1978, \mnras, 185, 69

\bibitem[\protect\citeauthoryear{{Paczynski}}{{Paczynski}}{1998}]{pac98}
{Paczynski} B.,  1998, \apjl, 494, L45

\bibitem[\protect\citeauthoryear{{Piran}}{{Piran}}{2005}]{pir05}
{Piran} T.,  2005, Reviews of Modern Physics, 76, 1143

\bibitem[\protect\citeauthoryear{{Popham}, {Woosley} \& {Fryer}}{{Popham}
  et~al.}{1999}]{pwf99}
{Popham} R.,  {Woosley} S.~E.,    {Fryer} C.,  1999, \apj, 518, 356

\bibitem[\protect\citeauthoryear{{Rhoads}}{{Rhoads}}{1997}]{rhoads_uniform_jet%
_1997}
{Rhoads} J.~E.,  1997, \apjl, 487, L1

\bibitem[\protect\citeauthoryear{{Rhoads}}{{Rhoads}}{1999}]{rhoads_uniform_jet%
_1999}
{Rhoads} J.~E.,  1999, \apj, 525, 737

\bibitem[\protect\citeauthoryear{{Rossi}, {Lazzati} \& {Rees}}{{Rossi}
  et~al.}{2002}]{rlr02}
{Rossi} E.,  {Lazzati} D.,    {Rees} M.~J.,  2002, \mnras, 332, 945

\bibitem[\protect\citeauthoryear{{Rossi}, {Beloborodov} \& {Rees}}{{Rossi}
  et~al.}{2006}]{rbr06}
{Rossi} E.~M.,  {Beloborodov} A.~M.,    {Rees} M.~J.,  2006, \mnras, 369, 1797

\bibitem[\protect\citeauthoryear{{Shafee}, {McClintock}, {Narayan}, {Davis},
  {Li} \& {Remillard}}{{Shafee} et~al.}{2006}]{shafee_bh_spin_2005}
{Shafee} R.,  {McClintock} J.~E.,  {Narayan} R.,  {Davis} S.~W.,  {Li} L.-X.,
   {Remillard} R.~A.,  2006, \apjl, 636, L113

\bibitem[\protect\citeauthoryear{{Shakura} \& {Sunyaev}}{{Shakura} \&
  {Sunyaev}}{1973}]{sha73}
{Shakura} N.~I.,  {Sunyaev} R.~A.,  1973, \aap, 24, 337

\bibitem[\protect\citeauthoryear{{Stephens}, {Shapiro} \& {Liu}}{{Stephens}
  et~al.}{2008}]{stephens_shapiro_collapse_2008}
{Stephens} B.~C.,  {Shapiro} S.~L.,    {Liu} Y.~T.,  2008, \prd, 77, 044001

\bibitem[\protect\citeauthoryear{Takiwaki, Kotake \& Sato}{Takiwaki
  et~al.}{2007}]{takiwaki_special_2007}
Takiwaki T.,  Kotake K.,    Sato K.,  2007, preprint (arXiv:0712.1949)

\bibitem[\protect\citeauthoryear{{Tchekhovskoy}, {McKinney} \&
  {Narayan}}{{Tchekhovskoy} et~al.}{2007}]{tch_wham07}
{Tchekhovskoy} A.,  {McKinney} J.~C.,    {Narayan} R.,  2007, \mnras, 379, 469

\bibitem[\protect\citeauthoryear{{Thorne}, {Price} \& {MacDonald}}{{Thorne}
  et~al.}{1986}]{tpm86}
{Thorne} K.~S.,  {Price} R.~H.,    {MacDonald} D.~A.,  1986, {Black holes: The
  membrane paradigm}.
Black Holes: The Membrane Paradigm

\bibitem[\protect\citeauthoryear{{Tomimatsu}, {Matsuoka} \&
  {Takahashi}}{{Tomimatsu} et~al.}{2001}]{tom01}
{Tomimatsu} A.,  {Matsuoka} T.,    {Takahashi} M.,  2001, \prd, 64, 123003

\bibitem[\protect\citeauthoryear{{Usov}}{{Usov}}{1992}]{usov_magnetar_grbs_199%
2}
{Usov} V.~V.,  1992, \nat, 357, 472

\bibitem[\protect\citeauthoryear{{Uzdensky} \& {MacFadyen}}{{Uzdensky} \&
  {MacFadyen}}{2007}]{um07}
{Uzdensky} D.~A.,  {MacFadyen} A.~I.,  2007, \apj, 669, 546

\bibitem[\protect\citeauthoryear{{Vlahakis} \& {K{\"o}nigl}}{{Vlahakis} \&
  {K{\"o}nigl}}{2003a}]{vla03a}
{Vlahakis} N.,  {K{\"o}nigl} A.,  2003a, \apj, 596, 1080

\bibitem[\protect\citeauthoryear{{Vlahakis} \& {K{\"o}nigl}}{{Vlahakis} \&
  {K{\"o}nigl}}{2003b}]{vla03b}
{Vlahakis} N.,  {K{\"o}nigl} A.,  2003b, \apj, 596, 1104

\bibitem[\protect\citeauthoryear{{Wang}, {Abel} \& {Zhang}}{{Wang}
  et~al.}{2007}]{wang_relativistic_2007}
{Wang} P.,  {Abel} T.,    {Zhang} W.,  2007, preprint (arXiv:astro-ph/0703742)

\bibitem[\protect\citeauthoryear{{Watson}, {Hjorth}, {Jakobsson}, {Xu},
  {Fynbo}, {Sollerman}, {Th{\"o}ne} \& {Pedersen}}{{Watson}
  et~al.}{2006}]{watson_short_grb_collimation_2006}
{Watson} D.,  {Hjorth} J.,  {Jakobsson} P.,  {Xu} D.,  {Fynbo} J.~P.~U.,
  {Sollerman} J.,  {Th{\"o}ne} C.~C.,    {Pedersen} K.,  2006, \aap, 454, L123

\bibitem[\protect\citeauthoryear{{Woosley}}{{Woosley}}{1993}]{woosley_gamma_ra%
y_bursts_1993}
{Woosley} S.~E.,  1993, \apj, 405, 273

\bibitem[\protect\citeauthoryear{{Zeh}, {Klose} \& {Kann}}{{Zeh}
  et~al.}{2006}]{zeh_long_grb_angles_2006}
{Zeh} A.,  {Klose} S.,    {Kann} D.~A.,  2006, \apj, 637, 889

\bibitem[\protect\citeauthoryear{{Zhang}, {Woosley} \& {Heger}}{{Zhang}
  et~al.}{2004}]{zwh04}
{Zhang} W.,  {Woosley} S.~E.,    {Heger} A.,  2004, \apj, 608, 365

\bibitem[\protect\citeauthoryear{Zhang, Woosley \& Heger}{Zhang
  et~al.}{2007}]{zhang_fallback_2007}
Zhang W.,  Woosley S.~E.,    Heger A.,  2007, preprint (arXiv:astro-ph/0701083)

\bibitem[\protect\citeauthoryear{{Zhang}, {Woosley} \& {MacFadyen}}{{Zhang}
  et~al.}{2003}]{zhang_relativistic_jets_2003}
{Zhang} W.,  {Woosley} S.~E.,    {MacFadyen} A.~I.,  2003, \apj, 586, 356

\end{thebibliography}

\appendix

\section{Approximate Analytical Description of Magnetodynamic Jets}
\label{sec_approximate_analytic_jet_solution}

\subsection{Approximate Solution for the Flux Function}
\label{appendix_fluxfunction}

An axisymmetric steady non-rotating force-free configuration is
described in spherical coordinates by a magnetic flux function
$P(r,\theta)$ which satisfies~\citep{nar07}
\begin{equation}
r^2 \frac{\p^2 P}{\p r^2}
   +\sin\theta\frac{\p}{\p\theta}
   \left(\frac{1}{\sin\theta}\frac{\p P}{\p\theta}\right) = 0.
   \label{eq_nonrotatingfluxfunction}
\end{equation}
Specializing to a self-similar configuration, we look for a solution
of the form
\begin{equation}
P(r,\theta) = r^\nu p(\theta),
\label{eq_Pvariablesplit}
\end{equation}
where we limit ourselves to $0\le\nu\le2$~\citep{nar07}.

Substituting~\eqref{eq_Pvariablesplit}
into~\eqref{eq_nonrotatingfluxfunction}, we obtain the following
differential equation for $p(\theta)$,
\begin{equation}
 p''(\theta) - \cot\theta \, p'(\theta) + \nu(\nu-1) p(\theta) = 0,
 \label{eq_hypergeometric}
\end{equation}
with boundary conditions, $p(0) = 0$ and $p(\pi/2) = 1$.
The general solution is
\begin{equation}
p(\theta) = c_1 f_1(\theta)
          - c_2 \cos\theta f_2(\theta),
          \label{eq_psolution}
\end{equation}
where $f_1$ and $f_2$ are hypergeometric functions,
\begin{align}
f_1(\theta) &= {}_2F_1\!\!\left(\frac{\nu }{2}-\frac{1}{2},-\frac{\nu}{2};
             \frac{1}{2};\cos ^2\theta\right), \\
f_2(\theta) &= {}_2F_1\!\!\left(\frac{1}{2}-\frac{\nu }{2},\frac{\nu}{2};
             \frac{3}{2};\cos ^2\theta\right),
\end{align}
and the constants $c_1$ and $c_2$ are determined from the boundary
conditions,
\begin{equation}
  c_1 = 1, \quad
  c_2 = \frac{\nu  \Gamma \left(3/2-\nu/2\right)
   \Gamma \left(\nu/2\right)}{\Gamma
   \left(1-\nu/2\right) \Gamma \left(\nu/2+1/2\right)}.
   \label{eq_pcoeffs}
\end{equation}

For the particular cases of $\nu=1$ and $\nu=0$, we find that
$f_1(\theta) \equiv f_2(\theta) \equiv 1$, and the solution is very
simple: $p(\theta)=1-\cos\theta$.  Using this result as a guide, we
write down the following {\it approximate solution} for the general
case:
\begin{equation}
P(r,\theta) \approx r^\nu(1-\cos\theta).
\label{eq_aphi0}
\end{equation}
This turns out to be a good approximation to the exact solution for
$0\le\nu\le 1.25$.  The relative error in $P$ is less than 10\% over
this entire range of $\nu$.

The most attractive feature of (\ref{eq_aphi0}) is its simplicity.
Furthermore, although we began by considering a non-rotating
configuration, the same solution turns out to be a reasonable
approximation even for the rotating case, at least for rotations up to
the maximum we consider, $\Omega_{\rm max}=0.25$.  Thus, using
(\ref{eq_aphi0}), we can obtain a variety of useful, though
approximate, results to describe force-free jets.

\subsection{Field Line Shape}
\label{appendix_fieldlineshape}

By definition, the flux function $P(r,\theta)$ is constant along a
field line.  Therefore, it is straightforward to obtain from equation
(\ref{eq_aphi0}) the shape of a field line as a function of the
position of its footpoint, $r = r_\fpt$, $\theta = \theta_\fpt$:
\begin{equation}
1-\cos\theta = \left(\frac{r}{r_\fpt}\right)^{-\nu} (1-\cos\theta_\fpt),
\end{equation}
which can be rewritten as
\begin{equation}
\sin\frac{\theta}{2}
    = \left(\frac{r}{r_\fpt}\right)^{-\nu/2} \sin\frac{\theta_\fpt}{2}.
\label{eq_fieldlineshape}
\end{equation}
We are primarily interested in the properties of the jet at large
distances from the compact object, where field lines collimate and
$\theta \ll 1$.  In this limit, the field line shape is given by
\begin{align}
\theta &\approx \left(\frac{r}{r_\fpt}\right)^{-\nu/2}
2\sin\frac{\theta_\fpt}{2},
\label{eq_fieldlineshapeapprox}
\intertext{or, inverting the formula,}
\frac{r}{r_\fpt} &\approx \theta^{-2/\nu} \left(2\sin\frac{\theta_\fpt}{2}\right)^{2/\nu}.
\label{eq_fieldlineshapeapproxinv}
\end{align}
In cylindrical coordinates, the field line shape is
\begin{align}
\frac{R}{r_\fpt}
   &\approx \left(\frac{z}{r_\fpt}\right)^{1-\nu/2}
            2\sin\frac{\theta_\fpt}{2},
   \label{eq_fieldlineshapecylRz}\\
\frac{z}{r_\fpt} &\approx \left(\frac{R}{r_\fpt}\right)^{2/(2-\nu)}
                 \left[2\sin\frac{\theta_\fpt}{2}\right]^{-2/(2-\nu)}.
\label{eq_fieldlineshapecylzR}
\end{align}

In the models we analyzed in the main text of the paper, we defined
the `jet' to include all field lines whose footpoints are on the
compact central object, $r_\fpt=1$, $0\le\theta_\fpt < \pi/2$, and the
`wind' to consist of the remaining field lines with footpoints on the
disc, $r_\fpt>1, \theta_\fpt=\pi/2$.  Thus, the field line that emerges at $r_\fpt=1$,
$\theta_\fpt=\pi/2$ defines the boundary between the jet and the wind.
The angle $\theta_j$ of this field line at a distance $r$ is
\begin{equation}
\theta_j \approx \sqrt{2} r^{-\nu/2}.
\label{eq_thetaj}
\end{equation}
This relation shows how collimation proceeds as a function of distance
for a given value of $\nu$.  Since the models of interest to us have
$0.5 < \nu < 1$, the collimation is gradual, going more slowly than
$1/\sqrt{r}$.

Field lines with $\theta<\theta_j$ connect back to the central object
($r_\fpt=1$), and their foot-points are located at
\begin{equation}
r_\fpt=1, ~\sin{\frac{\theta_\fpt}{2}} \approx \frac{1}{\sqrt2}\frac{\theta}{\theta_j},
\quad 0\le\theta\le\theta_j.
\end{equation}
Field lines at larger angles connect back to the disc
($\theta_\fpt=\pi/2$), and their foot-point radii are given by
\begin{equation}
r_\fpt \approx \left(\frac{\theta}{\theta_j}\right)^{2/\nu},
~\theta_\fpt=\pi/2,\quad \theta_j < \theta.
\label{eq_rfpt}
\end{equation}

\subsection{Poloidal Magnetic Field}
\label{appendix_magneticfieldandenclosedcurrent}

Given the magnetic flux function~\eqref{eq_aphi0}, we can determine
the poloidal components of the magnetic field according
to~\eqref{eq_deffluxfunction},
\begin{align}
B_r &= r^{\nu-2}, \label{eq_Br} \\
B_\theta &= - \nu r^{\nu-2} \tan(\theta/2). \label{eq_Btheta}
\end{align}
Both components of the field vary with radius as $r^{\nu-2}$, as they
should for the assumed form of $P(r,\theta)$.  Moreover, $B_r$ does
not vary with $\theta$, which is approximately the case for
self-similar force-free models~\citep{nar07}.  In the context of our
problem, we see that the normal component of the field at the surface
of the central compact object ($r=1$) is independent of $\theta$.
Thus we have a uniform field emerging from the object --- a split
monopole.  This is a nice property of the flux function
(\ref{eq_aphi0}).

For reference, we provide the poloidal components of the magnetic
field in cylindrical coordinates,
\begin{align*}
B_R &= B_r\sin\theta+B_\theta\cos\theta
     = r^{\nu-2} [1+(1-\nu)\cos\theta] \tan{\frac\theta2},\\
B_z &= B_r\cos\theta-B_\theta\sin\theta
     = r^{\nu-2}[\cos\theta+\nu\sin\theta\tan{\frac{\theta}{2}}],
\end{align*}
In the disc plane ($\theta=\pi/2$), we have
\begin{align}
B_R &= \phantom{\nu} r^{\nu-2}= B_r, \\
B_z &= \nu r^{\nu-2}= -B_\theta.
\end{align}
The normal component of the field on the disc varies as a power-law in
radius.

\subsection{Lorentz Factor}
\label{sec_lorentz_factor_scaling}

\subsubsection{General Formula for the Lorentz Factor}
\label{sec_lorentz_factor_general}

For an axisymmetric configuration in steady state, the angular
velocity $\Omega$ is constant on field lines.  This means that the
comoving frame of the fluid at any position rotates at the same
$\Omega$ as the angular velocity of the disc at the corresponding
footpoint.  In this rotating frame, there is no electric field (since
we assume infinite conductivity).  Therefore, in the inertial/lab
frame, the electric field $\myvec E$ is perpendicular to the magnetic
field $\myvec B$ and is equal to~\citep{tpm86, bes97,nar07}
\begin{equation}
E = \OmegaF R B_p,
\label{eq_electricfield}
\end{equation}
where $B_p$ is the magnitude of the poloidal magnetic field.

In the force-free approximation, the velocity of the fluid may be
conveniently taken to be the drift velocity \citep{nar07},
\begin{equation}
v_d = \Abs{\frac{\myvec E\times \myvec B}{B^2}} = \frac{E}{B}.
\label{eq_vdrift}
\end{equation}
This gives the Lorentz factor $\gamma$ of the fluid,
\begin{equation}
\frac{1}{\gamma^2}
  = \frac{ B^2 - E^2}{B^2}
  = \frac{B_p^2}{B^2} + \frac{B_\varphi^2 - E^2}{B^2}.
  \label{eq_gammatotraw}
\end{equation}

Consider now the far asymptotic region of the jet where \hbox{$\OmegaF
R \gg 1$} and the velocity is ultrarelativistic, \hbox{$v^2\approx1$},
$\gamma \gg 1$. In this limit we have \hbox{$E \gg B_p$} due
to~\eqref{eq_electricfield} and \hbox{$E \approx B$} due
to~\eqref{eq_vdrift}. These two relations, combined with the
definition~\eqref{eq_Btot} of $B$ in terms of $B_p$ and $B_\varphi$,
allow us to obtain the relative magnitudes of the electromagnetic
field components,
\begin{equation}
B \approx \abs{B_\varphi} \approx E \gg B_p
          \quad \text{for} \quad \OmegaF R \gg 1, \; v^2 \approx 1.
\label{eq_relative_fields}
\end{equation}
At small distances from the compact object the first term on the
\rhs\ of~\eqref{eq_gammatotraw} dominates, and we write its
inverse as
\begin{equation}
\frac{B^2}{B_p^2}
       = 1 + \frac{B_\varphi^2}{B_p^2}
       = \gamma_1^2,
\end{equation}
where from~\eqref{eq_electricfield} and~\eqref{eq_relative_fields} we
have
\begin{equation}
\gamma_1 \approx \left[1+ (\OmegaF R)^{2}\right]^{1/2},
\label{eq_asym1}
\end{equation}
In this near-zone, we have $\gamma\approx\gamma_1\approx\OmegaF R$.
We refer to this as the {\it first acceleration regime}.
Formula~\eqref{eq_asym1}, as we will see below, gives a good
approximation to the Lorentz factor at moderate distances from the
compact object and has been noted by many
authors~\citep{bz77,bes98,bes06,nar07}.

For certain field
geometries it is the second term on the \rhs\
of~\eqref{eq_gammatotraw} that determines the bulk Lorentz
factor.  The inverse of the second term can be written as
\begin{equation}
\frac{B^2}{B_\varphi^2-E^2} = \gamma_2^2,
\label{eq_gamma2viaBandE}
\end{equation}
where $\gamma_2$ is determined by the ratio of the local poloidal
radius of curvature of the field line~$R_c$ to the cylindrical radius
from the jet axis~$R$~\citep[and~\S\ref{sec_transversal_force_balance}]{beszak04},
\begin{equation}
\gamma_2 \approx C \; \left(\frac{R_c}{R}\right)^{1/2}.
\label{eq_asym2}
\end{equation}
Here~$C$ is a numerical factor of order unity which has to be
determined by a more careful study of the solution.
The value
of this factor depends on the spatial distribution of field
line rotation frequency $\Omega$ and as we show analytically in
Appendix~\ref{sec_transversal_force_balance},
in the jet region, where $\OmegaF = \const$,
we have $C=\sqrt3\approx1.73$.  We find that this single
value of $C$ does a very good job of explaining all our
numerical results, giving an error of less than $10$\%  in all cases we
have considered (see~\S\ref{sec_results}).
Equation~\eqref{eq_asym2} corresponds to the {\it second acceleration
regime}.  It illustrates that, in a relativistic electromagnetic flow,
collimation and acceleration are suppressed: for the fluid to have a
sufficiently large Lorentz factor, the field line to which it is
attached has to be sufficiently straight, \ie\ have a sufficiently
large value of $R_c/R$.

Combining equations~\eqref{eq_gammatotraw} -- \eqref{eq_asym2} we
obtain the following general formula for the Lorentz factor,
\begin{equation}
\frac{1}{\gamma^2} = \frac{1}{\gamma_1^2} + \frac{1}{\gamma_2^2},
\label{eq_asymall}
\end{equation}
where $\gamma_1$ and $\gamma_2$ are given by~\eqref{eq_asym1}
and~\eqref{eq_asym2}.  Although we derived this expression in the
asymptotic limit of large Lorentz factors, the formula is quite
accurate all the way from the foot-point of field lines to large
distances~(see \S\ref{sec_results}).

\subsubsection{Lorentz Factor of the Jet and the Wind}
\label{sec_lorentz_factor_jet_and_wind}

In order to numerically evaluate formula~\eqref{eq_asymall} for our
jet-wind solution, we adopt the field line
shape~\eqref{eq_fieldlineshapeapprox}.  In the jet all field lines
rotate at the same frequency,
\begin{equation}
\OmegaF^\text{jet} = \Omega_\co = \const,
\label{eq_omega_jet}
\end{equation}
whereas in the wind field lines rotate differentially, according to
the rotation profile in the disc,
\begin{equation}
\OmegaF^\text{wind}(r,\theta)
   = \Omega[r_\fpt(r,\theta)]
   \approx \Omega_\co  (r^\nu\theta^2/2)^{-\beta/\nu}.
\label{eq_omega_wind}
\end{equation}
We have used~\eqref{eq_rfpt} to substitute for $r_\fpt(r,\theta)$ and
equation~\eqref{eq_omegadisc} to substitute for~$\OmegaF(R_\fpt)$.
Combining expressions~\eqref{eq_omega_jet} and~\eqref{eq_omega_wind}
we have
\begin{equation}
\OmegaF(r,\theta) \approx
   \begin{cases}
      \Omega_\co, &\theta \le \theta_j \quad \text{(jet)}, \\
      \Omega_\co  (r^\nu\theta^2/2)^{-\beta/\nu},
                 &\theta > \theta_j \quad \text{(wind)}.
   \end{cases}
   \label{eq_omegafall}
\end{equation}

The first acceleration regime is straightforward.  For both the jet
and the wind we have
\begin{equation}
\gamma_1 = \Omega(r,\theta) \,  r \, \theta.
\label{eq_gamma1omegafall}
\end{equation}
For the second acceleration regime, we use the asymptotic field line
shape~\eqref{eq_fieldlineshapecylRz} to compute the curvature radius
of the field line,
\begin{equation}
R_c = \frac{\left[1+(dR/dz)^2\right]^{3/2}}{\Abs{\simplefrac{d^2R}{dz^2}}}
    \approx \frac{1}{\Abs{\simplefrac{d^2R}{dz^2}}},
\end{equation}
which together with the field line shape~\eqref{eq_fieldlineshapecylRz}
gives
\begin{equation}
\frac{R_c}{r_\fpt} \approx \left(\frac{z}{r_\fpt}\right)^{1+\nu/2}
                   \times\frac{1}{(1-\nu/2)\nu \sin(\theta_\fpt/2)}.
\end{equation}
Therefore, according to~\eqref{eq_fieldlineshapecylRz}
and~\eqref{eq_asym2}, for the jet field lines we have
\begin{equation}
\gamma_2 \approx \frac{1}{\theta}
         \times \frac{2C}{\sqrt{(2-\nu)\nu}}.
         \label{eq_asym2thetascaling}
\end{equation}
Note that the only explicit dependence of the Lorentz factor on
position in the second acceleration regime is the spherical polar
angle.  Even though this expression rather accurately describes the distribution of
the Lorentz factor in the jet with one single value of
$C=\sqrt3$~(\S\ref{sec_othermodels}),
we find that in the wind the values of $C$ are different for different field
lines because in the wind the field line shape deviates from the analytical
expectation~\eqref{eq_fieldlineshapecylRz}.

\subsubsection{Lorentz Factor Scaling Along Field Lines}
\label{sec_lorentz_factor_along_field_lines}

Consider a field line starting at a foot-point $r = r_\fpt$, $\theta =
\theta_\fpt$, and rotating at a frequency $\Omega_\fpt$.  For a jet
field line we have $r_\fpt = 1$ whereas for a wind field line we have
$\theta_\fpt = \pi/2$.  Let us evaluate the Lorentz factor
scalings~\eqref{eq_asym1} and \eqref{eq_asym2} along a given field
line as a function of distance from the compact object.  Using various
results derived earlier we obtain
\begin{align}
\gamma_1 &\approx \Omega_\fpt \, r_\fpt \left(\frac{r}{r_\fpt}\right)^{1-\nu/2}
                 \times 2\sin\frac{\theta_\fpt}{2},
         \label{eq_asym1scaling} \\
\gamma_2 &\approx C \left(\frac{r}{r_\fpt}\right)^{\nu/2}
                  \times \left[\sqrt{(2-\nu)\nu}\,\sin\frac{\theta_\fpt}{2}\right]^{-1}.
  \label{eq_asym2scaling}
\end{align}

Since $C=\sqrt3$ and $\Omega_\fpt$ is substantially less than
unity, it is easy to see that at small radii we have
$\gamma_1<\gamma_2$.  By equation~(\ref{eq_asymall}), $\gamma$ is
determined by the smaller of $\gamma_1$ and $\gamma_2$.  Therefore,
near the base of each field line the fluid is always in the first
acceleration regime.  As the field line goes to larger distances,
$\gamma_1$ and $\gamma_2$ increase at different rates.  If $\nu\ge1$,
$\gamma_1$ always remains smaller than $\gamma_2$, and the first
acceleration regime operates throughout the field line.  However, for
$\nu<1$, $\gamma_2$ grows more slowly than $\gamma_1$.  At a certain
distance along the field line, $\gamma_2$ then becomes smaller than
$\gamma_1$ and the flow makes a transition to the second acceleration
regime \citep{nar07}.  The transition between the two regimes happens
at a distance $r_{\rm tr}$ given by
\begin{equation}
\frac{r_\text{tr}}{r_\fpt}
   = \left[\frac{1}{\Omega_\fpt r_\fpt} \times
          \frac{C}{2 \sin^2(\theta_\text{fp}/2) \sqrt{(2-\nu)\nu}}
    \right]^{1/(1-\nu)}.
\label{eq_rtr}
\end{equation}

Equation (\ref{eq_rtr}) shows that the transition occurs sooner for
field lines with foot-points at lower latitudes ($\theta_\fpt$ closer
to $\pi/2$) on the compact object compared to field lines that emerge
near the pole.  This leads to the development of a fast jet core in
the polar region, where the Lorentz factor continues to be determined
by the first acceleration regime for a long distance, surrounded by a
slower sheath, where the acceleration switches to the second
less-efficient regime at a shorter distance (see
\S\ref{sec_results}).

\subsection{Toroidal Magnetic Field and Enclosed Poloidal Current}
\label{sec_toroidal_field}

To obtain the toroidal magnetic field~$B_\varphi$, we use the fact
that in ideal MHD the enclosed poloidal current,
\begin{equation}
I = R B_\varphi/2,
\label{eq_poloidalcurrentdef}
\end{equation}
is constant along each field line~\citep{tpm86,bes97,nar07}.  A
positive value of $I$ indicates a current in the positive $z$
direction.

Asymptotically far from the compact object, where the fluid motion is
ultrarelativistic ($\gamma \gg 1$), we have according
to~\eqref{eq_relative_fields},
\begin{equation}
B_\varphi \approx - \OmegaF R B_p,
\label{eq_bphiofbp}
\end{equation}
where the negative sign is because the field lines are swept back in
the opposite direction to rotation, \ie\ in the negative
$\varphi$-direction for a positive $\Omega$.  Since asymptotically,
$\theta\ll1$, according to~\eqref{eq_Br} -- \eqref{eq_Btheta} we have $B_p
\approx B_r = r^{\nu-2}$ and therefore
\begin{equation}
I = - \OmegaF R^2 r^{\nu-2}/2
  \approx - \OmegaF\, (r^\nu \theta^2/2),
\end{equation}
where we have approximated $R\approx r\theta$.  According
to~\eqref{eq_aphi0}, the expression in parentheses is approximately
equal to the flux function $P$.  Therefore we have
\begin{equation}
I \approx - \OmegaF P.
\label{eq_poloidalcurrent}
\end{equation}
This relation between two quantities that are each conserved along
field lines must hold throughout the solution (although we derived it
only asymptotically).  Therefore, according
to~\eqref{eq_poloidalcurrentdef} and~\eqref{eq_poloidalcurrent},
we obtain
\begin{equation}
B_{\varphi} \approx  - \frac{2 \OmegaF P}{R}
          =-\frac{2\OmegaF r^\nu (1-\cos\theta)}{r \sin\theta}
          = - 2 \OmegaF r^{\nu-1} \tan\left(\frac{\theta}{2}\right).
\end{equation}

\subsection{Magnetic Pressure}
\label{appendix_magneticpressure}

Because of the assumption of perfect conductivity, the electric field
in the comoving frame of the fluid vanishes and there is only a
comoving magnetic field $b$.  Given the electric and magnetic fields
in the lab frame, $E$ and $B$, the comoving magnetic field strength is
\begin{equation}
b = \sqrt{B^2 - E^2},
\label{eq_defb}
\end{equation}
and the comoving magnetic pressure is
\begin{equation}
p_\text{mag} = \frac{b^2}{8\pi}.
\label{eq_pmagdef}
\end{equation}
From~\eqref{eq_defb}, we can write
\begin{equation}
b^2 = B_p^2 + B_\varphi^2 - E^2
    \approx B_p^2 + \frac{B^2}{\gamma_2^2},
\label{eq_b2gam}
\end{equation}
where the approximate equality is due to~\eqref{eq_gamma2viaBandE}.
Further, due to~\eqref{eq_relative_fields} and~\eqref{eq_asym1},
we can rewrite~\eqref{eq_pmagdef} as
\begin{equation}
p_\text{mag}
   = \frac{b^2}{8\pi}
   \approx \frac{B_p^2}{8\pi} \left( 1 + \frac{\gamma_1^2}{\gamma_2^2} \right),
\end{equation}
where $B_p^2/(8\pi)$ is the pressure that the jet
would have if the compact object was not spinning (when $B_\varphi = E = 0$).

If $\nu\ge1$, we saw earlier that we always have $\gamma_1<\gamma_2$.
Therefore, using the expression for the radial magnetic
field~\eqref{eq_Br}, we obtain
\begin{align}
\nu\ge1:\qquad &p_\text{mag} \propto r^{2(\nu-2)}.
\intertext{The magnetic pressure follows a single power-law.  However, for
$\nu<1$, the term involving the Lorentz factors in \eqref{eq_b2gam}
cannot be neglected.  Using the expressions for
$\gamma_1$~\eqref{eq_asym1scaling}, $\gamma_2$~\eqref{eq_asym2scaling} and the
transition radius~$r_\text{tr}$~\eqref{eq_rtr}, we obtain}
&p_\text{mag} \propto r^{2(\nu-2)}
                     \left[
                        1+\left(\frac{r}{r_\text{tr}}\right)^{2(1-\nu)}
                    \right],
\intertext{and the magnetic pressure behaves as a broken power-law,}
\nu<1:\qquad &p_\text{mag} \propto
  \begin{cases}
  r^{2(\nu-2)}, \quad &r \lesssim r_\text{tr}, \\
  r^{-2}, \quad &r \gtrsim r_\text{tr}.
  \end{cases}
\end{align}
Note that the power-law index at large distances is independent of the
value of~$\nu$.

\subsection{Transverse Force Balance}
\label{sec_transversal_force_balance}

In this section we study the steady state transversal force balance
in force-free jets and
analytically determine the value of the prefactor $C$ that
scales the Lorentz factor in the second acceleration
regime (eq.~\ref{eq_asym2}).  The force balance equation is
\begin{equation}
\myvec{j} \times \myvec{B} + \rho \myvec{E} = 0,
\end{equation}
where $\rho$ and $\myvec j$ are the lab-frame electric charge and current density.
Using Ampere's law,
\begin{equation}
\myvec j = \frac{1}{4\pi} \myvec\nabla\times\myvec B,
\end{equation}
we have
\begin{equation}
4\pi\myvec j \times \myvec B
  = (\myvec\nabla \times \myvec B) \times \myvec B
  = (\myvec B \myvec \nabla) \myvec B - \frac{1}{2}\myvec\nabla(B^2).
\end{equation}
Let us introduce a local orthonormal tetrad
\begin{equation}
(\htau = \myvec B_p / B_p, \hphi, \hn = \myvec E/E),
\end{equation}
corresponding to local rotated cylindrical coordinates,
\begin{equation}
(\tau, R \varphi, n),
\end{equation}
where $\tau$ measures the distance along field lines, $n$ measures
distance perpendicular to field lines, and $R\varphi$ measures the
distance in the toroidal direction.
We have then
\begin{align}
(\myvec B \myvec \nabla) \myvec B
        = (B_\tau \frac{\p}{\p\tau} + B_\varphi \frac{\p}{R\p\varphi} + B_n \frac{\p}{\p n})
          (B_\tau \htau + B_\varphi \hphi + B_n \hn). \notag
\end{align}
Projecting the above relation along $\hn$, we obtain:
\begin{align}
\hn \cdot \left[(\myvec B \myvec \nabla)\myvec B\right]
        &=   \hn \cdot B_\tau \frac{\p}{\p\tau} (B_\tau \htau)
          + \hn \cdot B_\varphi \frac{\p}{R\p\varphi} (B_\varphi \hphi) \notag\\
        &=  B_\tau^2 \, \left(\hn\cdot\frac{\p\htau}{\p\tau}\right)
          + B_\varphi^2 \, \left(\hn\cdot\frac{\p\hphi}{R\p\varphi}\right) \notag\\
        &=  \frac{B_\tau^2}{R_c} - \frac{B_\varphi^2}{R} (\hn\cdot \hat R),
\end{align}
where $\hat R$ is a unit vector directed along the cylindrical radius.
Combining the above results, we have for the force balance in the
$\hn$-direction:
\begin{equation}
  \frac{B_p^2}{4\pi R_c} - \frac{B_\varphi^2}{4\pi R} (\hn\cdot \hat R)
- \frac{\p}{\p n}\left(\frac{B^2}{8\pi}\right)
+ \frac{(\myvec\nabla \cdot \myvec E)}{4\pi} E = 0.  \label{eq_nbalancenorho}
\end{equation}
We can write the divergence of the electric field in the local
rotated cylindrical coordinates as
\begin{align}
\myvec\nabla \cdot \myvec E
     &=  \frac{1}{R} \frac{\p}{\p\tau}(R E_\tau)
       + \frac{1}{R} \frac{\p}{\p n}    (R E_n)  \notag \\
     &=  \frac{\p E_\tau}{\p\tau}
       + \frac{1}{R} \frac{\p R}{\p n} E_n + \frac{\p E_n}{\p n}, \label{eq_divE3}
\end{align}
where we have used the fact that $E_\tau = E_\varphi = 0$.
The first term in \eqref{eq_divE3}
can be rewritten in terms of the curvature radius of the field line:
\begin{align}
    \frac{\p E_\tau}{\p\tau}
  = \htau \cdot \frac{\p (E \hn)}{\p\tau}
  = E \, \left(\htau \cdot \frac{\p \hn}{\p\tau}\right)
  = -\frac{E}{R_c}. \label{eq_dEdtau}
\end{align}

Finally, combining equations~\eqref{eq_nbalancenorho} -- \eqref{eq_dEdtau},
we obtain a form of the force balance equation that does not
explicitly contain any derivatives along field lines:
\begin{equation}
  \frac{B_p^2-E^2}{4\pi R_c}
  - (\hat R \cdot \hn) \frac{B_\varphi^2-E^2}{4\pi R}
  - \frac{\p}{\p\hn}\left(\frac{B^2-E^2}{8\pi}\right)
  = 0.
  \label{eq_force_balance_n}
\end{equation}
This equation is the analog of equation~(79) in \citet[they
have a sign typo]{bes06}.

At large distances from the compact object we have
$(\hat R \cdot \hn) \approx -1$ and $\p/\p\hn \approx -\p/\p R$
(eq.~\ref{eq_fieldlineshapecylRz}).
With this we can re-write~\eqref{eq_force_balance_n} in
projection along $\hat R$,
\begin{equation}
  -\frac{B_p^2-E^2}{4\pi R_c}
  - \frac{B_\varphi^2-E^2}{4\pi R}
  - \frac{\p}{\p R}\left(\frac{B_p^2 + B_\varphi^2-E^2}{8\pi}\right)
  \approx 0.
  \label{eq_force_balance_R}
\end{equation}
This form explicitly shows that the poloidal and toroidal
relativistic hoop stresses, which are the first two terms,
are balanced by the comoving pressure gradient, the last term.

Assuming that the field line rotation frequency varies as a power-law
in the cylindrical radius, $\OmegaF \propto R^\lambda$,
we have
$B_\varphi \approx -\Omega R B_p \propto R^{1+\lambda}$ (eq.~\ref{eq_relative_fields}),
where we took $B_p$ to be nearly independent of $R$ (eq.~\ref{eq_Br} -- \ref{eq_Btheta}).
With the help of~\eqref{eq_gamma2viaBandE} we have,
\begin{equation}
B_\varphi^2 - E^2 \approx \frac{B_\varphi^2}{\gamma_2^2} \propto R^{4+2\lambda},
\label{eq_2ndpressure}
\end{equation}
where we used eq.~\eqref{eq_asym2thetascaling} for the Lorentz factor in the second acceleration
regime, which gives $\gamma_2 \propto 1/R$.
We note that in the wind the poloidal magnetic field is not uniform,
and $\gamma_2$ follows a different power-law.

In the second acceleration regime we have $\gamma_1\gg\gamma_2$
and thus $B_\varphi^2-E^2\gg B_p^2$ (Appendix~\ref{sec_lorentz_factor_general}).
Therefore in this limit the terms involving $B_p$ in~\eqref{eq_force_balance_R} can be dropped,
\begin{equation}
  \frac{B_\varphi^2-E^2}{R}
  + \frac{\p}{\p R}\left(\frac{B_\varphi^2-E^2}{2}\right)
  \approx
  \frac{E^2}{R_c}.
  \label{eq_force_balance_R1}
\end{equation}
By combining equations~\eqref{eq_2ndpressure}
and~\eqref{eq_force_balance_R1} with~\eqref{eq_relative_fields}
and~\eqref{eq_gamma2viaBandE}, we can express the Lorentz factor
in the second acceleration regime as
\begin{equation}
  \gamma_2
  \approx \left(\frac{E^2}{B_\varphi^2-E^2}\right)^{1/2}
  \approx
  C \left(\frac{R_c}{R}\right)^{1/2},
  \label{eq_gamma2}
\end{equation}
where
\begin{align}
C = \sqrt{3+\lambda},  \quad \lambda = \frac{\p\log\OmegaF}{\p\log R}.
\label{eq_c_analytical}
\end{align}
Specializing to the jet region where $\OmegaF=\const$, or $\lambda = 0$,
we obtain $C = \sqrt{3} \approx 1.7$.
Numerically, however, we find a range of values  for $C$
rather than just a single value~(\S\ref{sec_othermodels}).
One of the reasons for this may be the fact that
instead of being exactly constant in the jet, $\Omega(R)$
slightly decreases with increasing $R$
near the jet-wind boundary due to numerical diffusion (see Fig.~\ref{fig_omegaitransversal}).
This causes $\lambda$ to vary from $0$
at the jet axis to $\approx-1$ near the jet-wind boundary
(Fig.~\ref{fig_omegaitransversal}).  According to~\eqref{eq_c_analytical},
this variation in $\lambda$ translates into a variation in
$C = \sqrt{2} - \sqrt{3} \approx 1.4 - 1.7$ across the jet,
consistent with the
numerical simulations~(\S\ref{sec_othermodels}, Fig.~\ref{fig_bnufit}).

\subsection{Power Output}
\label{appendix_poweroutput}
The power output in a magnetodynamic jet is equal to the Poynting
flux, whose magnitude is given by
\begin{equation}
S = \frac{|\myvec E \times \myvec B|}{4\pi}.
\end{equation}
At asymptotically large distances, according
to~\eqref{eq_relative_fields} we can approximately write
\begin{equation}
B\approx E \approx \OmegaF R B_p.
\end{equation}
Therefore, using~\eqref{eq_omegafall}, we have
\begin{equation}
S \approx \frac{\OmegaF(r,\theta)^2 R^2 B_p^2}{4\pi}
  \approx \frac{\OmegaF(r,\theta)^2 r^{2\nu-2} \theta^2}{4\pi},
\end{equation}
where we have approximated $B_p \approx B_r \approx r^{\nu-2}$ with
the help of~\eqref{eq_Br}.  We can now evaluate the lateral dependence
of $S$ in the jet and the wind,
\begin{equation}
S(r,\theta) \approx \frac{1}{4\pi} \times
   \begin{cases}
      \Omega_\co^2 r^{2\nu-2} \theta^2, &\theta \le \theta_j, \\
      \Omega_\co^2 r^{2\nu-2} \theta^2 (r^\nu\theta^2/2)^{-2\beta/\nu},
                 &\theta > \theta_j.
   \end{cases}
   \label{eq_poynting}
\end{equation}

Let us define the energy flux per unit solid angle as
\begin{equation}
\frac{dP}{d\omega} = r^2 S(r,\theta).
\end{equation}
For small angles the energy flux grows quadratically with the polar
angle inside the jet, $\propto \theta^2$, and then falls off rapidly
in the wind, $\propto \theta^{2-4\beta/\nu}$.  The maximum power
output occurs at the jet-wind boundary, $\theta=\theta_j$,
\begin{equation}
\frac{dP}{d\omega}\biggr|_{\theta=\theta_j}
         = \frac{\Omega_\co^2}{\pi\theta_j^2}.
\end{equation}
These results are confirmed in Fig.~\ref{fig_gammathetaprofile}.
Keplerian discs have $\beta = -3/2$, so the power output in the wind
falls off extremely rapidly (\eg~$\propto \theta^{-6}$ for our
fiducial model~A which has $\nu = 0.75$).  However, the contribution
to the total energy output from the disc may still be significant.

The total power output of the jet is
\begin{align}
P^{\text{jet}}(r) &= \int\limits_0^{\theta_j} 2\pi\sin\theta
  \frac{dP}{d\omega}d\theta
  \approx \frac{\Omega_\co^2}{2}
  \left(\frac{r^\nu\theta^2}{2}\right)^2\biggr|_0^{\theta_j}
  \notag\\
  &\approx \frac{\Omega_\co^2}{2} = \frac{1}{32},
\intertext{
  The region
  $\theta < \theta_j = \sqrt2r^{-\nu/2}$ defines the jet region, and
  the numerical evaluation was performed for the fiducial model~A.
  For the disc wind,}
P^{\text{disc}}(r) &=
  \int\limits^\infty_{\theta_j} 2\pi\sin\theta
  \frac{dP}{d\omega}d\theta
  \approx
  \frac{\Omega_\co^2}{2(1-\beta/\nu)}
  \left(\frac{r^\nu\theta^2}{2}\right)^{2-2\beta/\nu}\biggr|_{\theta_j}^\infty
  \notag\\
  &\approx \frac{\Omega_\co^2}{2(\beta/\nu-1)} =
  \frac{1}{32},
\end{align}
where the numerical evaluation, again, was performed for model~A.
For this model the disc wind provides the same energy
output as the jet.

Converting the results to physical units, the total power output of a
jet in an astrophysical system with BH mass $M$, radial magnetic field
strength near the BH $B_r$, and angular rotation frequency
$\Omega_\co$, is
\begin{align}
P^\text{jet}
  &= \frac{1}{2} \Omega_\co^2 \, B_r^2 \, r_0^2 \, c
  \notag\\
  &\approx 1.8\times10^{50} \left(\frac{\Omega_\co}{\Omega_\text{max}}\right)^2
                     \left(\frac{B_r}{10^{15}G}\right)^2
                     \left(\frac{M}{3M_\odot}\right)^2
                     \left[\frac{\text{erg}}{\text{s}}\right],
\end{align}
where $r_0\approx r_g = GM/c^2$ for a rapidly spinning BH, and
$\Omega_\text{max} = 0.25c/r_g$ is
the maximum angular rotation frequency that magnetic field lines
threading a spinning compact object can have in a stationary state.

\section{Fluid Speed in Force-Free Equilibrium}
\label{sec_fluid_speed}

In this section we would like to answer the following question: In the
limit of infinite magnetization, i.e., $\sigma\to\infty$, are we
guaranteed that a force-free solution will give a physically
meaningful solution for any fluid that is carried along with the
field?  Specifically, if we have a force-free solution which has drift
speed $v_d=E/B$ (eq.~\ref{eq_vdrift}) less than $c$ everywhere of interest, and if we add some
fluid with negligible inertia ($\sigma\to\infty$) which is carried
along with the field, will the flow speed $v_f$ of the fluid be
physical, i.e., will we have $v_f<c$ everywhere?  The answer to this
question is not obvious.  A fluid possesses additional conserved
quantities, e.g., the Bernoulli constant, which are not relevant for
the electromagnetic field.  Could the additional constraints give
inconsistent results for the fluid motion even though the fields
behave physically?

Consider a steady axisymmetric force-free solution in which the
poloidal field is $B_p\hat{n}_p$ along a poloidal unit vector
$\hat{n}_p$ and the toroidal field is $B_\varphi\hat{n}_\varphi$ along the
toroidal unit vector $\hat{n}_\varphi$.  Assume that $B_p$ is positive,
and write the total field strength as
\begin{equation}
B=(B_p^2+B_\varphi^2)^{1/2} \equiv \alpha B_p,
\quad \alpha\ge1.
\end{equation}
Assume further that the rotation is in the positive sense, which means
that the toroidal component of the field will be negative:
\begin{equation}
B_\varphi = -(B^2-B_p^2)^{1/2}=-\frac{(\alpha^2-1)^{1/2}}{\alpha}B
=-(\alpha^2-1)^{1/2}B_p.
\end{equation}

Let us write the comoving magnetic field in terms of the lab-frame magnetic
field,
\begin{equation}
b = (B^2-E^2)^{1/2} = \delta B, \quad 0 < \delta \le 1.
\end{equation}
We will see below that $\delta$ determines the drift velocity of
magnetic field.  The electric field is given by
\begin{equation}
\myvec E = -\frac{\Omega R}{c}\,\hat n_\varphi \times \myvec B =
-\frac{\Omega RB_p}{c}\,\hat n_\varphi \times \hat n_p.
\end{equation}
From this it follows that
\begin{equation}
\Omega R = \alpha(1-\delta^2)^{1/2} c.
\label{OmegaR}
\end{equation}
We then obtain the drift speed,
\begin{equation}
\myvec v_d=c\,\frac{\myvec E\times\myvec B}{B^2} =
-\frac{\Omega RB_p}{B^2}\,\left(-B_p\hat n_\varphi +B_\varphi \hat n_p\right)
\equiv v_{dp}\hat n_p + v_{d\varphi}\hat n_\varphi,
\end{equation}
where the two velocity components are given by
\begin{align}
v_{dp} &= \frac{(\alpha^2-1)^{1/2}(1-\delta^2)^{1/2}}{\alpha}\,c, \\
v_{d\varphi} &= \frac{(1-\delta^2)^{1/2}}{\alpha}\,c.
\end{align}
The Lorentz factor corresponding to the drift speed is then
\begin{equation}
\gamma = \frac{1}{(1-v_d^2/c^2)^{1/2}} = \frac{1}{\delta}.
\label{gammad}
\end{equation}
We now see that $\delta$ is simply the inverse of the drift Lorentz
factor.

In \S\ref{sec_lorentz_factor_general} we showed that there are two distinct
regimes of acceleration in the jet.  In the first regime, we have
$\gamma = (1+\Omega^2R^2/c^2)^{1/2}$.  Making use of equations
(\ref{OmegaR}) and (\ref{gammad}), we obtain
\begin{equation}
{\rm First~regime:}\quad \delta=\frac{1}{\alpha}.
\end{equation}
In the second regime, the Lorentz factor depends on the shape of the
field line, i.e., it is no longer a local quantity but depends on
spatial derivatives of the field components.  Nevertheless, we
know that $\gamma$ in this regime is smaller than for the
first regime.  Therefore, we have
\begin{equation}
{\rm Second~regime:}\quad \delta > \frac{1}{\alpha}.
\end{equation}
Combining the two, we have the condition
\begin{equation}
{\rm Either~regime:}\quad \delta \geq \frac{1}{\alpha}.
\label{either}
\end{equation}

The fluid velocity $\myvec v_f$ is constrained to be equal to $\Omega
R\,\hat n_\varphi$ plus an arbitrary velocity parallel to $\myvec B$.  This
is expressed by the following condition on the poloidal and toroidal
components of the fluid velocity
\begin{equation}
v_{f\varphi}=\Omega R+\frac{B_\varphi}{B_p}\,v_{fp} =
\alpha(1-\delta^2)^{1/2}c-(\alpha^2-1)^{1/2}v_{fp}.
\end{equation}
Note in particular that the two components of the drift velocity
satisfy this condition.  Let us now rewrite the fluid velocity
in terms of the drift velocity as follows,
\begin{align}
v_{fp} &= v_{dp}+\epsilon c = \left[\frac{(\alpha^2-1)^{1/2}
(1-\delta^2)^{1/2}}{\alpha} + \epsilon\right]\,c, \\
v_{f\varphi} &= \left[\frac{(1-\delta^2)^{1/2}}{\alpha} -
(\alpha^2-1)^{1/2} \epsilon\right]\,c,
\end{align}
where $\epsilon$ is a small parameter which determines the deviation of the fluid
velocity from the drift velocity.
The fluid Lorentz factor is then
\begin{equation}
\gamma_f = \frac{1}{[1-(v_{fp}^2+v_{f\varphi}^2)/c^2]^{1/2}} =
\frac{1}{(\delta^2-\alpha^2\epsilon^2)^{1/2}}.
\label{gammaf}
\end{equation}
Thus, if we want the fluid motion to be physical ($v_f<c$), the following
condition must be satisfied,
\begin{equation}
|{\epsilon}| < \frac{\delta}{\alpha} = \frac{1}{\alpha\gamma}.
\label{epsiloncondition}
\end{equation}

We now use the Bernoulli equation to fix the value of the parameter
$\epsilon$.
According to the Bernoulli and angular momentum conservation equations, the following quantity must be
conserved along each field line~\citep{bekenstein_new_conservation_1978,mestel_shibata_pulsar_1994,con95,ckf99}:
\begin{equation}
\mu=\gamma_f\left(1-\frac{\Omega Rv_{f\varphi}}{ c^2}\right) = {\rm constant}.
\end{equation}
Near the base of the jet the flows that we consider in this paper
are sub-Alfv\'enic, so
we expect the fluid motion to be not very
relativistic.  Therefore, we expect $\mu$ to be $\sim 1$.  For
instance, if the fluid moves at the drift velocity at the base of the
jet, then we have
\begin{equation}
\mu = \left[\gamma_f\left(1-\frac{\Omega Rv_{f\varphi}}{ c^2}\right)
\right]_0 = \delta_0 = \frac{1}{\gamma_{0}},
\end{equation}
where the subscript 0 refers to values at the base of the jet.  For
most problems of interest to us, $\gamma_{0}\sim1$, so we have
$\mu\sim1$.

Now let us apply the $\mu$ constraint at a general point in the
force-free jet.  This gives the condition
\begin{equation}
\mu=\frac{\delta^2+\alpha(1-\delta^2)^{1/2}(\alpha^2-1)^{1/2}\epsilon }{
(\delta^2-\alpha^2\epsilon^2)^{1/2}},
\end{equation}
which is a quadratic equation for $\epsilon$.  The quadratic is easily
solved and gives the somewhat messy result,
\begin{equation}
\epsilon = \frac{\delta(\mu^2-\delta^2) }{ \alpha\mu(\alpha^2+\mu^2-1-
\alpha^2\delta^2)^{1/2}+\alpha\delta(\alpha^2-1)^{1/2}(1-\delta^2)
^{1/2}}.
\label{exact}
\end{equation}

So far, we have not made any approximations; all the results are exact.
Now, for simplicity, let us assume that we are in an asymptotic region
of the jet where $\alpha\gg1$ and $\gamma\gg1$.  In this limit, the
first term in the denominator of~(\ref{exact}) dominates over the
second.  We thus have
\begin{equation}
\epsilon \approx \frac{\mu}{\alpha^2\gamma},
\end{equation}
which clearly satisfies the condition (\ref{epsiloncondition})
that implies $v_f<c$.  Thus if $v_d<c$ in the force-free solution, then the
total fluid velocity satisfies $v_f<c$ even in the limit of $\sigma\to\infty$.
Substituting this solution into (\ref{gammaf}), we find that the
deviation of the fluid Lorentz factor from the drift Lorentz factor is
\begin{equation}
\Delta\gamma = \gamma_f-\gamma\approx
\frac{\gamma}{2\alpha^2}\,\mu^2 \ll 1.
\label{Deltagamma}
\end{equation}
The inequality on the right follows from the fact that $\alpha \geq
1/\delta=\gamma\gg1$ (see eq. \ref{either}) and $\mu\sim1$.

Equation (\ref{Deltagamma}) shows that any fluid which is carried
along with an infinitely magnetized force-free flow has only a
slightly modified Lorentz factor relative to the drift speed.  In
other words, the perturbation to the plasma motion is negligibly small
and we can, for all practical purposes, assume that the fluid has the
same speed as the drift speed of the electromagnetic fields.

\label{lastpage}
\end{document}